\documentclass[aps,prc,reprint,floatfix,groupedaddress,nofootinbib]{revtex4-1}

\usepackage{graphicx}
\usepackage{hyperref}

\begin{document}

%\preprint{}

\title{Improved determination of the $\beta$-$\overline{\nu}_e$ angular correlation coefficient $a$\\ in free neutron decay with the $a$SPECT spectrometer}


\author{M.~Beck}
\author{F.~Ayala~Guardia}
\author{M.~Borg}
\author{J.~Kahlenberg}
\author{R.~Mu\~noz Horta}
\author{C.~Schmidt}
\author{A.~Wunderle}
\author{W.~Heil}
\email[]{Corresponding author, email: wheil@uni-mainz.de}
\affiliation{Institut f\"ur Physik, Johannes Gutenberg-Universit\"at Mainz, 55128 Mainz, Germany}
\author{R.~Maisonobe}
\author{M.~Simson}
\author{T.~Soldner}
\author{R.~Virot}
\author{O.~Zimmer}
\affiliation{Institut Laue-Langevin, CS 20156, 38042 Grenoble Cedex 9, France}
\author{M.~Klopf}
\author{G.~Konrad}
\affiliation{Technische Universit\"at Wien, Atominstitut , 1020 Wien, Austria}
\author{S.~Bae\ss{}ler}
\affiliation{Department of Physics, University of Virginia, Charlottesville, VA 22904, USA}
\affiliation{Oak Ridge National Lab, Bethel Valley Road, Oak Ridge, TN 37831, USA}
\author{F.~Gl\"uck}
\affiliation{Institut f\"ur Kernphysik (IKP), Karlsruhe Institute of Technology (KIT), 76344 Eggenstein-Leopoldshafen, Germany}
\author{U.~Schmidt}
\affiliation{Physikalisches Institut, Universit\"at Heidelberg, 69120 Heidelberg, Germany}


\date{\today}

\begin{abstract}
We report on a precise measurement of the electron-antineutrino angular correlation ($a$ coefficient) in free neutron beta-decay from the $a$SPECT experiment. The $a$ coefficient is inferred from the recoil energy spectrum of the protons which are detected in 4$\pi$ by the $a$SPECT spectrometer using magnetic adiabatic collimation with an electrostatic filter. Data are presented from a 100 days run at the Institut Laue Langevin in 2013. The sources of systematic errors are considered and included in the final result. We obtain $a = -0.10430(84)$ which is the most precise measurement of the neutron $a$ coefficient to date. From this, the ratio of axial-vector to vector coupling constants is derived giving $|\lambda| = 1.2677(28)$.
\end{abstract}

%%\pacs{...}
%\keywords{}

\maketitle

\section{Introduction}

The free neutron presents a unique system to investigate the standard model of particle physics (SM). Its $\beta$-decay into a proton, an electron and an electron-antineutrino is the prototype semileptonic decay. The low decay energy allows a simple theoretical interpretation within the Fermi theory, which is a very good approximation of the underlying field theory at low energies. Due to the absence of nuclear structure this decay is easy to interpret with only minor theoretical corrections compared to nuclear $\beta$-decays. 

While the neutron lifetime gives the overall strength of the weak semileptonic decay, neutron decay correlation coefficients depend on the ratio of the coupling constants involved, and hence determine its internal structure. Today, neutron $\beta$-decay gives an important input to the calculation of semileptonic charged-current weak interaction cross sections needed in cosmology, astrophysics, and particle physics. With the ongoing refinement of models, the growing requirements on the precision of these neutron decay data must be satisfied by new experiments.

The $a$SPECT experiment \cite{bae2008,glu2005,zim2000} has the goal to determine the ratio of the weak axial-vector and vector coupling constants $\lambda = g_{\text{A}}/g_{\text{V}}$ from a measurement of the $\beta$-$\overline{\nu}_e$ angular correlation in neutron decay. The $\beta$-decay rate when observing only the electron and neutrino momenta and the neutron spin and neglecting a T-violating term is given by \cite{jtw1957}
\begin{eqnarray}
d^3\Gamma & \sim & G_F^2 V_{\text{ud}}^2 (1+3\lambda^2) p_e E_e (E_0 - E_e)^2 \nonumber \\
 & &\times \left( 1 + a \frac{\vec{p_e} \cdot \vec{p_\nu}}{E_e E_\nu} + b \frac{m}{E_e} + \frac{\vec{\sigma_n}}{\sigma_n} \cdot \left[ A\frac{\vec{p_e}}{E_e} + B\frac{\vec{p_\nu}}{E_\nu}  \right]\right) \nonumber\\
 & &\times dE_e d\Omega _e d\Omega _\nu
\label{eqn:equations}
\end{eqnarray}

with $\vec{p_e}$, $\vec{p_\nu}$, $E_e$, $E_\nu$ being the momenta and energies of the beta electron and the electron-antineutrino, $m$ the mass of the electron, $G_F$ the Fermi constant, $V_{\text{ud}}$ the first element of the Cabibbo-Kobayashi-Maskawa (CKM) matrix, $E_0$ the endpoint decay energy and $\vec{\sigma}_n$ the spin of the neutron. $b$ is the Fierz interference coefficient. It vanishes in the purely vector axial-vector ($V-A$) interaction of the SM since it requires scalar ($S$) and tensor ($T$) interaction (see {\it \textit{e.g.}} \cite{severijns2006,vos2015}). The correlation coefficients $a$ and $A$ are most sensitive to $\lambda$ and can be used for its determination. The SM dependence of the beta-neutrino angular correlation coefficient $a$ on $\lambda$ is given by \cite{jtw1957, abele2008}
\begin{equation}
\label{eq:lambda}
a = \frac{1-\left|\lambda\right|^2}{1+3\left|\lambda\right|^2}
\end{equation}
To date, the most accurate value of $\lambda$ has been extracted from measurement of the $\beta$-asymmetry parameter $A$ \cite{perkeo2013, maerk2018, brown2018}. However, determining $\lambda$ from $a$ yields complementary information since the experimental systematics are different and systematic effects are relevant in this type of high precision experiments. 

 $\lambda$ together with the neutron lifetime $\tau_n$ can be used to test the unitarity of the top row of the CKM matrix \cite{abele2002, townerhardy2015} since it yields its first element $V_{\text{ud}}$ according to~\cite{czar2018, marc2006, seng2018, seng2019, czarnecki2019}:
\begin{equation}
\label{eq:unitarity}
\left|V_{\text{ud}}\right|^2 = \frac{(4905.7 \pm 1.7) \; \text{s}}{\tau_n \left( 1+3\left|\lambda\right|^2 \right)}\quad ,
\end{equation}
with the recent updates on the radiative corrections from \cite{czarnecki2019}.

The neutron decay determination of $V_{\text{ud}}$ is compelling as it is free of isospin breaking and nuclear
structure corrections. Within the SM, neutron beta decay is described by two parameters only, \textit{i.e.,} $V_{\text{ud}}$ and $\lambda$. Since more than two observables are accessible, the redundancy inherent in the SM description allows  uniquely sensitive checks of the model's validity and limits \cite{DUBBERS1991173,profumo07,Konrad2010,Bhattacharya2012,Cirigliano2013,gonz2019}, with strong implications in astrophysics \cite{dubbers2011}. Of particular interest in this context are the search for right-handed currents and for $S$ and $T$ interactions where the various correlation coefficients exhibit different dependencies. These investigations at low energy in fact are complementary to direct searches for new physics beyond the SM in high-energy physics (see \textit{e.g.} \cite{Bhattacharya2012,Cirigliano2013,gupta2018}).

The present precision of $a$ measurements is $\Delta a/a \approx 3\%$ taking the PDG value $-0.1059(28)$ \cite{tanabashi2018, stratowa1978, byrne2002, darius2017}. The work with $a$SPECT presented here improved the measurement of the $\beta$-$\overline{\nu}_e$ angular correlation $a$ to $\Delta a/a \approx 1\%$.

\section{The Experiment}

\begin{figure}
    \includegraphics[width=\linewidth]{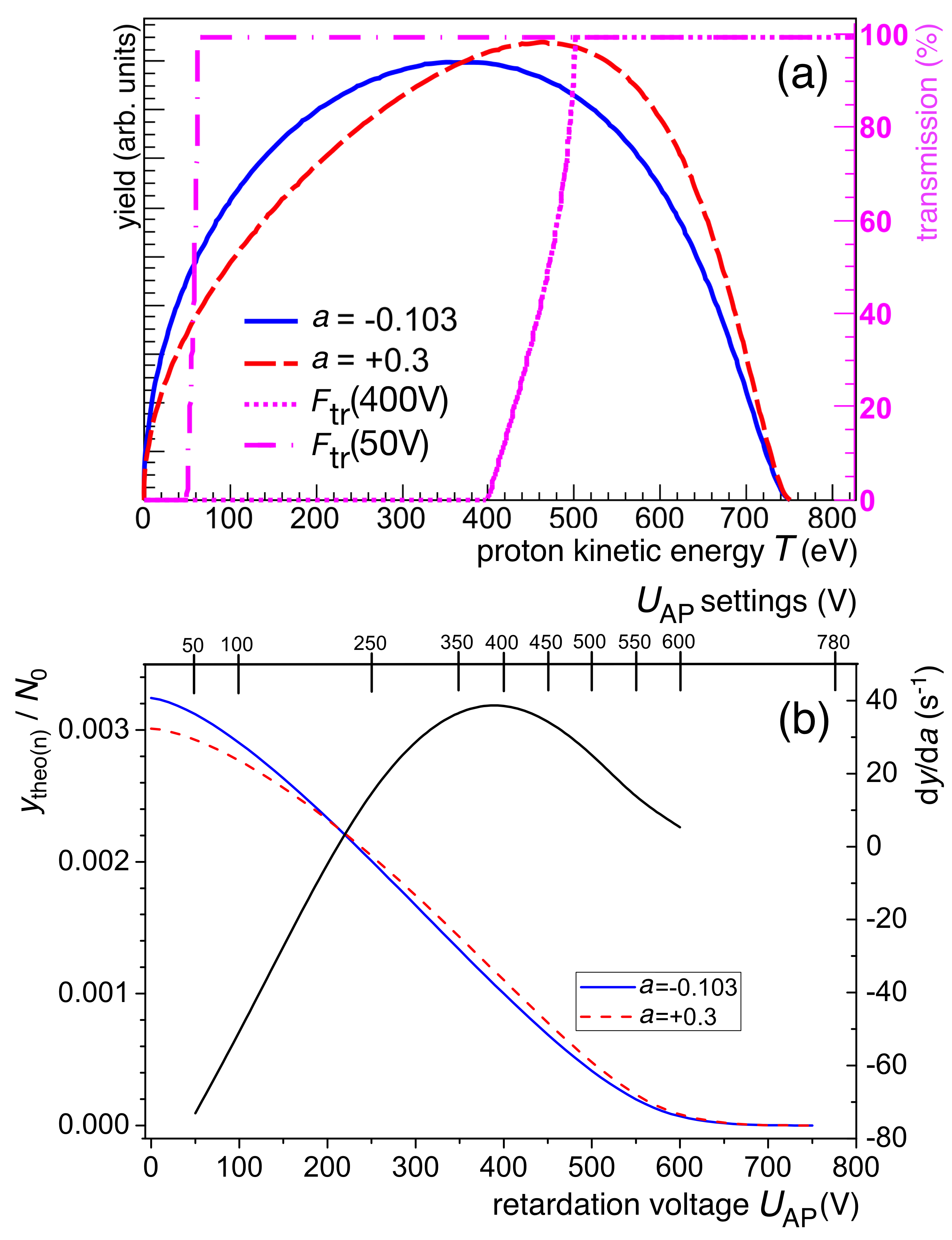}
    \caption{(a) Expected proton recoil spectrum for $a = -0.103$ (blue solid line) which we use in the following as reference value ($a_{\text{ref}}$) and for an extreme value of $a=+0.3$ (red dashed line). The decay proton has its maximum energy at $T_{\text{max}} = 751~\mathrm{eV}$. Also shown are the respective transmission functions $F_{\text{tr}}$ for the retardation potentials $U_{\text{AP}} = 50~\mathrm{V}$ and $U_{\text{AP}} = 400~\mathrm{V}$ and $r_{\text{B}} = 0.203$ (magenta lines). (b) The corresponding (normalized) integral proton spectra. Normalization means (see Eqs. (\ref{eq:fitn}), (\ref{eq:diffprot}), (\ref{eq:normfac})) that the area under the respective curve is 1, \textit{\textit{i.e.}}, does not depend on $a$. The derivative $dy/da$ (black solid curve) expresses the sensitivity of the yield $(y_{\text{theo(n)}}/N_0)$ to changes in $a$ which (in absolute numbers) is maximal at $U_{\text{AP}}$ voltage settings of $\approx 50$~V and $\approx 400$~V (top horizontal axis). We do not use a lower voltage setting ($U_{\text{AP}} < 50$ V), as it is potentially sensitive to background from the ionization of residual gas.\label{fig:spectrum}}
\end{figure}

At $a$SPECT the $\beta$-$\overline{\nu}_e$ angular correlation is inferred from the energy spectrum of the recoiling protons from the $\beta$-decay of free neutrons. The shape of this recoil energy spectrum is sensitive to $a$, due to energy and momentum conservation: the proton gains a large recoil energy when the electron and neutrino are emitted in the same direction (dominant process for positive $a$) and only a small recoil energy when they are emitted in opposite directions (dominant for negative $a$). The resulting differential energy spectrum is shown for two different values of $a$ in Fig.~\ref{fig:spectrum} (a). The recoil energy spectrum in turn is measured with a spectrometer using magnetic adiabatic collimation with an electrostatic filter (MAC-E filter) \cite{MAC-E-filter,picard-nimb,lobashev85}. Such a MAC-E filter collimates the momenta of charged particles, protons in the case of $a$SPECT, into the direction of the magnetic field by guiding them from a high magnetic field $B_0$ into a low magnetic field region $B_{\text{A}}$. The inverse magnetic mirror effect provides for a conversion of their transversal energy into longitudinal energy. In the low magnetic field most of the kinetic energy of the proton therefore resides in its longitudinal motion, which is then probed by an applied retardation voltage $U_{\text{AP}}$. A variation of the retardation voltage yields a measurement of the integral proton energy spectrum (Fig.~\ref{fig:spectrum} (b)). This technique in general offers a high luminosity combined with a high energy resolution at the same time. In order to extract a reliable value of the $\beta$-$\overline{\nu}_e$ angular correlation coefficient any effect that changes the shape of the integral proton energy spectrum has to be understood and quantified precisely. Examples are a.o. the transmission function of the MAC-E filter and background that depends on the retardation voltage.

\subsection{The transmission function}
\label{sec:transmissionfunc}

As long as the protons move adiabatically through the MAC-E filter, the ratio of radial energies at emission and retardation points is given by $1/r_{\text{B}}$, with $r_{\text{B}} := \frac{B_{\text{A}}}{B_0}$, where $B_0$ and $B_{\text{A}}$ are the magnetic fields at the place of emission and retardation, respectively. This amounts to the energy resolution of $a$SPECT. Hence, the transmission function $F_{\text{tr}}$ for isotropically emitted protons of initial kinetic energy $T$ is a function both of $U_{\text{A}}$ and $r_{\text{B}}$ \cite{glu2005,bae2008,kon2011}:
  \begin{equation}
    F_{\text{tr}} = \left\{ \begin{array}{ll} 0 & \textmd{if}~ T \le eU_{\text{A}} \\
      1 - \sqrt{1-\left( 1-\frac{eU_{\text{A}}}{T} \right) /r_{\text{B}}} & \textmd{if } eU_{\text{A}} < T < \frac{eU_{\text{A}}}{1-r_{\text{B}}} \\
      1 & \textmd{if}~ T \ge \frac{eU_{\text{A}}}{1-r_{\text{B}}}
    \end{array} \right. 
    \label{eq:tfunc}
  \end{equation} 

with $e$ the elementary charge and $U_{\text{A}} = \phi_{\text{A}} - \phi_0$, the potential difference between the place of retardation ($\phi_A$) and emission ($\phi_0$). The place of retardation, the so-called analysing plane (AP), is defined as the plane, in which the kinetic axial energy of the protons in the magnetic flux tube from the decay volume (DV) to the detector becomes minimal. The AP of \text{a}SPECT is a surface in $R^3$. It can be determined by particle tracking simulations given the known electric and magnetic field configurations. In case of homogeneous electric and magnetic fields inside the DV and AP electrode, the AP is nearly the midplane of the AP electrode.

In the ideal case $U_{\text{A}}$ is just the applied retardation voltage $U_{\text{AP}}$ between the DV and AP electrode (see Fig.~\ref{fig:aspecta}). In reality, the electric potentials $\phi_{\text{A}}$ and $\phi_0$ get slightly shifted and distorted by field leakage and locally different work functions of the electrodes creating these potentials. For the magnetic field ratio $r_{\text{B}}$, variations are caused by locally inhomogeneous $B$ fields in the DV and AP region. Hence, $U_{\text{A}}$ and $r_{\text{B}}$ depend on the individual proton trajectories $P_i$. Therefore, they get replaced in Eq.~(\ref{eq:tfunc})  by their averages $\langle U_{\text{A}} \rangle$ and $\langle r_{\text{B}} \rangle$, where the averages are over all trajectories of those protons that reach the detector\footnote{To be precise, one would have to find $\langle F_{\text{tr}} \rangle$ for an applied retardation voltage and initial kinetic energy $T$. Access to $\langle F_{\text{tr}} \rangle$ including $\langle U_{\text{A}} \rangle$ and $\langle r_{\text{B}} \rangle$ is provided by particle tracking simulations, where we find with sufficiently high accuracy the following relation to Eq. \ref{eq:tfunc}: $\langle F_{\text{tr}} \rangle = F_{\text{tr}}\left(T, \langle U_{\text{A}} \rangle, \langle r_{\text{B}} \rangle \right)$.}. For details on the determination of $\langle r_{\text{B}} \rangle$ and $\langle U_{\text{A}} \rangle$, see sections \ref{sec:rb} and \ref{sec:ua}. For more details on the transmission through MAC-E filters and the influence of the field configuration, see \cite{glu2005,glu2013}.

The uncertainties of $\langle U_{\text{A}} \rangle$ and $\langle r_{\text{B}} \rangle$ form the principal systematic uncertainties of $a$SPECT, albeit not the only ones. Two examples of transmission functions for $a$SPECT are included in Fig.~\ref{fig:spectrum} (a). Simulations show \cite{glu2005,kon2011} that the sensitivity of the measured $a$ values on $\langle U_{\text{A}} \rangle$ and $\langle r_{\text{B}} \rangle$ is given by $\Delta a/a \approx$ $1.4\times{}10^{-4} \times \Delta \langle U_{\text{A}} \rangle /\mathrm{mV}$ and $\Delta a/a \approx$ $5.5 \times \Delta \langle r_{\text{B}} \rangle / \langle r_{\text{B}} \rangle$. Therefore, a shift of $\Delta \langle U_{\text{A}} \rangle$ $\approx 80~\mathrm{mV}$ or $\Delta \langle r_{\text{B}} \rangle / \langle r_{\text{B}} \rangle \approx{}10^{-3}$ corresponds to a shift $\Delta a/a$ $\approx 1~\%$. 

\subsection{Experimental set-up}
\label{sec:setup}

In 2013 $a$SPECT was set-up for a production beam time at the cold neutron beam line of PF1b \cite{abele2006} at the Institut Laue Langevin in Grenoble, France. Here we present the basic layout of the $a$SPECT experiment. Details are discussed in \cite{bae2008,glu2005,zim2000} and \cite{csh2017,wun2016,mai2014,kon2011,gua2011,bor2010,sim2010,mun2011}. Modifications of the experimental arrangement used for the measurement in 2013 with respect to the ones presented in the previous articles are shortly mentioned at the relevant places.

\begin{figure}
    \includegraphics[width=\linewidth]{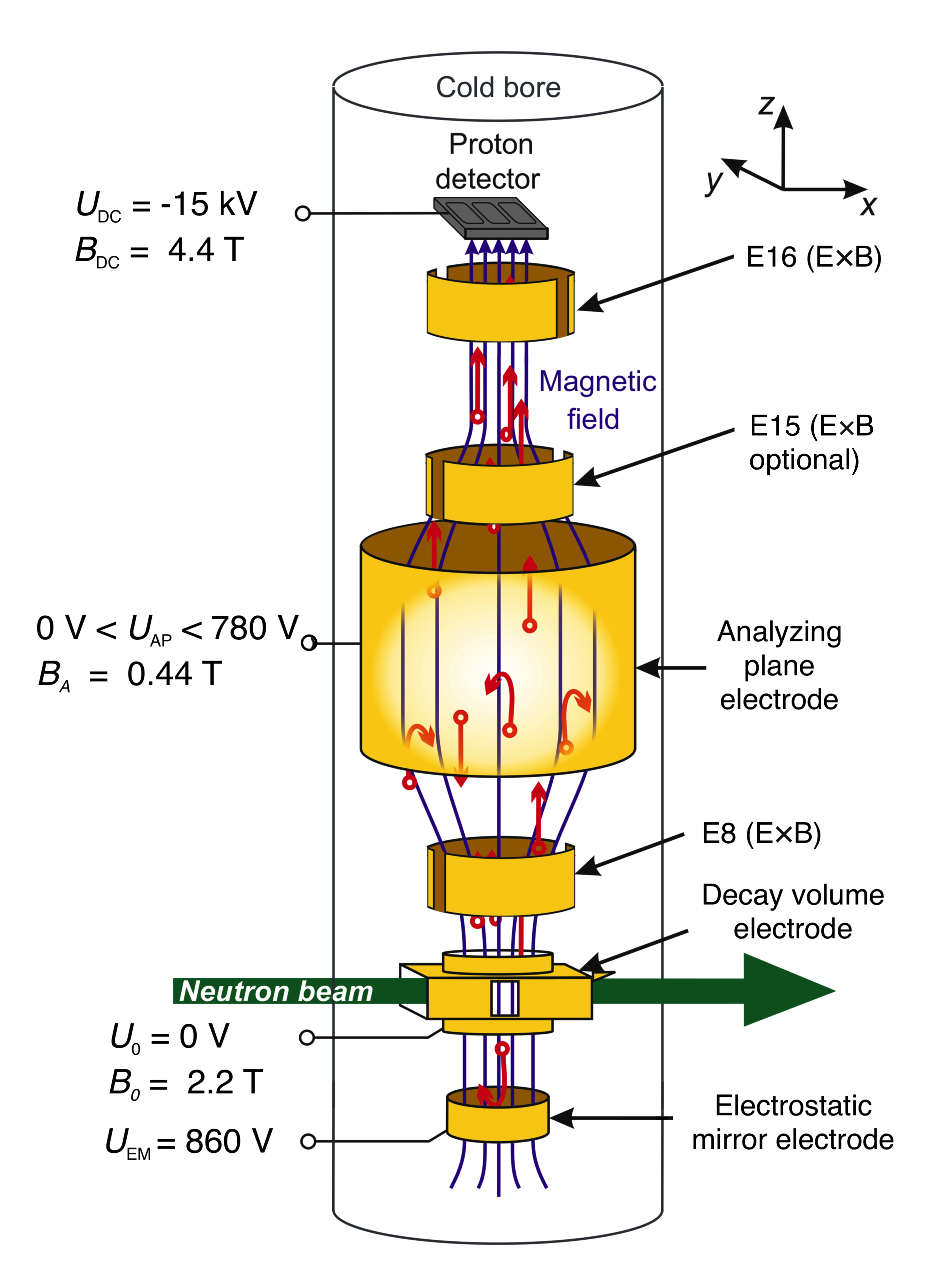}
    \caption{Schematic of $a$SPECT. Only the most important electrodes are shown. The magnetic field is oriented in vertical direction (blue lines). The whole set-up is under ultra-high vacuum conditions.\label{fig:aspecta}}
\end{figure}

\begin{figure}
\includegraphics[width=\linewidth]{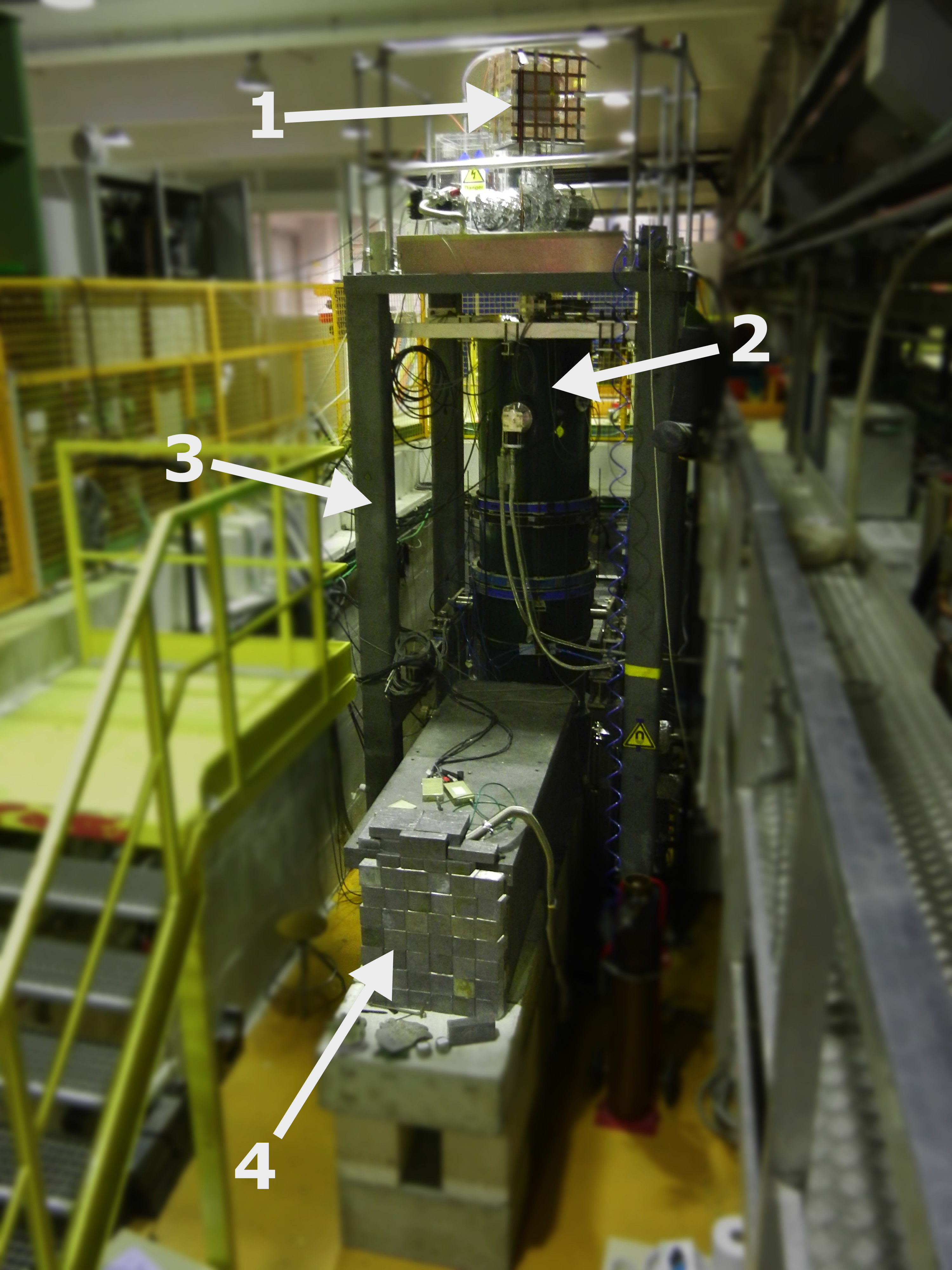}
\caption{The $a$SPECT set-up at the cold neutron beam line of PF1b at the ILL in 2013. Clearly visible are (1) the detector electronics on top, (2) the vertically aligned $a$SPECT cryostat system, (3) the massive magnetic field return yoke for the magnetic shielding and (4) the beam dump in front.\label{fig:pf1b}}
\end{figure}

A schematic of the 2013 $a$SPECT spectrometer is shown in Fig.~\ref{fig:aspecta}. The longitudinal magnetic field of the MAC-E filter is created by a superconducting multi-coil system oriented in vertical direction \cite{bae2008}. The neutron beam enters horizontally in the lower part of the $a$SPECT spectrometer at the height of the high magnetic field $B_0$ and is guided through the DV electrode towards the beam dump further downstream. Protons and electrons from neutron decays inside the DV electrode are guided adiabatically along the magnetic field lines. Downgoing protons are converted into upgoing protons by reflection off an electrostatic mirror electrode (EM) at $U_{\text{EM}} = 860~\mathrm{V}$ (Table~\ref{tab:potentials}) below the DV electrode, providing a $4\pi$ acceptance of $a$SPECT. The protons are guided magnetically towards the AP inside the main AP electrode (E14 in Table~\ref{tab:potentials}). Protons with sufficient energy pass through the AP and are focused onto a silicon drift detector (SDD) both magnetically and electrostatically. A reacceleration voltage of $U_{\text{DC}} = -15~\mathrm{kV}$ applied to an electrode surrounding the detector, the so-called detector cup (DC) electrode, is used in order to be able to detect the protons. A photograph of the set-up at PF1b is shown in Fig.~\ref{fig:pf1b}.

\begin{table}[h]
  \flushright
  \caption{Typical voltage settings with respect to the DV during the 2013 beam time. In the case shown here E15 was not used as E$\times$B drift electrode (symmetric). When operated in E$\times$B mode to reduce background, side L of E15 was switched to the same voltage as E11 (asymmetric). For details of the electrodes see~\cite{bae2008}.}
  \begin{center}
    \begin{tabular}{lcl}
    \hline
    \hline
    Electrode & Voltage [V] & Comments \\
    \hline
    E1 & +860 & EM mirror electrode \\
    E8 (L/R) & -1/-200 & Lower E$\times$B drift electrode \\
    E10 & $0.4 \times U_{\text{AP}}$ & Variable \\
    E11 & $0.7125 \times U_{\text{AP}}$ & Variable \\
    E12 & $0.9 \times U_{\text{AP}}$ & Variable \\
    E13 & $0.9925 \times U_{\text{AP}}$ & Variable \\
    E14 & $U_{\text{AP}}$ & Main AP electrode, variable \\
    E15 (L\&R) & $0.9875 \times U_{\text{AP}}$ & E$\times$B drift electrode\\
    & & (optional), variable\\
    E16 (L/R) & $-1750/-2250$ & Upper E$\times$B drift electrode\\
    E17 & -15000 & DC electrode\\
    \hline
    \hline
    \end{tabular}%
  \label{tab:potentials}%
  \end{center}
\end{table}%

\begin{figure}
     \includegraphics[width=\linewidth]{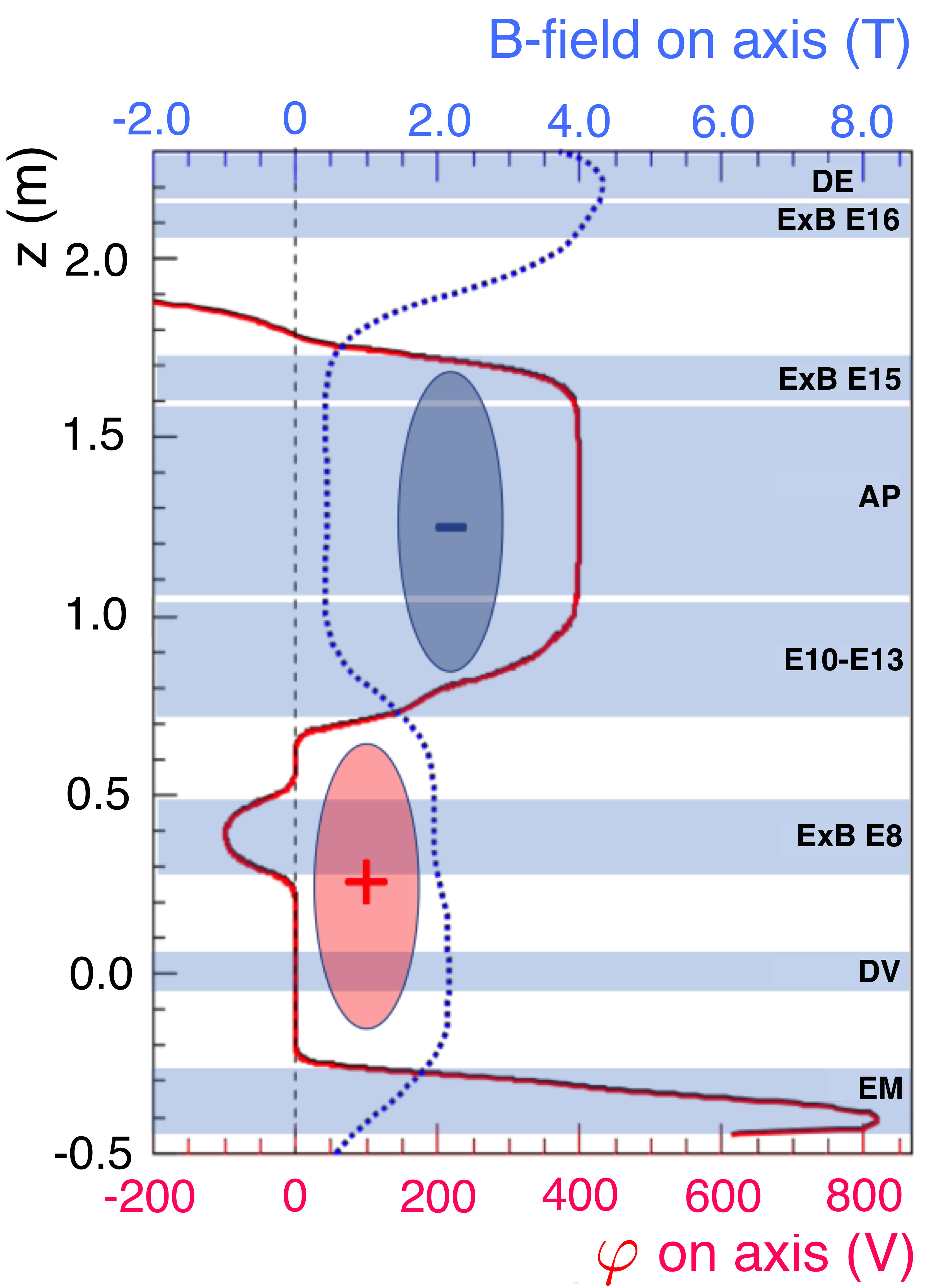}
     \caption{Fields inside $a$SPECT along the vertical direction (z-axis). The magnetic field is shown in blue (dotted curve), the electric potential in red. The position of the electrodes as mentioned in the text and listed in Table~\ref{tab:potentials} are indicated by horizontal bars (in blue). The most important Penning-like traps for  positively (+) and negatively (-) charged particles inside $a$SPECT are indicated as ellipses.\label{fig:aspectb}}
\end{figure}

The main superconducting coils are operated in persistent mode. Additionally, there are two superconducting correction coils in driven mode to create a small magnetic field gradient across the DV, as well as a combination of external air-cooled coils in Helmholtz and Anti-Helmholtz configuration in the AP region. 
For more details regarding the magnetic fields and the $a$SPECT magnet system, see \cite{glu2005,bae2008,gua2011,wun2016}. The whole set-up is surrounded by a magnetic field return yoke to reduce the stray magnetic field (see Fig.~\ref{fig:pf1b}), but does not affect significantly the internal magnetic field and its homogeneity \cite{kon2014}. During the beam time in 2013 the magnetic field was $B_0 \approx$ $2.2~\mathrm{T}$ in DV region, $B_{\text{A}} \approx$ $0.44~\mathrm{T}$ around the AP and $B_{\text{DC}} \approx$ $4.4~\mathrm{T}$ at the position of the detector.

The electrode system creating the electric potentials has been described in detail in \cite{glu2005,bae2008,gua2011,kon2011}. Between the DV and the main AP electrode the electrode system contains cylindrical electrodes with subsequently higher potential (electrodes E10 to E13 in Table~\ref{tab:potentials}). Their purpose is to avoid steep gradients of the electric potential to achieve a sufficiently adiabatic motion of the decay protons from the DV to the AP \cite{glu2005}. They also help to minimize field leakage into the main AP electrode (E14). The resulting electric potential along the vertical axis of the $a$SPECT spectrometer is shown in Fig.~\ref{fig:aspectb} together with the course of the magnetic field strength (for more details on field- and potential measurements/simulations in particular in the DV and AP region, see section~\ref{sec:rb}, as well as sections~\ref{sec:sim} and \ref{sec:ua} (3.)). Between the AP and DV electrode, the voltage $U_{\text{AP}}$ is applied. The applied voltage $U_{\text{AP}}$ is supplied by a precision power supply\footnote{FuG Elektronik GmbH model HCN 0,8M-800 (custom-modified for higher precision).}. A voltage divider further provides the voltages for the electrodes above and below the main AP electrode, see Table \ref{tab:potentials}. $U_{\text{AP}}$ is measured with a precision of $<13~\mathrm{mV}$ at a second connection to the main AP electrode using a precision voltmeter\footnote{Agilent model 3458A multimeter.} (section~\ref{sec:ua}). Typical voltages applied to the relevant electrodes during the 2013 beam time are shown in Table~\ref{tab:potentials}. The nomenclature is from \cite{bae2008}. Besides the new DV and AP electrodes major differences compared to \cite{bae2008} are the omission of the diaphragm electrode E7, the segmentation of the mirror electrode E1 into two parts for improved adiabatic motion during reflection of the protons \cite{kon2011}, and the change of E15 above the main AP electrode to a dipole electrode, cf.~Fig.~\ref{fig:aspecta}.

\begin{figure}
\includegraphics[width=\linewidth]{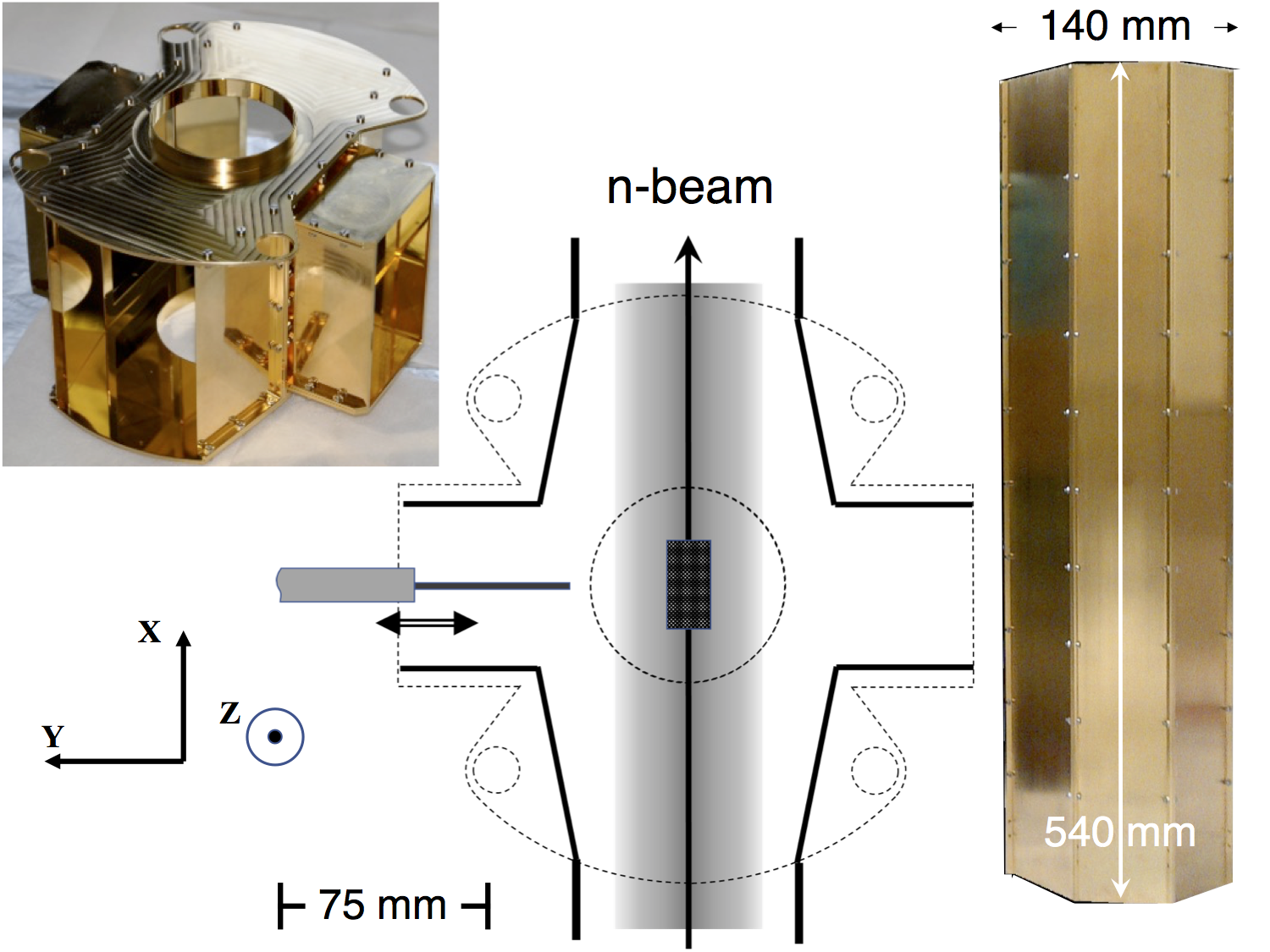}
\caption{Photograph of the DV electrode (left) and the main, octagonally shaped AP electrode (right) as used during the 2013 beam time. The sketch shows the collimated neutron beam (in gray) as it passes through the DV electrode. The dark area indicates the projection (to $z = 0$) of the fiducial decay volume in which the protons are magnetically focussed along the flux tube onto the two pads (2, 3) of the SDD. The side ports of the DV electrode are used for pumping and lateral access, \textit{e.g.}, beam profile measurements.\label{fig:APDVpicture}}
\end{figure}

The DC electrode as well as the upper E$\times$B drift electrode E16 are made of stainless steel (316LN), which has been electropolished to reduce field emission. Furthermore, the thickness of about 3 cm of the DC electrode housing the SDD reduces the environmental background seen by the detector. All other electrodes are made of OFHC copper (mostly CW009A). They are gold-coated galvanically with a thickness of $1~\mathrm{\mu m}$ and an underlayer of $10~\mu$m silver. Most electrodes have got a cylindrical shape. The DV and AP electrodes, in contrast, are made from flat segments (cf. Fig.~\ref{fig:APDVpicture}). This is one difference to previous set-ups of $a$SPECT. Flat electrodes lead to a more homogeneous workfunction on the electrode surface during manufacture \cite{kon2011,csh2017}. In addition, they allow a measurement of the work function of the electrodes using a scanning Kelvin probe, see Appendix~\ref{app:wf}. The DV and AP electrodes were made from the same slab of copper and the electrode surfaces were machined and treated identically\footnote{Except for the bottom plate of the DV electrode: this plate had a mechanical defect, a deep scratch. To remove this the plate was remachined some time after manufacture. This led to slightly different surface properties, visible in the work function measurements, see Appendix~\ref{app:wf}.}. Both the DV and AP electrodes were polished before coating using a non-magnetic polish.

In between beam times aging of the surfaces was observed due to diffusion of Cu into the Ag layer and to some extend into the final top layer of Au \cite{tom1976,pin1979}, leading to increased surface roughness contributing to increased field emission and as a result to an increased background during a beam time in 2011. As a consequence, the Au coating with its underlayer of Ag was simply renewed shortly before a scheduled beam time. Prior to the assembly all electrodes were cleaned in an ultrasonic bath using the cleaning sequence soap (P3 Almeco 36), deionized water, solvent (isopropyl), and again deionized water. Before final installation any visible dust that had accumulated on the electrodes was removed using lint-free tissue. Using the identical material, identical production procedures like machining, polishing and coating and handling the electrodes identically resulted in similar properties of the work function and its dependence on environmental conditions like the formation of surface adsorbates with their dependence on temperature and pressure. 
After the production beam time in 2013 and until the measurement of the work function of the electrodes with the Kelvin probe, the electrodes were stored in a commercial deep freeze at a temperature of $<-18^\circ$C. Since the diffusion coefficient follows an Arrhenius equation, the lower temperature effectively suppresses the aforementioned diffusion processes \cite{pin1972}. Additionally, the electrodes were enclosed individually in plastic bags filled with Argon to avoid contamination. The measured long-term stability of the work function of the electrodes after the beam time shows that these measures effectively suppressed the deterioration of the surfaces. Consequently,  no significant change of their work function was the finding, see Appendix~\ref{app:wf}.

Inside $a$SPECT, the neutron beam is shaped in front of the DV and further downstream towards the beam dump by several $^6$LiF apertures \cite{bor2010}. These apertures have been mounted originally on non-conductive Borosilicate glass plates. To avoid any potential charge up effect and therefore field leakage into the DV, the glass plates have been replaced by conductive plates made out of BN and TiB$_2$\footnote{ESK, DiMet Type 4.} \cite{mai2014,wun2016}. For the same reason the $^6$LiF apertures have been sputtered with Ti.

\begin{figure}
\centering
\includegraphics[width=\linewidth]{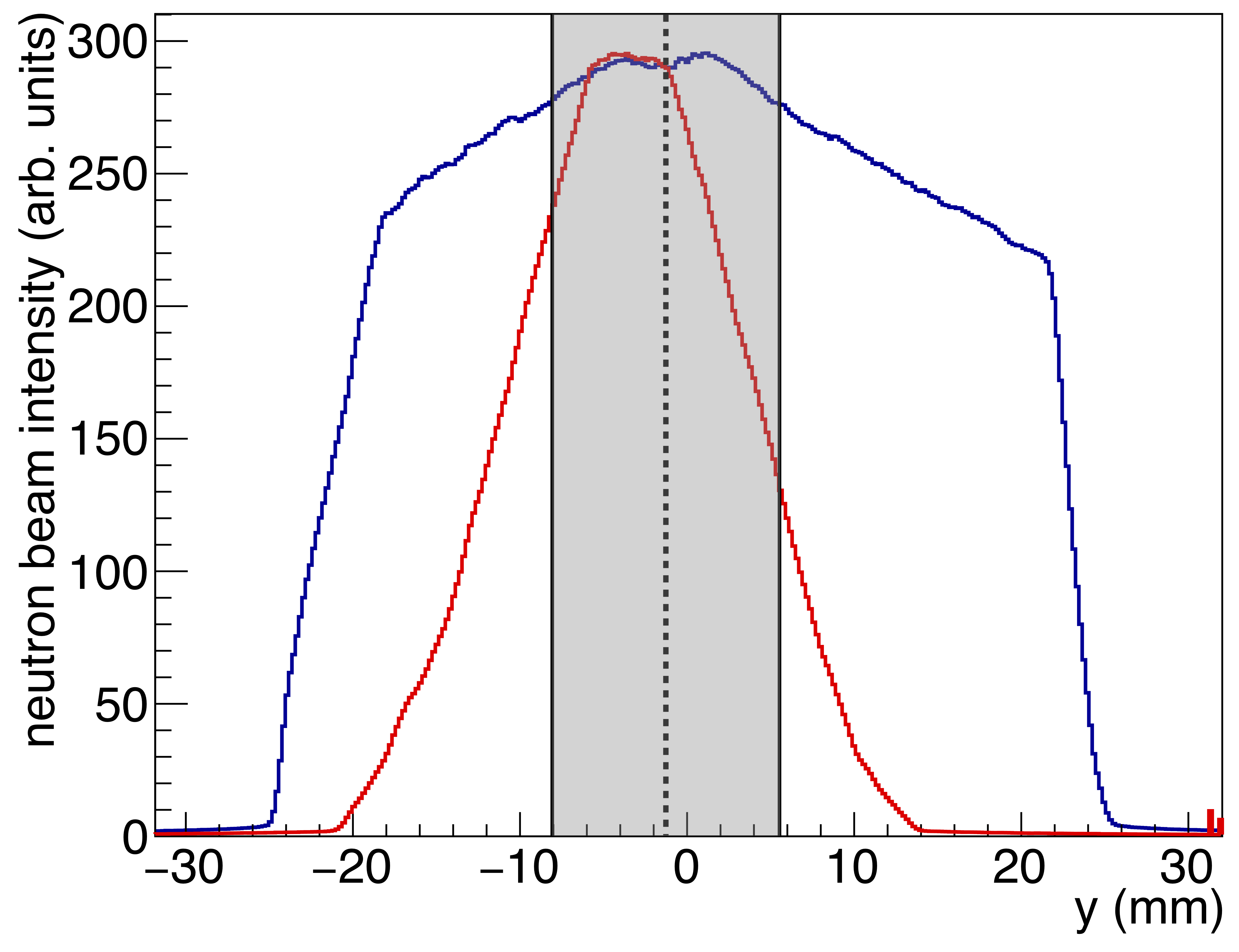}
\caption{Measured neutron beam profiles (projection onto the horizontal y-axis) for standard (blue) and reduced (red) beam size. The latter was used to investigate systematic effects. The shaded area indicates the magnetic projection of the central pad (2) of the SDD onto the y-axis in the DV (cf.~Eq.~(\ref{eq:radialdisp})).\label{fig:beamprofile}}
\end{figure}

\begin{figure}[h]
\includegraphics[width=\linewidth]{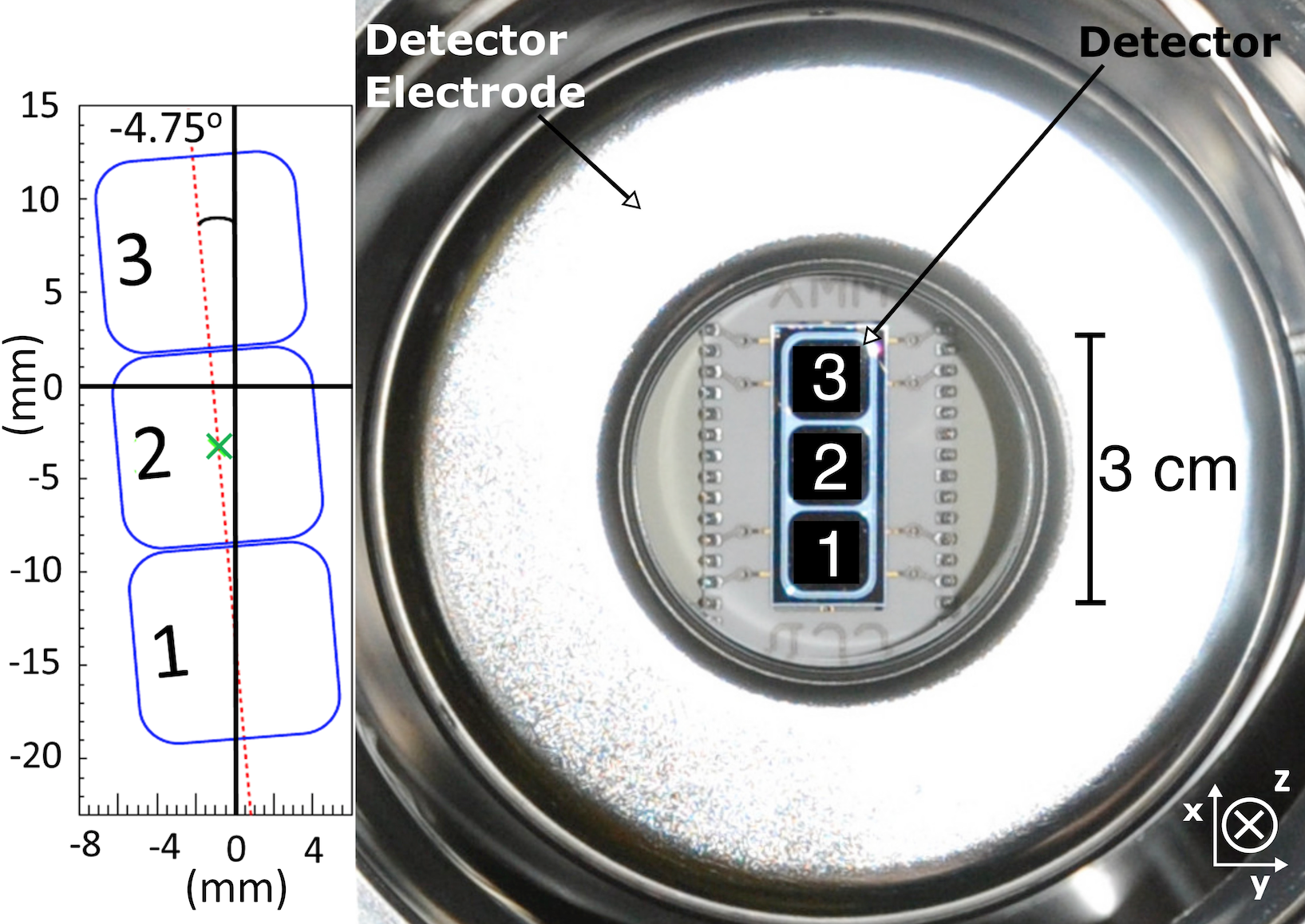}
\caption{The SDD with its three detector pads is mounted inside the detector cup electrode featuring a wall thickness of 3 cm, which is at the reacceleration potential of  $-15~\mathrm{kV}$. The alignment with respect to the $a$SPECT coordinate system was determined by dedicated measurements \cite{mai2014} and is shown on the left plot. Data from detector pad 1 could not be used for the analysis of $a$ (details see text).\label{fig:sdd}}
\end{figure}

A manipulator installed at the cross-piece on a side port of the spectrometer at the height of the DV electrode provides the possibility to insert probes into the center of the DV electrode (Fig.~\ref{fig:APDVpicture}). It was used, among others, to insert Cu foils for measurements of the neutron beam intensity profile inside the DV, removing the necessity to extrapolate from beam profile measurements further up- and downstream of the DV, which had introduced a significant uncertainty in the past. $^{63}$Cu and $^{65}$Cu of the foil are activated by neutrons from the beam with half-lives of $12.7~\mathrm{h}$ of $^{64}$Cu and $5.1~\mathrm{min}$ of $^{66}$Cu. The X-rays and $\beta^+$ particles of $^{64}$Cu in the activated Cu foil are imaged using a X-ray imaging plate and an image plate scanner. In Fig.~\ref{fig:beamprofile} the horizontal projection (y-axis) of the measured neutron beam profile is shown. Along the incident neutron beam the beam profile does not change, at least not across the effective neutron decay length of $\approx$~3 cm. This section is defined by the magnetic projection (in $x$-direction) of the decay protons along the flux tube onto the two detector pads (2, 3) of the SDD (cf.\ Fig.\ \ref{fig:sdd}).
A flux tube is a generally tube-like (cylindrical) region of space which fulfils $\int B\cdot{}dA = \text{const}$. Both the cross-sectional area ($A$) of the tube and the field contained may vary along the length of the tube, but the magnetic flux is always constant. Therefore, for the radial displacement ($r$) of the decay protons along the symmetry axis ($z$) of the $a$SPECT cryostat it follows to a good approximation:
\begin{equation}
r(z) = r_{\text{DV}}\cdot{}\sqrt{B_0/B(z)}
\label{eq:radialdisp}
\end{equation}
Also shown in Fig.~\ref{fig:beamprofile} is a distribution measured using a reduced beam profile. The latter was used to investigate an important systematic effect of $a$SPECT, the edge effect, see section~\ref{sec:ee}.

Inside the $a$SPECT system an ultra-high vacuum is maintained by means of cascaded turbomolecular pumps, one at the height of the DV electrode and two at the detector. The cold bore of the cryostat, with temperatures locally reaching down to $\approx$ $50~\mathrm{K}$, is acting as a cryo\-pump. Furthermore, good vacuum conditions are maintained by internal getter pumps\footnote{SAES type CapaciTorr C 400-2 DSK.} at the height of the lower E$\times$B electrode E8 and just below the DV electrode as well as an external getter pump\footnote{SAES type CapaciTorr C 500-MK5.} at the height of the DV electrode. With this vacuum set-up a pressure of $p_{\text{DV}} \approx$ $5\times10^{-10}~\mathrm{mbar}$ was achieved close to the DV electrode after several weeks of pumping. This is far below the critical pressure for proton scattering off residual gas (cf.~section \ref{sec:misc}, \cite{glu2005}). Despite the very good vacuum of $a$SPECT, the remaining residual gas gets ionised and trapped in Penning-like traps, created by the B- and E-fields of the spectrometer. The most prominent ones are indicated by ellipses in Fig.~\ref{fig:aspectb}. Stored protons, ions and electrons are removed to a large extent from these traps by two longitudinally split dipole electrodes, above the DV electrode (E8) and above the main AP electrode (E15) by their E$\times$B drift motion\footnote{Charged particles moving in crossed E- and B-fields have a drift motion perpendicular to both fields \cite{jackson}. Due to this E$\times$B drift, stored charged particles move outside of their storage volume, where they usually hit the electrode walls and are of no longer concern.}, see Fig.~\ref{fig:aspecta} and \ref{fig:aspectb}. Hence, the low vacuum level (the vacuum gradually improved during the whole production run) and the removal of stored particles by E$\times$B drifts reduces the retardation voltage-dependent background as one of the potential sources of systematics to an acceptable level. This background stems from positively charged rest gas ions ionized in the AP region (section~\ref{sec:bg}, \cite{mai2014,wun2016}). The E$\times$B electrode E16 does not serve for trap cleaning but is used to pre-accelerate protons which have passed the AP (ensuring that they overcome the increasing magnetic field) and to tune their alignment onto the detector.

The SDD for proton counting consists of an array of three detector pads of an area of $10 \times 10~\mathrm{mm^2}$ each\footnote{pnSensor UM-141101.}, see Fig.~\ref{fig:sdd} (\cite{sim2007,sim2010}). It has an entrance window of $30~\mathrm{nm}$ thickness made from aluminium. Use of a SDD with its intrinsic low electronic noise compared with Si PIN diodes, combined with a thin deadlayer, permits to lower the reacceleration voltage to $U_{\text{DC}} = -15~\mathrm{kV}$\footnote{With a kinetic energy of $\approx$ $15~\mathrm{keV}$, protons passing the 30~nm aluminium deadlayer (manufacturer specified) have a range of $\approx$ $200~\mathrm{nm}$ in silicon (section~\ref{sec:lld}).}. This significantly reduces field emission. The reacceleration voltage is provided by a high-voltage power supply\footnote{Type: FuG HCN 35-35000.}.

\begin{figure}[h]
\includegraphics[width=\linewidth]{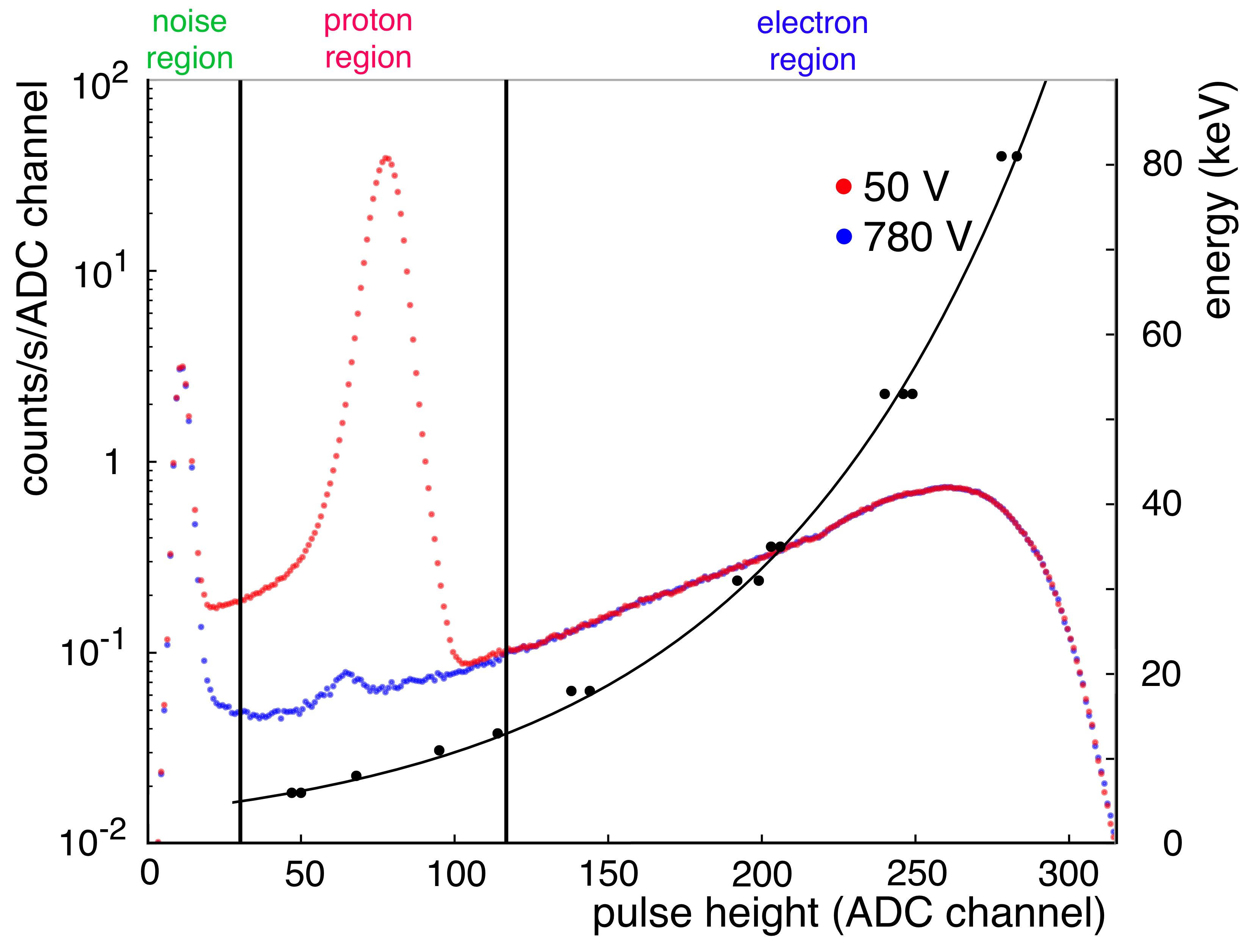}
\caption{Pulse height spectrum of protons and electrons from neutron decay (in red) measured at ILL in 2013 (config 1, cf. Table \ref{tab:configurations}). The proton peak is well separated from the noise. A background measurement at $U_{\text{AP}} = 780~\mathrm{V}$ is shown in blue. The small peak visible in the 780 V spectrum is caused by ionized rest gas and reduced in later configurations, see section \ref{sec:bg}. The two vertical lines denote the chosen lower (ADC channel: 29) and upper (ADC channel: 120) integration limits for the proton region. Demonstration of the logarithmic amplification of the SDD electronic (black solid curve) using characteristic X-rays of energy $E$ (black circles). The right axis indicates the X-ray energy.\label{fig:detspec}}
\end{figure}

Signals from the SDD are read out by a custom-built preamplifier and spectroscopy amplifier with logarithmic amplification (shaper). The shaped signals are digitized with a sampling ADC ($12~\mathrm{bit}$, $50~\mathrm{ns}$ resolution) \cite{bae2008,man2006,sim2009}. Figure~\ref{fig:detspec} shows a pulse height spectrum (cf. section~\ref{sec:uld}) taken during the beam time. The proton peak is well separated from the electronic noise. The SDD is also sensitive to the $\beta$-particles from the decay of the neutron. They are clearly visible above the proton region in Fig.~\ref{fig:detspec} and steadily continue into the proton region, as can be deduced from a measurement at $U_{\text{AP}} = 780~\mathrm{V}$, where all decay protons are blocked by the potential barrier. Low energetic $\beta$-particles, indeed, form the dominant background in the proton region, see Fig.~\ref{fig:detspec}. On the other hand, the highest energy $\beta$-particles from neutron decay will not lose all their energy in the active region of only $\approx$ $450~\mathrm{\mu m}$ (depending on their impact angle). Therefore and because of the logarithmic amplification, the $\beta$ spectrum trails off at intermediate $\beta$ energies. 

\begin{table*}[t]
  \flushright
  \caption{Typical voltage settings of the $a$SPECT spectrometer during the 2013 beam time. When operated in dipole mode to reduce background, different voltages are applied on side L and R of E15, with side L set to the same voltage as E11 (asymmetric case, see Table \ref{tab:potentials}). Configurations 3 to 7 use -200 V/-5 V instead of -1 V/-200 V for the lower E$\times$B electrode. In configuration 6, the electric field direction of the dipole electrode E8 was repeatedly inverted.}
  \begin{center}
    \begin{tabular}{ccl}
    \hline
    \hline
    Configuration name & Settings & Effect to be studied\\
    \hline
    Config 1 & Equal to Table~\ref{tab:potentials} & Standard data taking \\
    Config 2 & Repeatedly switched mirror on/off & Proton traps in DV, section~\ref{sec:mirror}\\
    Config 3 & E15L = E11 & $U_{\text{AP}}$-dependent background, section~\ref{sec:bg}\\
    Config 4 & Config 3 with reduced beam profile & Edge effect and $U_{\text{AP}}$-dependent background, section~\ref{sec:ee}\\
    Config 5 & Config 1 with reduced beam profile & Edge effect, section~\ref{sec:ee}\\
    Config 6 & Config 5, E8 repeatedly interchanged & Influence of lower E$\times$B on edge effect, section~\ref{sec:ee}\\
    Config 7 & Config 3 with E3=+4 V, E6=-4 V & To prevent proton traps in the DV, section~\ref{sec:mirror}\\
    \hline
    \hline
    \end{tabular}%
  \label{tab:configurations}%
  \end{center}
\end{table*}%

To determine the exact position of the detector with respect to the DV electrode, a copper wire of length $l\approx$ 8 cm aligned along the z-axis was mounted on the manipulator and then inserted into the DV electrode from the side ports. This wire was first activated in the neutron beam and then moved perpendicularly to the beam direction (beam off). By detecting the emitted electrons from the activated copper with the SDD, the magnetic projection of the detector in y direction onto the DV electrode was determined. In order to measure the corresponding magnetic projection of the detector in x direction, \textit{i.e.} along the beam direction, a second activated Cu wire ($l \approx$ 15 mm) placed parallel to the y-axis was scanned along the x-axis \cite{mai2014}.
These measurements showed that the DC electrode was not fully centred in the cryostat (cf.~Fig.~\ref{fig:sdd}). As a consequence, the magnetic flux tube from one of the detector pads, pad 1, was partially crossing one of the electrodes (E12) of $a$SPECT. This was confirmed off-line by particle tracking simulations. On the one hand, this pad therefore experienced a significantly higher and also fluctuating background. On the other hand, some of the decay protons  would scatter off this electrode, whereby they will lose an unspecified amount of energy. Therefore, the data from this detector pad could not be used for the analysis of $a$.

In a beam time in 2008 \cite{sim2009} saturation effects in the detector electronics caused by the high energetic $\beta$-particles from neutron decay were observed \cite{sim2010,kon2011}. This was solved by a reduction of the amplification of the preamplifier and a new spectroscopy amplifier with logarithmic amplification, see Fig.~\ref{fig:detspec}. The logarithmic amplification was checked using a $^{133}$Ba source and characteristic X-rays from Cu, Fe and Pb excited by the radiation from the $^{133}$Ba source. This improvement also allowed to measure the energy spectrum of the $\beta$-particles during the beam time in 2013 (see Fig.~\ref{fig:detspec}), limited at higher energies only by the thickness of the sensitive area of the detector of $450~\mathrm{\mu m}$.

Two systematic effects are associated with the proton detection: first, even though the proton energy at the detector varies only from $15~\mathrm{keV}$ to $15.75~\mathrm{keV}$, the energy-dependence of the backscattering of the protons at the SDD has to be taken into account at the precision needed for $a$SPECT (section~\ref{sec:lld}). Second, since the diaphragm E7 described in \cite{bae2008} has been omitted in the electrode system, the beam profile is much wider than the detector, see Fig.~\ref{fig:beamprofile}. Since the profile is non-uniform and asymmetric over the projected area of the detector, protons close to the edges of the detector may be falsely detected or lost depending on their radius of gyration and azimuthal phase with which they arrive at the SDD. This energy-dependent so-called edge effect has to be taken into account in the analysis (section~\ref{sec:ee}).

\section{Measurement with $a$SPECT}
\label{sec:meas}

Several beam times were taken with $a$SPECT at the cold neutron beam line of PF1b \cite{abele2006} at ILL. The beam time in 2008 showed that the spectrometer was fully operational but the aforementioned saturation effect of the detector prevented a result on $a$. This saturation effect was solved for a beam time in 2011. However, strong discharges, mostly inside the AP trap (Fig.~\ref{fig:aspectb}), again foiled a successful beam time: Temporal fluctuations of the measured background count rate, as well as their strong dependence on the retardation voltage precluded a meaningful data analysis. At times, an exponential increase in the background events was seen. To prevent saturation of the detector and to empty the trap, the retardation voltage had to be prematurely zeroed. Such Penning discharges in systems with good vacuum and crossed magnetic and electric fields can be initiated by field emission and may be self-amplifiying due to a feedback from secondary ionization of the residual gas under a range of specific conditions (see {\it \textit{e.g.}} \cite{art_trap}). Such discharges of similar high-voltage induced background have been observed at other experiments in the past \cite{witch2016,penning_katrin,perkeo2005}.
For $a$SPECT it was found that degradation of some electrode surfaces had caused increased field emission leading to these discharges. The above-mentioned improvements eliminated that problem. This was shown with measurements in 2012 in an offline zone in the ILL neutron hall, see \cite{mai2014}. The beam time of 100 days in 2013 then constituted the production measurement for a new determination of $a$.

\subsection{The measurement procedure}
\label{sec:proced}
The 2013 beam time consisted of measurement runs with a typical duration of half a day. Initially, the experimental settings were tuned and optimized. This included finding the settings for the E$\times$B electrodes to minimize the background and to optimize the steering of the protons onto the detector\footnote{The E$\times$B electrodes can steer the protons by  $\mathcal{O}$ (mm) at the place of the detector.} with respect to count rate, edge effect, etc.. After this optimization procedure the experimental settings were kept constant for several days in a row for measurements of $a$. Measurements runs with the same settings of electrodes and magnetic fields are grouped into a so-called configuration for the data analysis (see Table~\ref{tab:configurations}). In order to study the major systematic effects (section~\ref{sec:systematics}), dedicated measurements were taken at detuned settings of the electrodes and/or different beam profiles to study the enhanced effect.

\begin{figure}
\includegraphics[width=\linewidth]{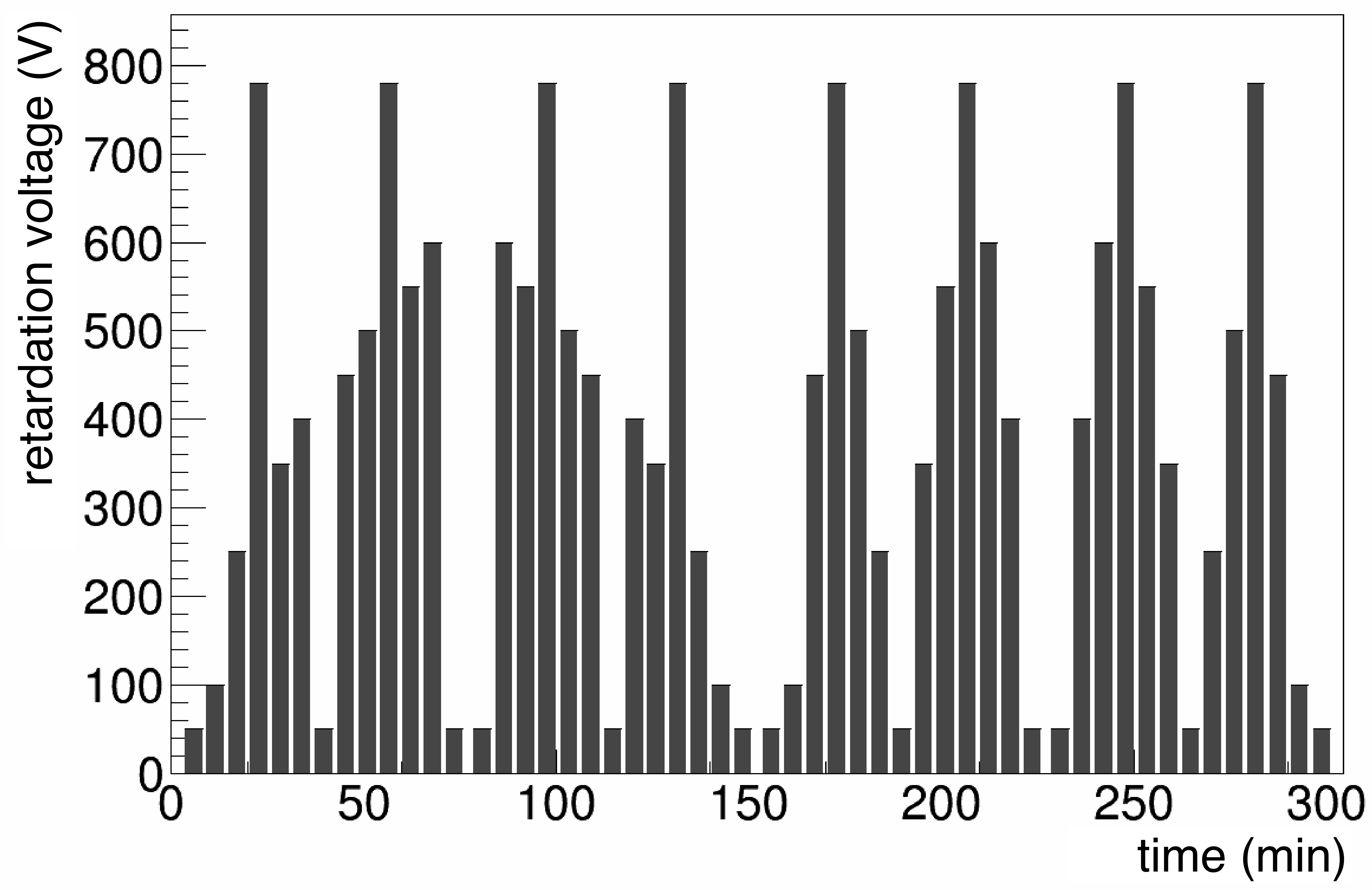}
\caption{Measurement sequence of the 2013 beam time which repeats after 300 min until data taking is stopped. Each bar in the diagram corresponds to a measurement cycle of $\approx$ 5 min duration for the respective value of $U_{\text{AP}}$ and has a time structure as shown in Fig.~\ref{fig:tempcycle} of section~\ref{sec:bg}.\label{fig:sequence}}
\end{figure}

Within a measurement run measurements were organized in sequences of applied voltages $U_{\text{AP}}$ that were repeated until a run was stopped. A typical measurement sequence used is shown in Fig.~\ref{fig:sequence}. In order to eliminate first order temporal drifts (time scale $>$ 30 min) during the measurements, {\it \textit{e.g.}} due to a variation of the neutron flux, the measurement sequence was not in ascending or descending order of $U_{\text{AP}}$ but alternated the voltage as shown.
Each measurement at a given voltage $U_{\text{AP}}$ in the measurement sequence consists of its own measurement cycle:
\begin{small}
\begin{itemize}
\item{Initially the neutron beam is blocked and $U_{\text{AP}}$ is set to $U_{\text{AP}}= 0$ V. Data taking starts at $t=0$. After $t_{\text{AP, on}} = 10$ s, the AP electrode is ramped up to $U_{\text{AP}}$ (cf. Table \ref{tab:configurations})\footnote{The time to ramp up (down) to $\approx$ 97\% of the full potential difference is about 5 s. The measurement cycle was only continued after reaching sufficient stability of $U_{\text{AP}}$ (using the feedback from the precision voltmeter \cite{sim2010}.}}.
\item{Between 20 s $\le{}t\le{}$ 40 s, instrumental- and environmetal-related background is measured.}
\item{At $t_{\text{n,on}}\approx$ 40 s , the neutron beam is switched on by means of a fast neutron shutter (B$_4$C) placed in the neutron beam line about 5 m upstream of the DV electrode \cite{mai2014}.}
\item{For pre-defined shutter opening times $t_{\text{op}}$ of 50 s, 100 s, and 200 s, the decay protons are counted (see Table~\ref{tab:cr50V}). After $t_{\text{n,off}}$, background is measured again for about $\Delta{}t_{\text{int}}\approx$ 20 s in order to extract a possible retardation voltage-dependent background (section~\ref{sec:bg}).}
\item{Approximatively 30 s after closing the shutter, $U_{\text{AP}}$ is ramped down again to ensure that stored particles in Penning-like traps (cf.~Fig.~\ref{fig:aspectb}) are definitely gone.}
\item{After another $\approx$ 50 s, data taking is completed for that particular measurement cycle. The individual sections of data acquisition add up to a total duration of about 5 min. The timing diagram of such a cycle is shown in Fig.~\ref{fig:tempcycle} of section~\ref{sec:bg} in which background contributions are discussed in more detail.}
\end{itemize}
\end{small}
Each measurement sequence contains an above-average number of 50 V and 780 V measurement cycles. The 50 V measurements with the highest proton count rate are needed with good statistics in order to normalize the integral proton spectrum and are also used  to check the temporal stability of the incoming neutron flux. The 780 V measuring cycles (cf.~Fig.~\ref{fig:detspec}) together with the recorded background measurements during shutter off serve for a complete background analysis (section~\ref{sec:bg}).

\subsection{Data analysis}

The measurements of Table \ref{tab:configurations} were used for the analysis of $a$. They include measurement configurations ($c =$ 1, 2 (ON), 3, 7) with changes of the optimal parameter settings in order to investigate their influence on $a$. In configurations $c =$ 4, 5, and 6, the neutron beam profile has been reduced to considerably enhance a major systematic effect, \textit{i.e.,} the edge effect. With $c=$ 2 (OFF) - mirror off in config 2 - the 4$\pi$ symmetry of proton detection was broken, increasing the sensitivity to trapped protons in the DV region as well as to non-isotropic emission of the protons with respect to the spin of the decaying neutron in case of a finite beam polarization.

The data analysis was performed for each detector pad ($p$) individually. For a given configuration ($c$), the pulse-height spectra of the individual measurement cycles with the same retardation voltage settings $i$ ($i = 1, \cdots, 10$ , in total) were added (counts) to a sum spectrum (cf.~Fig.~\ref{fig:detspec}). From these sum spectra the integral count rates  within the proton region can be calculated by dividing them by the measuring time accordingly. The proton region encloses the proton peak, which is located around pulse height channel 80. The lower integration limit was set at ADC channel 29 to exclude low energy electronic noise. The upper integration limit was set to safely include the high energy tail of the proton peak while minimizing the amount of $\beta$-electron events (background) in the proton region. Consequently, some fraction of the protons, tail events below the lower integration limit and backscattered protons, as well as some pile-up events above the upper integration limit are not counted but lost. How these loss effects have been taken care of is discussed in sections \ref{sec:lld} and \ref{sec:uld}, respectively. In the proton region, typical count rates for $a$SPECT are $\approx$ 450 cps at $U_{\text{AP}}$ = 50 V and $\approx$ 6 cps without protons ($U_{\text{AP}}$ = 780 V). Above the upper integration limit, the count rate of $\beta$-electron events is $\approx$ 70 cps independent of voltage settings, see Appendix~\ref{sec:ebg}.

\subsection{Fit procedure}
\label{sec:fit}

To simplify expressions, the indexing $c$ and $p$ for a given configuration and detector pad is omitted hereinafter. For the analysis of $a$ from the integral proton recoil spectra, a fit is performed to the measured data, with $a$ as one of the free fit parameters. In the ideal case without any systematic effect, this fit would be a $\chi^2$ minimization of the fit function $f_{\text{fit}} (U_{\text{AP}}, r_{\text{B}}; a, N_0)$ to the measured integral proton spectrum. $f_{\text{fit}}$, \textit{i.e.,} the integral of the product of two functions, would only consist of the theoretical recoil energy spectrum $\omega_{\text{p}}(T, a)$ and the transmission function $F_{\text{tr}}(T, U_{\text{AP}}, r_{\text{B}})$ (Eq.~(\ref{eq:tfunc})) as well as an overall pre-factor $N_0$ in units of cps (the second fit parameter) which serves to match the measured count rate spectrum:
\begin{eqnarray}
f_{\text{fit}}(U_{\text{AP}}, r_{\text{B}}; a, N_0) &=& N_0 \int_0^{T_{\text{max}}} \omega_{\text{p}}(T, a) \nonumber \\
& \cdot & F_{\text{tr}}(T, U_{\text{AP}}, r_{\text{B}}) \; dT \nonumber \\
&=:& y_{\text{theo}}(U_{\text{AP}}, r_{\text{B}}; a, N_0) \; .
\label{eq:countrate}
\end{eqnarray}
The $\chi^2$ function is then given by
\begin{equation}
\label{eq:chisquareplain}
\chi^2=\sum_{i=1}^n \frac{\left(y_{\text{exp}, i} - f_{\text{fit}}(U_{\text{AP}}, r_{\text{B}}; a, N_0) \right)^2}{\left(\Delta{}y_{\text{exp}, i} \right)^2}
\end{equation}
where $U_{\text{AP}}$ is the applied retardation voltage at measurement point $i$. The dead time-corrected count rates in the proton region are denoted by $y_{\text{exp}, i}$ (cf.~section \ref{sec:uld}) with $\Delta{}y_{\text{exp}, i}$ as their statistical uncertainties. The theoretical proton recoil spectrum $\omega_{\text{p}}(T, a)$ is given by Eq. (3.11) in \cite{glu1993}. This spectrum includes relativistic recoil and higher order Coulomb corrections, as well as order-$\alpha$ radiative corrections. These corrections are precise to a level of $\Delta{}a/a\approx$ 0.1 \%. In Appendix~\ref{sec:prrec} is given the analytical expression of $\omega_{\text{p}}(T, a)$ were recoil-order effects and radiative corrections are neglected: $\omega_{\text{p}}^{*} (T,a)$.

The $\chi^2$ fit of Eq.~(\ref{eq:chisquareplain}), however, shows a strong correlation ($> 0.9$) among the fit parameters $N_0$ and $a$ with a correspondingly large correlated error on the extracted value of the $\beta$-$\overline{\nu}_e$ angular correlation coefficient $a$. In order to reduce this correlation considerably, the proton integral count rate spectrum is fitted by a distinctly better fit function largely orthogonalized with respect to the fit parameters $N_0$ and $a$ according to
\begin{tiny}
\begin{eqnarray}
\label{eq:fitn}
f_{\text{fit(n)}}(U_{\text{AP}}, r_{\text{B}}; a, N_0) & = & N_0 \int_0^{T_{\text{max}}} \omega_{\text{p(n)}}(T, a, r_{\text{B}})\cdot{}F_{\text{tr}}(T, U_{\text{AP}}, r_{\text{B}}) \; dT \nonumber \\
& =: & y_{\text{theo(n)}} \; .
\end{eqnarray}
\end{tiny}
Here, a normalized differential proton recoil spectrum is used with
\begin{equation}
\label{eq:diffprot}
w_{\text{p(n)}}(T, a, r_{\text{B}})= \text{norm}(a, r_{\text{B}})\cdot{}\omega_{\text{p}}(T, a) \; .
\end{equation}
The normalization factor $\text{norm}(a, r_{\text{B}})$ given by
\begin{footnotesize}
\begin{eqnarray}
\label{eq:normfac}
\text{norm}(a, r_{\text{B}}) = \left(\int\limits_{0}^{751 \; \text{V}} \frac{y_{\text{theo}}(U_{\text{AP}}, r_{\text{B}}; a, N_0)}{N_0} \; dU_{\text{AP}} \right)^{-1}
\end{eqnarray}
\end{footnotesize}
provides an integral value of $\int_0^{751 \; \text{V}}y_{\text{theo(n)}} \; dU_{\text{AP}}$ of area $N_0$ which does no more depend on $a$ in contrast to $\int_0^{751 \; \text{V}}y_{\text{theo}} \; dU_{\text{AP}}$ (cf. Eq.~(\ref{eq:countrate})). 

In the actual conduction of the experiment one has to deal with systematic effects, like shifts and inhomogeneities of the applied electric and magnetic fields or background and its possible dependency on the retardation voltage, etc., which alter the measured integral proton spectrum. This can be taken into account by additional functions $f_{\text{sys}}$ which modify the spectrum accordingly. For each systematic effect ($j$) the function depends on a set of fit parameters $\{ fpar_j \}$ representing the coefficients of a polynominal expansion up to order 4 of the quantities $U_{\text{AP}}$\footnote{In the argument of $f_{\text{sys}}^j$ we have set $\langle{}U_{\text{A}}\rangle = U_{\text{AP}}$ since corrections on the applied retardation voltage $U_{\text{AP}}$ are of 2$^{\text{nd}}$ order here.}, $T$, or $y_{\text{theo}}$. The polynomial approach with these variables (including the constant function as zero order polynomial function) is sufficient to describe all possible modifications on the spectrum's shape by the investigated systematic effects listed in section~\ref{sec:systematics}.

The corresponding fit function is then given by
\begin{tiny}
\begin{eqnarray}
\label{eq:chisquare}
&f_{\text{fit(n)}}\left(U_{\text{AP}}, r_{\text{B}}, y_{\text{theo(n)}}; a, N_0, \{ fpar_{j = 1} \}, \{ fpar_{j = 2} \}, \cdots{}\right) \nonumber \\
& = N_0\cdot{}\left(\int\limits_0^{T_{\text{max}}} \omega_{\text{p(n)}}(T, a, \langle r_{\text{B}} \rangle)\cdot{}F_{\text{tr}}(T, \langle U_{\text{A}} \rangle, \langle r_{\text{B}} \rangle) \; dT\right)_{f_{\text{sys}}^{j'}} \nonumber \\
& + \sum_{j''\neq{}j'} f_{\text{sys}}^{j''}\left(U_{\text{AP}}, r_{\text{B}}, y_{\text{theo(n)}}; \{fpar_{j''}\}\right)
\end{eqnarray}
\end{tiny}
with $j', j'' \in \{j\}$. The integral expression indexed by $f_{\text{sys}}^{j'}\left(U_{\text{AP}}, r_{\text{B}}; \{fpar_{j'}\}\right)$ means that for certain systematic errors ($j'$) the corresponding function is included as a modification of the integral expression: Concerning the transmission function $F_{\text{tr}}\left(T, \langle{}U_{\text{A}}\rangle, \langle{}r_{\text{B}}\rangle\right)$, one has to describe the average retardation potential $\langle{}U_{\text{A}}\rangle$ as a function of $U_{\text{AP}}$, \textit{i.e.,} $\langle{}U_{\text{A}}\rangle = f_{\text{sys}}^{\langle{}U_{\text{A}}\rangle}\left(U_{\text{AP}}; \{fpar_{\langle{}U_{\text{A}}\rangle}\} \right)$ (cf.~section \ref{sec:ua}) and to replace the magnetic field ratio $\langle{}r_{\text{B}}\rangle\rightarrow{}f_{\text{sys}}^{\langle{}r_{\text{B}}\rangle}\left(\{fpar_{\langle{}r_{\text{B}}\rangle} \} \right)$, a zero order polynomial function (cf.~section \ref{sec:rb}).

The fit parameters we introduce in $f_{\text{sys}}^j$ may have correlations with the value of $a$ as a result of the $\chi^2$ minimization. To get a statistically meaningful handle on these correlations, we combine the data acquired for the determination of $a$ with \textit{supplementary} measurements and simulations of the different systematic effects to an overall data set. From the now more comprehensive fit to this overall data set we can determine the value and uncertainty of $a$ including correlations with the respective parameters used to correct for systematic effects.
In general the additional measurements/simulations of systematic effects ($j$) are described by $n_j$ measured values $y_{\text{sys}, k}^j$ with $k= 1, \dots, n_j$. Together with their functional descriptions $g_{\text{sys}}^j\left(U_{\text{AP}}, T, r_{\text{B}}, y_{\text{exp}}, y_{\text{theo(n)}}; \{gpar_j\} \right)$, they are implemented in the $\chi^2$-fit of the overall data set as
\begin{widetext}
\begin{eqnarray}
\label{eq:chisquareoverall}
\chi^2&=&\sum_{i=1}^n\frac{(y_{\text{exp}, i} - f_{\text{fit(n)}}(U_{\text{AP}}, r_{\text{B}}, y_{\text{theo(n)}}; a, N_0, \{fpar_{j = 1}\}, \{fpar_{j = 2}\}, \cdots ))^2}{(\Delta{}y_{\text{exp}, i})^2} \nonumber \\
&+& \sum_j \sum_{k=1}^{n_j}\frac{(y_{\text{sys}, k}^j - g_{\text{sys}}^j(U_{\text{AP}}, T, r_{\text{B}}, y_{\text{exp}}, y_{\text{theo(n)}}; \{gpar_j\} ))^2}{(\Delta{}y_{\text{sys}, k}^j)^2} \; .
\end{eqnarray}
\end{widetext}

The first term on the right hand side of Eq.~(\ref{eq:chisquareoverall}) is the original $\chi^2$ (cf. Eq.~(\ref{eq:chisquare})) now including all systematic corrections in the fit function to describe the measured count rate spectrum at the measurement points $i$.

The second term - the double sum - describes the fit $g_{\text{sys}}^j\left(U_{\text{AP}}, T, r_{\text{B}}, y_{\text{exp}}, y_{\text{theo(n)}}; \{gpar_j\} \right)$ on the supplementary measurements or simulations $y_{\text{sys}, k}^j$ with error bars $\Delta{}y_{\text{sys}, k}^j$, where the sum over $j$ encompasses all systematic investigations applied. As in the case of $f_{\text{sys}}^j$, we have set $\langle{}U_{\text{A}}\rangle = U_{\text{AP}}$ in the argument of $g_{\text{sys}}^j$. $f_{\text{sys}}^j$ and $g_{\text{sys}}^j$ may or may not be the same function. This depends on how we get access to the relevant systematic effect through the supporting measurements/simulations and on how these results have to be transferred to $f_{\text{sys}}^j$ in order to make the appropriate correction on the systematic effect ($j$) in the integral proton spectrum. That is why the parameter set $\{fpar_j\}$ and $\{gpar_j\}$ which enter into the fit may be different for a given systematic effect. This, for example, is the case when describing the background with its retardation voltage-dependent part (cf.~section \ref{sec:bg}).

Since the systematic effects may vary between pads $(p)$ and configurations $(c)$, the $\chi^2$ function of Eq.~(\ref{eq:chisquareoverall}) has to be indexed by $\chi^2_{c, p}$.
For the final result, both detector pads and all selected configurations have to be included in the global fit with $a$ being the same fit parameter for all, but all other systematics individually for the respective pad and configuration. Formally, the so-called global $\chi^2$ fit can be expressed as 
\begin{equation}
\label{eq:finalfit}
\chi^2_{\text{global}} = \sum_c\sum_p\chi^2_{c, p}
\end{equation}
by adding up the $c$ and $p$ dependency of the expressions on the right-hand side of Eq.~(\ref{eq:chisquareoverall}) accordingly.

The routine we employed is based on Wolfram Mathe\-matica and has been used for other experiments in the past  \cite{hoyle2004,tullney2013,allmendinger2014}. In Appendix \ref{sec:markovmc}, the treatment of statistical and systematic uncertainties using a Bayesian averaged (\textit{i.e.,} marginal) likelihood as well as a comparative approach using the profile likelihood is discussed.

\subsection{Field and particle tracking simulations}
\label{sec:sim}

In order to understand the behaviour of the experimental set-up and to determine several systematic uncertainties quantitatively, simulations of the electric and magnetic fields were performed, as well as particle tracking simulations. For this purpose, the open-source KASPER simulation framework is used, containing the KGeoBag, KEMField, and KASSIOPEIA packages \cite{fur2017, cor2014, fur2015}. The EM field and particle tracking simulation routines of KASPER were originally developed and used for $a$SPECT, then modified and hugely improved at KIT and MIT for the KATRIN experiment to determine the neutrino mass. The $a$SPECT coils and electrodes geometry is implemented using the KGeoBag software package for designing generic 3-dimensional models for physics simulations. This geometry is forwarded to KEMField, a high-performance field simulation software which incorporates a Boundary Element Method (BEM) solver for electromagnetic potential and field calculations. We checked KEMField versus COMSOL Multiphysics, a finite element analysis, solver and simulation software and found excellent agreement in the accuracy required for $a$SPECT ($\approx$mV ). The computation of the magnetic field is less elaborate and challenging due to the fact that the magnetic sources are known and the coils are arranged axially symmetric.

At that point the applied currents, voltages (see Table \ref{tab:potentials}) as well as the measured work functions of the particular electrode segments have to be set as input parameters. The different methods used for charge density and field calculation are described in \cite{lazic2006, hilk2017, glu2011c}. The calculated fields together with the geometrical arrangement are then used for the particle tracking, performed with the KASSIOPEIA package \cite{fur2017}. In KASSIOPEIA, the track contains the initial particle state (position, momentum vector, and energy) as well as the current state which is consecutively updated as the simulation progresses. The equation of motion is solved at each step using an 8$^{\text{th}}$ order Runge-Kutta algorithm. KASSIOPEIA also stores parameters like path length, elapsed time, number of steps in the trajectory calculation and exit condition identification containing the reason why track calculation was stopped, \textit{i.e.,} particle hits the detector plane, an electrode surface or is trapped in Penning-like field configurations. In the particle tracking simulation the weighting with the measured beam profile is taken into account.

To achieve the required precision on the simulated systematic corrections, $\approx$ $10^{10}$ protons had to be tracked with KASSIOPEIA resulting in a multi-core CPU computing time of $\approx$ 0.5 y\footnote{Mogon high performance cluster of Mainz university \cite{mogon}.}. In addition, 40 weeks of single GPU computation time with KEMField was necessary to solve the charge density distribution for the different electrostatic configurations. For details of this simulation see \cite{csh2017}.

\section{Quantitative determination of the systematic effects}
\label{sec:systematics}

The systematic uncertainties relevant in this analysis lie in the knowledge of the transmission function and any effect that shows a dependence on the recoil energy or the retardation voltage. The relevant experimental systematic effects in no order of strength are 
\begin{itemize}
\item[A.] Temporal stability and normalization
\item[B.] Magnetic field ratio $\langle r_{\text{B}} \rangle$
\item[C.] Retardation voltage $\langle U_{\text{A}} \rangle$
\item[D.] Background
\item[E.] Edge effect
\item[F.] Backscattering and below-threshold losses
\item[G.] Dead time and pile-up
\item[H.] Proton traps in the DV region
\item[I.] Miscellaneous effects
\end{itemize}
In the following we explain each effect, show with which method it was investigated and what its influence on the proton spectrum or on $a$ is. Systematic effects are taken into account down to $\Delta{}a/a \le 0.1$ \%.
In addition to these major systematics there are some minor systematics which have been shown to be small enough to not significantly influence the experimental result at the present level of precision. These are the adiabatic motion of the proton that has been taken care of in the design of the spectrometer, electron backscattering at the electrodes below the DV and higher order corrections in the fit function.

\subsection{Temporal stability and normalization}
\label{sec:tempstab}
\begin{figure}
\includegraphics[width=\linewidth]{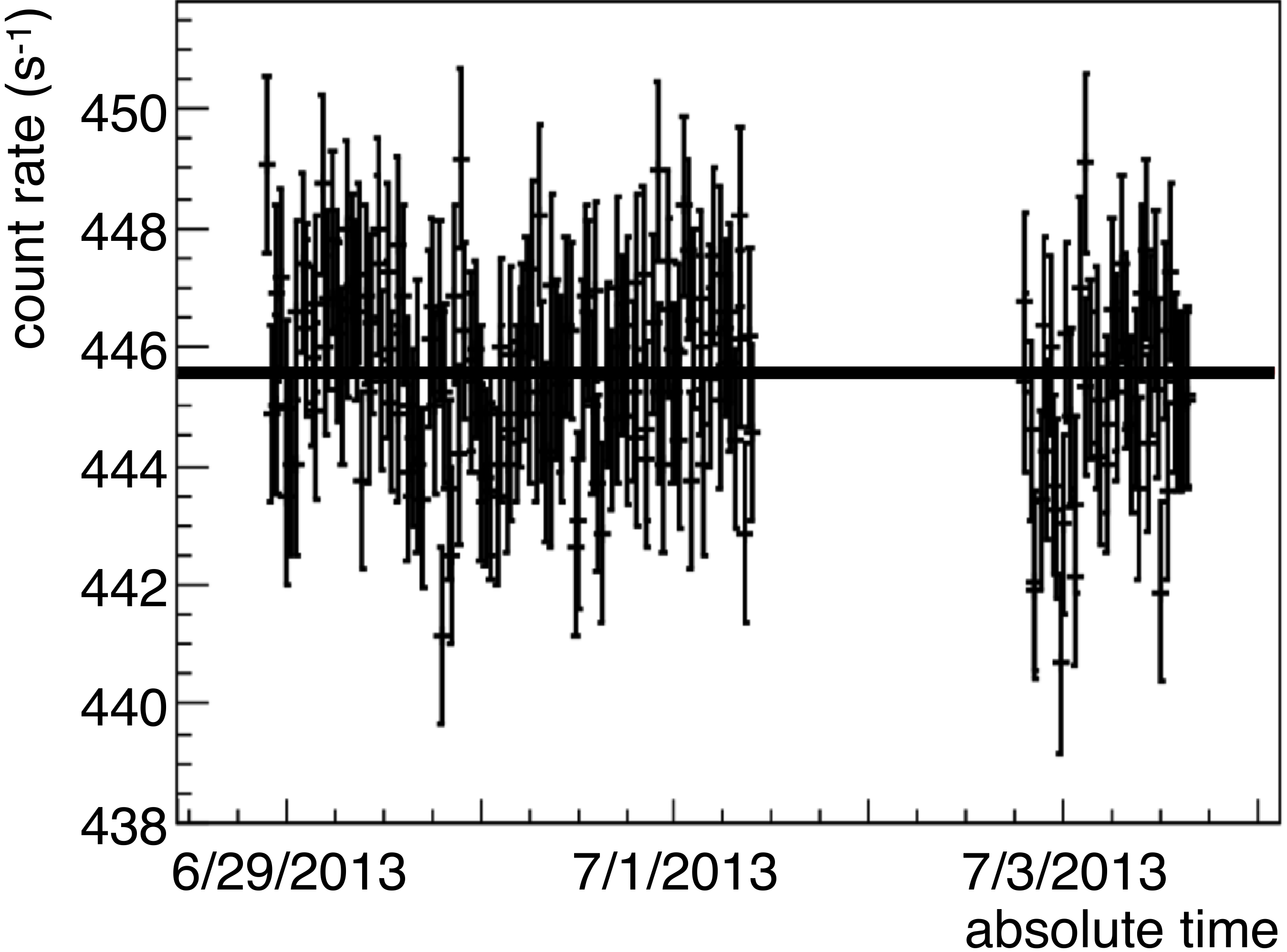}
\caption{Temporal sequence of 50 V runs for config 1. Plotted are the average count rates of the central detector pad for the individual measurement runs of 200 s duration. The distribution of data points around their common mean(solid line) corresponds to the drawn error bars resulting from pure counting statistics. A constant fit to the data gives: $\chi^2/\nu = 1.07$.\label{fig:tempstab}}
\end{figure}

The temporal stability of the measurement was checked via the measured count rates in the proton region at 50 V retardation voltage where we have the highest event rates. The resulting good statistics can be utilized to trace possible systematic drifts and non-statistical fluctuations. Figure~\ref{fig:tempstab} shows the sequence of count rates (central pad) for the 50 V measurement runs in config 1 according to the scheme depicted in Fig.~\ref{fig:sequence}. The individual 50 V runs were 200 s long (shutter opening time), resulting in a relative statistical accuracy of $\approx$~0.34 \% per pad at an average count rate of about 445 Hz.  The distribution of the count rates around their common mean (standard deviation) essentially reproduces the expected error from pure counting statistics. In config 1, for example, a total of 193 runs at 50 V were conducted within 3.5 days including an interruption of about 30 h. For the central pad the average count rate is 445.65(11) Hz which after dead-time correction enters as data point $y_{\text{exp}}^{\text{config1, pad2}}(50 \; \text{V})$ in the integral proton spectrum (see Fig.~\ref{fig:spectrum} (b). Table \ref{tab:cr50V} shows the average count rates at 50 V for the seven measurement configurations and the results of the respective $\chi^2$ fits (constant fit). The distribution of count rates in all measurement configurations clearly indicate the absence of drifts $>$ 1 Hz/day (estimated conservatively). The influence of linear drifts on $a$ exactly cancels as long as the drift period $T_{\text{D}}$ is an integer multiple (n) of $t_0 \approx$ 150 min as can be deduced from Fig.~\ref{fig:sequence} with the worst case scenario when the drift kinks at a half-integer multiple of $t_0$. For the latter case we estimated the influence on $a$ to be less than 0.1\% (relative) assuming a drift period of one day.

For the other retardation voltage settings, the average count rates and their associated error bars were extracted in a similar manner. They then provide the remaining data points $y_{\text{exp}}^{c, p}(U_{\text{AP}})$ to determine the shape of the integral proton spectra differentiated according to configuration ($c$) and detector pad ($p$).

In the final global fit (cf.~Eq.~(\ref{eq:finalfit})) the counting statistics of the total number of events ($\approx 2\times 10^{8}$) enter. The latter are more than a factor of 10 higher than the corresponding events from the sub-data sets, where checks were made for possible deviations from pure counting statistics (cf.~Table~\ref{tab:cr50V}). Non-statistical count rate fluctuations, e.g., due to the ILL reactor power fluctuations \cite{vesna2011} will show up more prominently with better statistics (see section~\ref{sec:globalfit}).

\begin{table}[t]
  \flushright
  \caption{Average count rates in the proton region at 50 V for the different configurations (central detector pad). The statistical error bars were scaled with $\sqrt{\chi^2/\nu}$ whenever the condition $p = \int_{\chi^2_\nu}^{\infty} f_\nu \left( \chi^2\right) \; d\chi^2 < 0.05$ was met with $f_{\nu} \left(\chi^2 \right)$ being the $\chi^2$ distribution function with $\nu$ degrees of freedom. We took the significance level $\alpha = 0.05$ according to the PDG guidelines \cite{beringer2012}. Configuration runs where measurement sequences (cf. Fig.~\ref{fig:sequence}) have also been performed at other shutter opening times are noted.}
  \begin{center}
    \begin{tabular}{ccccc}
    \hline
    \hline
    Configuration & Average count & a $= 50$ s & $\chi^2/\nu$ & $p$-value $(\nu)$ \\
    (start time in & rate & b $= 100$ s & &  \\
    2013) & at 50 V & c $= 200$ s & &  \\
    \hline
    Config 1 (06/28) & 445.65(11) & c & 1.07 & 0.241(192) \\
    Config 2\footnote{The mirror electrode was alternated between ON and OFF. Numbers are for mirror ON only.} (07/05) & 445.57(27) & c & 1.06 & 0.378(29) \\
    Config 3 (07/26) & 452.33(15\footnote{\label{doublenote}Error bars scaled with $\sqrt{\chi^2/\nu}$, where the reactor power noise \cite{vesna2011} presumably led to non-statistical count rate fluctuations, which are reflected in an increased $\chi^2$ value.}) & a, b, c & 1.27 & 0.011(162) \\
    Config 4 (07/30) & 395.82(21) & a, b, c & 0.97 & 0.544(63) \\
    Config 5 (08/01) & 393.07(25$^{\text{b}}$) & a, b, c & 1.51 & 0.015(45) \\
    Config 6 (08/04) & 389.79(32) & c & 0.61 & 0.887(17) \\
    Config 7 (08/05) & 443.36(16) & a, b, c & 0.95 & 0.642(127) \\
    \hline
    \hline
    \end{tabular}%
  \label{tab:cr50V}%
  \end{center}
\end{table}%

\subsection{Magnetic field ratio $\langle r_{\text{B}} \rangle$}
\label{sec:rb}

The fields inside $a$SPECT were scanned with a Hall probe sufficient to bridge the dynamic field range along the entire flux tube and to measure magnetic fields with a relative accuracy of $\approx 5\times10^{-3}$ (see Fig.~\ref{fig:aspectb}). 

\begin{figure}
\includegraphics[width=\linewidth]{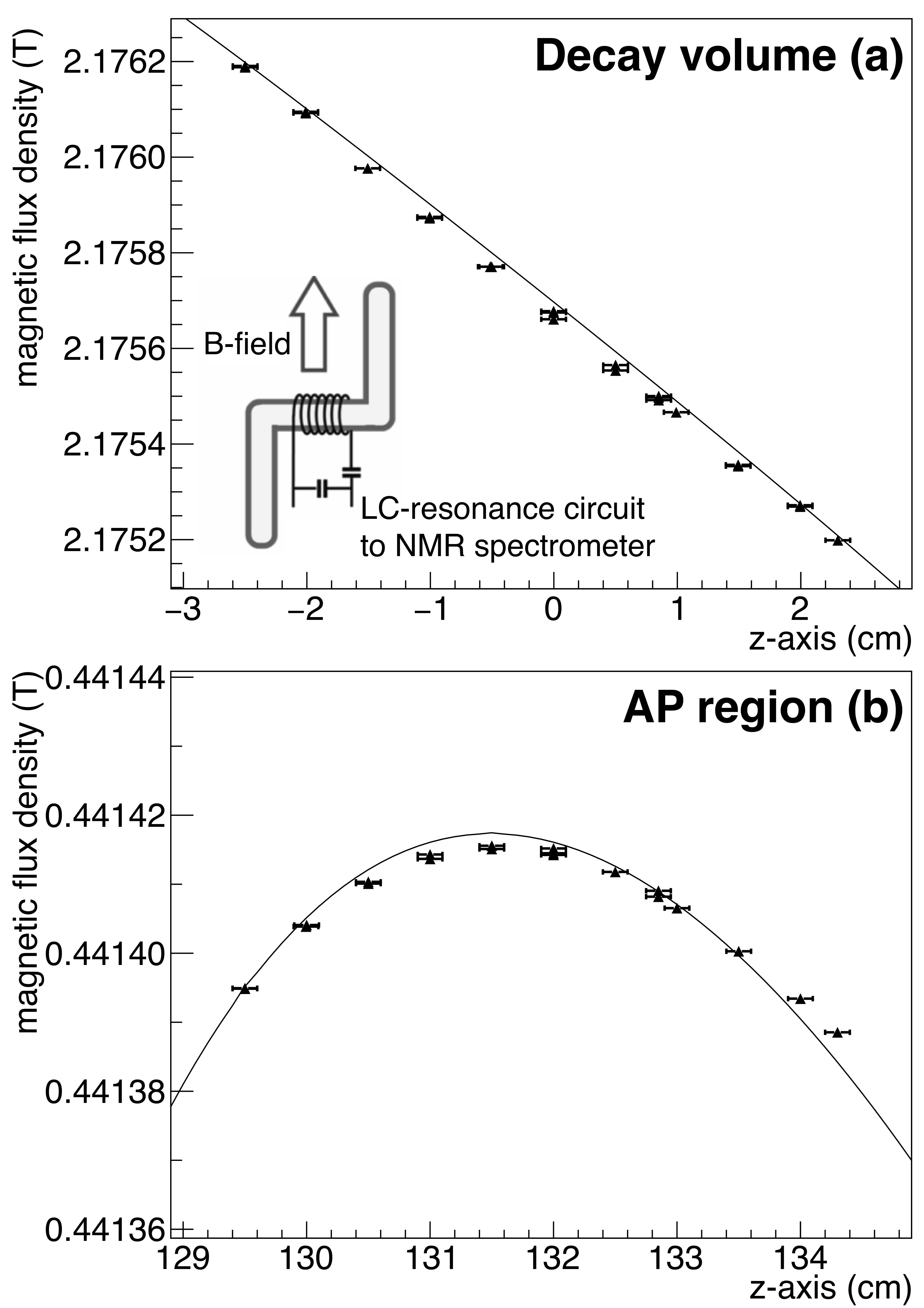}
\caption{NMR measurements of the magnetic field on axis around the center position of the DV and the AP electrodes. The uncertainties in the field measurement correspond to the symbol size and mainly reflect the measurement reproducibility. The positioning error of the NMR probe was estimated to be $\pm$ 1 mm. The solid lines are the results from KEMField field simulations based on the known $a$SPECT coil configuration as well as the current settings used in the 2013 beam time.  a) The B-field inside the DV exhibits a small axial gradient of $\approx$ $2\times{}10^{-4}$ T/cm to ensure that no decay protons get trapped by the magnetic mirror effect between the DV and EM. Inset: Sketch of NMR probe used to measure the fields. b) In the AP region the B-field has a tiny, local maximum for optimal transmission condition \cite{glu2005, glu2013}.\\\hspace{\textwidth}
Deviations of the NMR field measurements from the KEMField simulations are mainly caused by the positioning error of the NMR probe. A total offset error common to all $\langle r_{\text{B}} \rangle$ values ($< 2.4\times10^{-5}$, relative) takes this matter into account (see text).
}
\label{fig:nmrDVAP}
\end{figure}

To precisely determine $\langle r_{\text{B}} \rangle$, a proton-based NMR system has been developed \cite{gua2011, csh2017}. It consists of two z-shaped glass tubes of inner diameter 2.5 mm and outer diameter 4 mm. Each glass tube is filled with a 1:1 mixture of acetone and ethanol which stays liquid down to 150 K. The central part of the z-shape is surrounded by a solenoidal NMR coil of $\approx$ 1 cm length, which is oriented horizontally in the B-field of $a$SPECT (see inset of Fig.~\ref{fig:nmrDVAP} (a)).

The resonant circuits ($Q\approx$ 150) were tuned to the respective resonance frequencies of $\approx$ 92 MHz and $\approx$ 18 MHz of the local B-fields inside the DV and AP electrode and finally matched to the standard impedance of the connecting lines (50 $\Omega$).
 
Shortly after the 2013 beam time, the $a$SPECT spectrometer was brought to room temperature, and the whole electrode system including the detector setup was removed. To provide both free access to the inner part of the spectrometer and the necessary temperature conditions for the NMR probe measurements, an inverted, non-magnetic Dewar was built and fitted inside the bore tube of the spectrometer. After cooling down and ramping the magnetic field up again with the same current settings as before, the field along the $z$-axis was measured\footnote{The field measurements with the Hall probe were also carried out with this measurement setup.}. The two probes measured simultaneously at fixed distance, with the lower probe positioned around the center of the DV electrode and the upper probe at the place of the local field maximum at the height of the AP electrode. The measured fields are shown in Fig.~\ref{fig:nmrDVAP}. They are used to confirm the quality of field simulations with KEMField for the given coil configuration of $a$SPECT and the respective current settings. Minor adaptations due to the influence of the return yoke on the internal magnetic field \cite{kon2014} as well as environmental fields were taken into account.

The field simulations were used to determine the off-axis fields inside the DV and AP electrode. From the known field configuration and the beam profile measurements the magnetic field ratio $\langle r_{\text{B}} \rangle$ as result of the particle tracking simulation was determined.

When electrode E15 was used as dipole electrode (config 3, config 4, config 7), the local magnetic field maximum in the AP region had to be slightly shifted ($\approx$ -3 cm) by means of the external anti-Helmholtz coils (AHC). The resulting field changes in the DV and AP region were considered with their impact on $\langle r_{\text{B}} \rangle$. Table \ref{tab:rbsimulated} shows the $\langle r_{\text{B}} \rangle$ values from particle-tracking simulations differentiated by detector pad and configuration.

\begin{table}[t]
  \flushright
  \caption{Simulated $\langle r_{\text{B}} \rangle$ values. The uncertainty ($\approx 0.6\times10^{-6} $ relative) is dominated by the particle tracking simulation-based error. The error given for the fit results is the uncertainty in $c_{0}^{\langle r_{\text{B}} \rangle}$ (see text). Possible influences of beam position variation ($\pm1$~mm) and differences due to standard and reduced beam profile on $\langle r_{\text{B}} \rangle$ (the main contributions) result in an offset error common to all $\langle r_{\text{B}} \rangle$ values: The value in the last line ('offset') indicates $c_{0, \text{offset}}^{\langle r_{\text{B}} \rangle}$ and its uncertainty.}
  \begin{center}
    \begin{tabular}{lcc}
    \hline
    \hline
    & $\langle r_{\text{B}} \rangle$: Config 1, 2, 5, 6 & $\langle r_{\text{B}} \rangle$: Config 3, 4, 7 \\
    \hline
    Pad 2 input & 0.2028870(12) & 0.2028897(12) \\
    Pad 2 fit result & 0.2028870(14) & 0.2028897(14) \\
    Pad 3 input & 0.2028930(12) & 0.2029000(12) \\
    Pad 3 fit results & 0.2028930(14) & 0.2029001(14) \\
    Pad 2/3 offset & $-1.4\times{}10^{-8} \pm 4.8\times{}10^{-6}$ & \\
    \hline
    \hline
    \end{tabular}%
  \label{tab:rbsimulated}%
  \end{center}
\end{table}%

This simulation-based error analysis must be extended by an offset error common to all $\langle r_{\text{B}} \rangle$ values. The main contribution comes from the uncertainty of the exact position ($\pm$ 1 mm) of the two NMR samples in axial direction (cf.~Fig.~\ref{fig:nmrDVAP}) with $\left(\Delta{}\langle r_{\text{B}} \rangle / \langle r_{\text{B}} \rangle \right)_{\text{pos.}} = 1.7 \times{} 10^{-5}$. The field ratio is quite insensitive to repeatedly ramping the superconducting magnets down and up\footnote{The superconducting magnet shows a kind of hysteresis, which is a small, but known, effect \cite{scott1968}). It disappears after the coils are warmed up above their critical temperature of $T_{\text{crit}} = 9$ K, which was applied systematically for field changes.}, moving the detector mechanics, changing the status of nearby valves, etc. Possible influences of these were estimated conservatively and are included in the error budget (cf. Table~\ref{tab:rbsimulated}) resulting in a total offset error of $< 2.4\times{}10^{-5}$ (relative).

To include these results into the fit procedure of Eq.~(\ref{eq:chisquareoverall}) we have to set $y_{\text{sys}}^{\langle r_{\text{B}} \rangle} = \langle r_{\text{B}} \rangle$ and $\Delta y_{\text{sys}}^{\langle r_{\text{B}} \rangle} = 1.2\times{10^{-6}}$ (cf. Table \ref{tab:rbsimulated}) and further $g_{\text{sys}}^{\langle r_{\text{B}} \rangle} = c_0^{\langle r_{\text{B}} \rangle}$ with $c_0^{\langle r_{\text{B}} \rangle}$ as free fit parameter. In the fit function of Eq.~(\ref{eq:chisquare}) one has to replace $\langle r_{\text{B}} \rangle \rightarrow{}f_{\text{sys}}^{\langle r_{\text{B}} \rangle} = c_0^{\langle r_{\text{B}} \rangle} + c_{\text{0, offset}}^{\langle r_{\text{B}} \rangle}$. The parameter $c_{\text{0, offset}}^{\langle r_{\text{B}} \rangle}$ is a restricted fit parameter in the fitting procedure which is Gaussian distributed around zero mean with standard deviation $\sigma = 2.4\times{}10^{-5}\cdot{}\bar{\langle r_{\text{B}} \rangle} = 4.8\times{}10^{-6}$. This way the offset error on $\langle r_{\text{B}} \rangle$ is taken into account. In Table \ref{tab:rbsimulated} the corresponding fit results for $\langle r_{\text{B}} \rangle$ including error bars are listed.

\subsection{Retardation voltage $\langle U_{\text{A}} \rangle$}
\label{sec:ua}

Like $\langle r_{\text{B}} \rangle$ , the retardation voltage $\langle U_{\text{A}} \rangle$ directly enters the transmission function (Eq.~(\ref{eq:tfunc})). Sources of uncertainties of $\langle U_{\text{A}} \rangle$ are
\begin{enumerate}
\item{the measurement precision of the applied voltage,}
\item{inhomogeneities and instabilities of the potential in the DV and the AP region due to spatial and temporal variations of the work function of the DV and AP electrodes, and}
\item{inhomogeneities of the potential in the DV and AP region due to field leakage}
\end{enumerate}

\subsubsection{Measurement precision of the applied voltage}

The retardation voltage $U_{\text{AP}}$ was measured continuously at the readback connections of the AP and the DV electrode using the Agilent 3458A multimeter. Each voltage reading was integrated for 4 s to achieve the required precision. The multimeter was calibrated at least annually and was working within specification during the beam time 2013, \textit{i.e.,} the corresponding precision of each measurement of the retardation voltage was $\Delta{}U_{\text{AP, Agilent}} < 13$ mV for all voltages. The short-time voltage stability was found to be better than 1.5 mV on the 1000V scale.

\subsubsection{Impact of spatial and temporal variations of the work function}

$a$SPECT utilizes gold-coated electrodes to obtain inert electrode surfaces, to achieve a high temporal stability of the surface properties, and to avoid any potential surface charges on an electrically insulating oxide layer \cite{dobr1974}. The work function of these electrodes modifies the actual retardation voltage measured between the DV and AP electrode. The work function (WF) of gold varies by up to $\Delta{}\text{WF}^{\text{Au}}/e \approx$ 500 mV depending on its crystalline structure and orientation \cite{Haynes2015}. Besides, a WF decrease of as much as one volt may occur on exposure of gold electrodes to water vapor (humidity) \cite{wells1972}. All in all, this is significantly larger than the desired uncertainty of $\Delta{}\langle U_{\text{A}} \rangle <$ 30 mV needed to keep retardation voltage related uncertainties of $a$ below 0.3 \%. Since only WF differences are relevant, the problem is largely relaxed if only common drift modes are present. Furthermore, WF differences may be greatly compensated if the electrodes have passed the same manufacturing process. This particularly applies for the DV and AP electrodes where we used the measures as described in section \ref{sec:rb} for the production process, cleaning procedures and depositary. Nonetheless, great efforts were made to measure precisely the WF of the individual electrode segments by means of a scanning Kelvin Probe.  The WF investigations were conducted after the 2013 beam time in extensive measuring campaigns in the years 2014 and 2015. The time span of almost two years was also important to trace possible drifts and fluctuations of the WF. The safe knowledge about the actual WF during the 2013 run under the given measuring conditions in $a$SPECT was a cornerstone to meet the required accuracies in the specification of the potential distribution inside the DV and AP electrodes. (details are presented in Appendix \ref{app:wf}).

\subsubsection{Field leakage}

Both the DV electrode and its surroundings are on ground potential to prevent possible field leakage into the DV. However, the WF of the DV electrode and those of the materials in immediate vicinity, \textit{i.e.,} bore tube (stainless steel), BN (TiB$_2$ enriched) collimation guide, and Ti-coated LiF frames are different, leading to field leakage into the DV through the large openings of the DV electrode (cf.~Fig.~\ref{fig:APDVpicture}). The WF of these materials were measured and are shown in Table \ref{tab:wfmeasured}. The maximal WF difference between materials is $\Delta{}\text{WF}/e \approx$ 500 mV, with the bore tube and collimation materials more negative than the DV electrode, leading to a small potential bump for the protons inside the DV electrode. The potential distributions in the DV and AP region were finally simulated using re-scaled WF, \textit{i.e.,} from the measured relative WF the WF average of all DV and AP electrode segments was subtracted. Since only potential differences are relevant this measure is of no relevance.

\begin{figure}
\includegraphics[width=\linewidth]{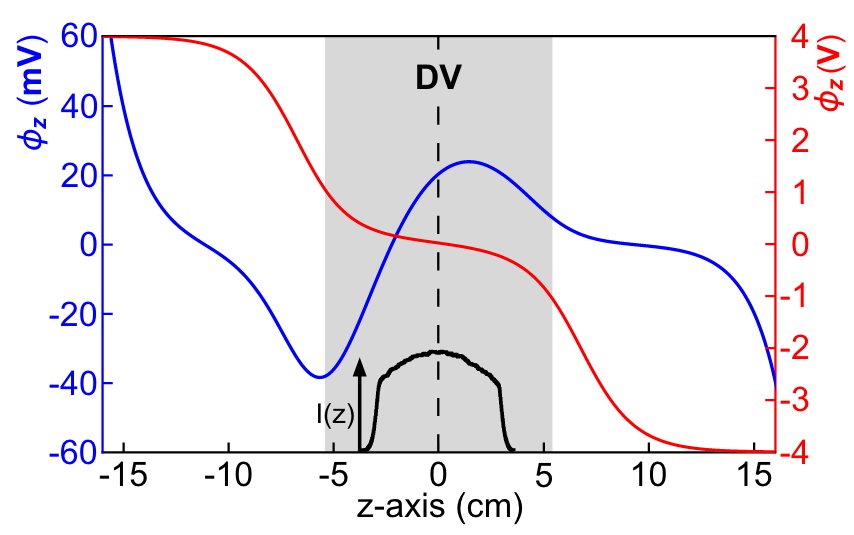}
\caption{Simulation of the potential distributions along the z-axis in the DV region. The blue curve (left axis) shows the potential for configurations 1-6 based on the measured work functions of the DV electrode and surrounding materials. The red curve (right axis) shows the resulting potential for config 7 where the electrodes below and above the DV were set to +4 V and -4 V, respectively. Inset $I(z)$: Measured beam profile along the z-axis (intensity in arbitrary units).\label{fig:simpotentialDV}}
\end{figure}

Figure~\ref{fig:simpotentialDV} shows the potential distribution along the z-axis inside the DV electrode and the adjacent electrodes which like the DV electrode are kept at ground potential. The distribution simulated by KEMField is essentially a superimposition of the potential drop between top and bottom plate of the DV electrode (cf.~Fig.~\ref{fig:APDVpicture}) caused by the measured WF differences of $\approx$ 100 meV and the potential bump due to field leakage. For config 7, the red curve is the relevant one, since the adjacent electrodes were put at $\pm$ 4 V to prevent protons from being trapped in the DV region.

\begin{table}[t]
  \flushright
  \caption{Measured WF differences between materials at $a$SPECT and the Kelvin Probe tip: $\text{WF}_{\text{rel}}:=\text{WF}_{\text{tip}} - \text{WF}_{\text{mat}}$. The individual measurements have a measurement uncertainty of $\pm$ 30 meV, whereas the average WF differences of the DV and AP electrode segments could be determined more precisely on a statistical basis (see Appendix \ref{app:wf}). The fact that the titanium-containing materials for the collimation show a higher WF than the gold-coated electrodes can be attributed to titanium oxide layers which lead to a significant increase of the WF of the substrate \cite{giordano2006}.}
  \begin{center}
    \begin{tabular}{ccc}
    \hline
    \hline
    Location & Surface material & Relative work function\\
    \hline
    DV electrode & Au & (113 $\pm$ 12) meV \\
    (average) & & \\
    AP electrode & Au & (127.4 $\pm$ 12) meV \\
    (average) & & \\
    Bore tube & Stainless steel 316L & (-85 $\pm$ 30) meV \\
    Collimation & BN with TiB$_2$ & (-240 $\pm$ 30) meV \\
    Collimation & Ti-coated LiF & (-394 $\pm$ 30) meV \\
    \hline
    \hline
    \end{tabular}%
  \label{tab:wfmeasured}%
  \end{center}
\end{table}%

\begin{figure}
\includegraphics[width=\linewidth]{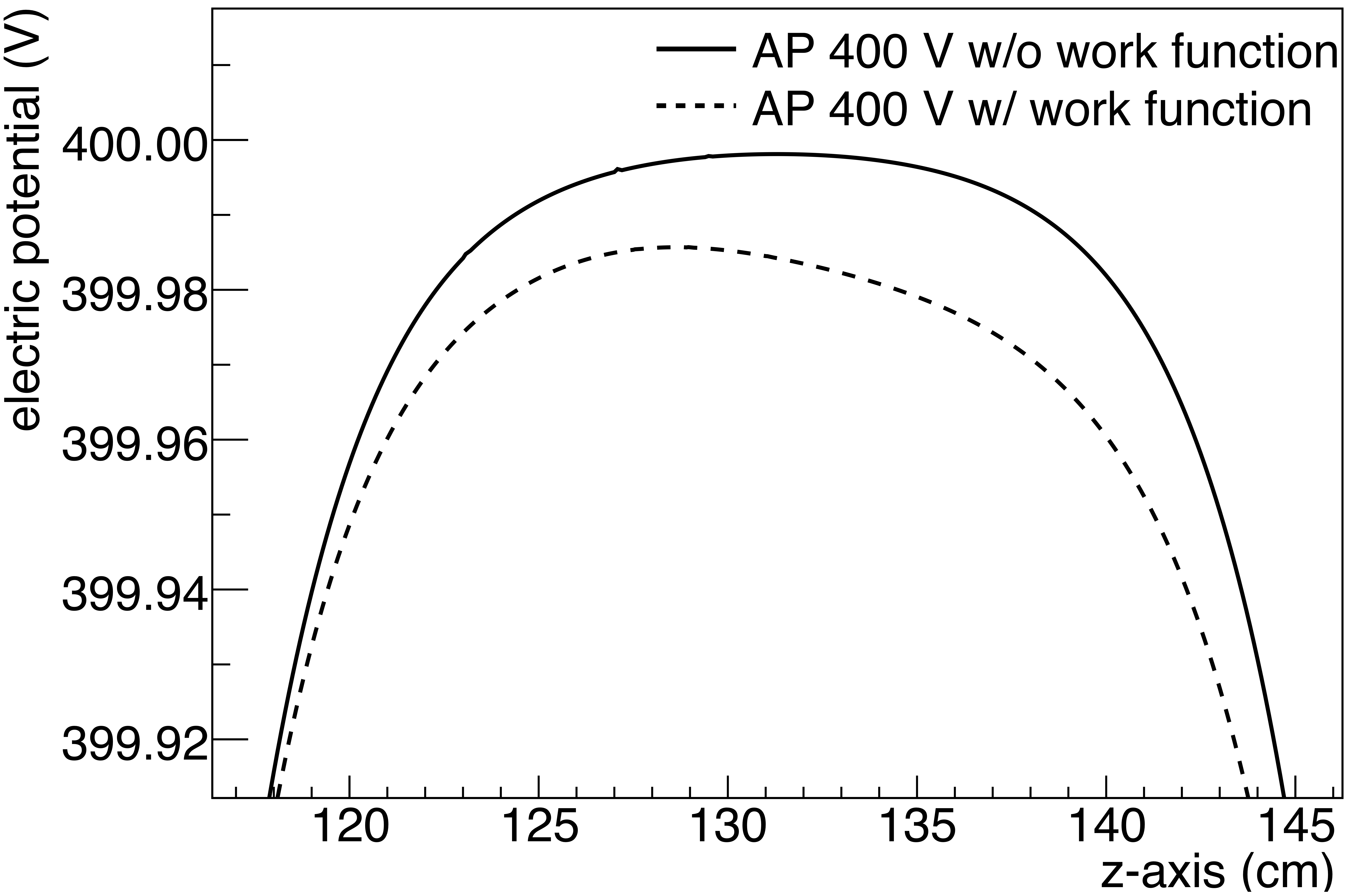}
\caption{Simulated potential distribution along the z-axis in the AP region with the retardation voltage set to 400V. A symmetric distribution around $z\approx$ 131 cm (solid curve) is the result if only field leakages are considered. The inclusion of the electrodes’ WF leads to an asymmetric shape with somewhat lower potential values and a slightly shifted position of the maximum (dashed curve).\label{fig:simpotentialAP}}
\end{figure}

The AP electrode with an aspect ratio of 3.6 : 1 is long compared to its diameter and shielded at both ends by overlapping electrodes with only slightly lower potential (cf. Table \ref{tab:potentials}). Field simulations show that the residual field leakage results in a homogeneity of the potential in the AP region of better than 2 mV. This can be deduced from Fig.~\ref{fig:simpotentialAP} where the shallow potential maximum is plotted for an applied retardation voltage of 400 V. It peaks at z $\approx$ 131 cm, \textit{i.e.,} it ideally overlaps with the position of the local B-field maximum (see Fig.~\ref{fig:nmrDVAP}). However, the inclusion of the electrodes’ WF which were only accessible to measurement after the 2013 beam time somewhat lowers the actual potential values inside the AP electrode and makes the distribution slightly asymmetric. Still, sufficient overlap with the local B-field maximum is given. Similar results were obtained for config 3 and config 4 (E15 dipole electrode used in E$\times$B mode), where both the E- and B-field maxima were shifted by $\approx$~3~cm towards the DV region.

\subsection*{The effective retardation voltage $\langle U_{\text{A}} \rangle$}

\begin{figure}[h]
\includegraphics[width=\linewidth]{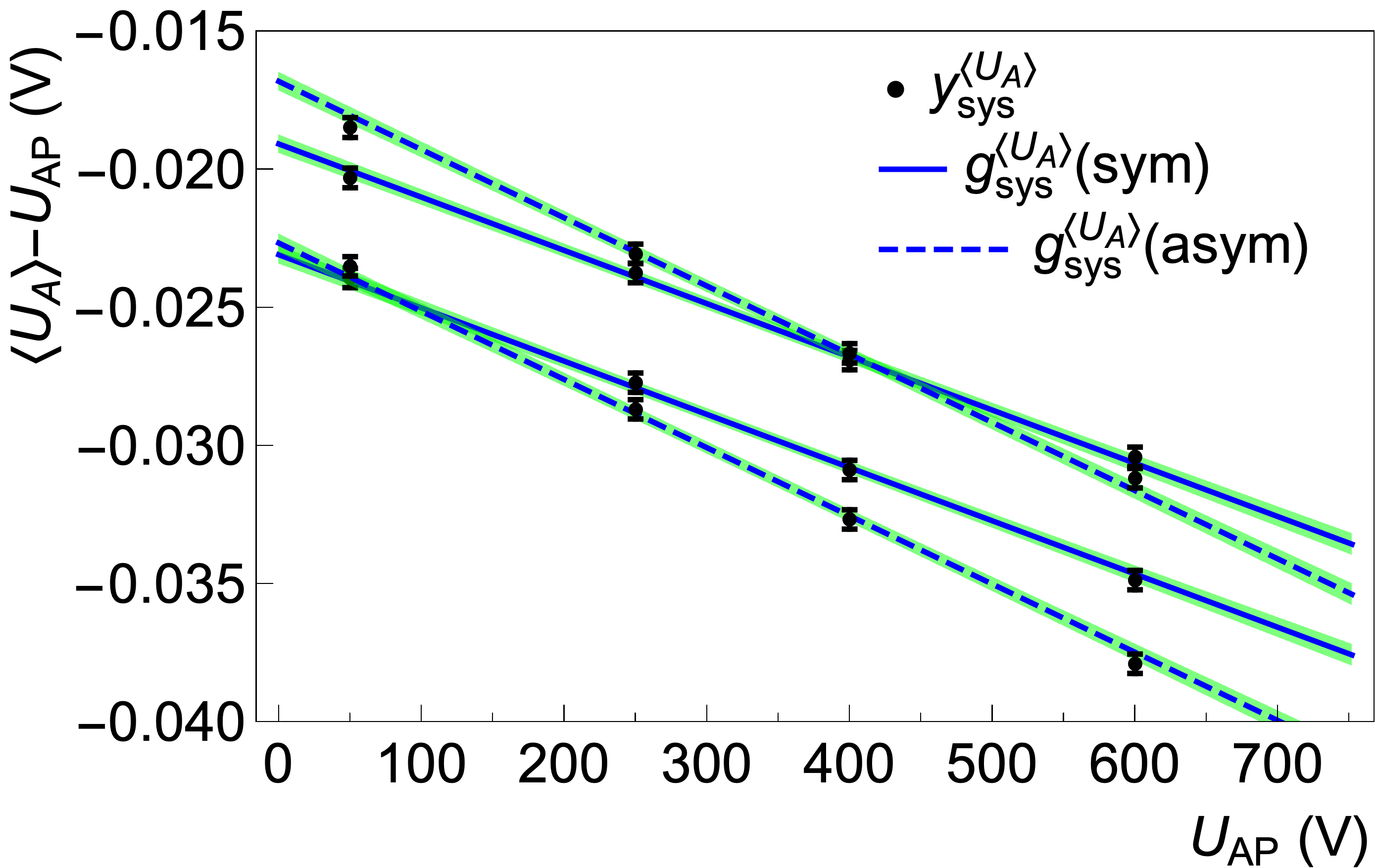}
\caption{Deviation $\Delta{}U_{\text{AP}} = \langle U_{\text{A}} \rangle - U_{\text{AP}}$ extracted from particle tracking simulations for $U_{\text{AP}}$ = 50, 150, 400, and 600 V. We find that $\Delta{}U_{\text{AP}}$ depends linearly on $U_{\text{AP}}$. However, there are differences in slope and intercept for the respective detector pad 2/3 (upper/lower pair of curves) and symmetric/asymmetric settings of the E15 electrode. The error bars are dominated by the statistics of the particle tracking simulation. Further drawn are the global fit results of $g_{\text{sys}}^{\langle U_{\text{A}} \rangle}$ (cf.~Eqs. (\ref{eq:chisquareoverall}), (\ref{eq:finalfit})) for the overall dataset.\label{fig:devparticletrack}}
\end{figure}

The inhomogeneities of the potential in the DV and AP region lead to a slight shift of the effective retardation voltage $\langle U_{\text{A}} \rangle$ from the applied voltage $U_{\text{AP}}$. Figure~\ref{fig:devparticletrack} shows the corresponding deviations $\Delta{}U_{\text{AP}} = \langle U_{\text{A}} \rangle - U_{\text{AP}}$ determined from particle tracking simulation for a total of four selected voltages. The error bars give the statistics of the MC simulation and include the uncertainties from a $\pm$ 1 mm variation of the true beam position as well as changes of the beam profile (standard/reduced). The functional dependence can be described by a straight line; however, a distinction must be made between the individual detector pads and configuration runs with symmetrical or asymmetrical setting of the E15 electrode. 

The corresponding assignment in the fit procedure according to section \ref{sec:fit} is then:
\begin{footnotesize}
\[
y_{\text{sys, k}}^{\langle U_{\text{A}} \rangle}(U_{\text{AP, k}}) = \left( \langle U_{\text{A}} \rangle_k - U_{\text{AP, k}} \right) \; \; ; \; k = 1, \cdots, 4
\]
\[
\Delta{}y_{\text{sys, k}}^{\langle U_{\text{A}} \rangle} \text{corresponding error bars from Fig.~\ref{fig:devparticletrack}}
\]
\begin{eqnarray}
\label{eq:fitproc}
g_{\text{sys}}^{\langle U_{\text{A}} \rangle}\left(U_{\text{AP}}; \left\{ c_0^{\langle U_{\text{A}} \rangle}, c_1^{\langle U_{\text{A}} \rangle} \right\}  \right) & = & c_0^{\langle U_{\text{A}} \rangle} + c_1^{\langle U_{\text{A}} \rangle}\cdot{}\left(U_{\text{AP}} - 320\right) \nonumber \\
f_{\text{sys}}^{\langle U_{\text{A}} \rangle} & = & U_{\text{AP}} + g_{\text{sys}}^{\langle U_{\text{A}} \rangle}(U_{\text{AP}}) + c_{\text{AP, offset}}^{\langle U_{\text{A}} \rangle} \nonumber \\
\end{eqnarray}
\end{footnotesize}

\begin{table}[h]
  \flushright
  \caption{Uncertainties from WF measurements and $U_{\text{AP}}$ reading. For details, see Appendix \ref{app:wf}.}
  \begin{center}
    \begin{footnotesize}
    \begin{tabular}{lc}
    \hline
    \hline
    Effect & Uncertainty \\
    \hline
    \footnotesize{Temporal changes of the WF diff-} & $\Delta{}\text{WF}_{\text{DV/AP}}/e \le$ 20 mV \\
    \footnotesize{erences between DV and AP electrode} & \\
    \footnotesize{Transferability of WF measurements} & $\Delta{}\text{WF}_{\text{UHV}}/e \le$ 10 mV \\
    \footnotesize{to UHV conditions inside $a$SPECT} & \\
    \footnotesize{Influence of temperature gradients} & $\Delta{}\text{WF}_{\text{gradT}}/e \le$ 10 mV \\
    \footnotesize{inside $a$SPECT on WF differences} & \\
    \footnotesize{Measurement precision} & $\Delta{}\text{U}_{\text{AP}}^{\text{Agilent}} \le$ 13 mV \\
    \footnotesize{of applied voltage} & \\
    \footnotesize{Influence of WF measurement uncer-} & $\Delta{}\text{U}_{\text{AP}}^{\text{p-tracking}} \le$ 10 mV \\
    \footnotesize{tainties on particle tracking results} & \\
    \hline
    $\Delta{}U_{\text{AP, offset}}$ (quadratic sum) & 30 mV \\
    \hline
    \hline
    \end{tabular}%
    \end{footnotesize}
  \label{tab:wfuncertain}%
  \end{center}
\end{table}%

As in case of $\langle r_{\text{B}} \rangle$ (cf.~section \ref{sec:rb}) the simulation-based errors must be extended by an offset error $c_{\text{AP, offset}}^{\langle U_{\text{A}} \rangle}$ common to all $\langle U_{\text{A}} \rangle$ values. In the fit procedure, $c_{\text{AP, offset}}^{\langle U_{\text{A}} \rangle}$ is again a restricted fit parameter which is Gaussian distributed around zero mean with standard deviation $\sigma_{\text{AP, offset}} := \Delta{}U_{\text{AP, offset}}$ = 30 mV.  In Table \ref{tab:wfuncertain}, the different contributions to $\sigma_{\text{AP, offset}}$ are listed. Details are discussed in Appendix \ref{app:wf}).

\subsection{Background}
\label{sec:bg}

The measured background in the proton region for the most part stems from electrons from neutron $\beta$-decay. Further contributions to the detected background are instrumental/environmental background, \textit{i.e.,} background measured with beam off\footnote{This also includes the tail of the electronic noise leaking into the proton integration window (cf.~Fig.~\ref{fig:detspec}).}, and other beam induced background, like $\gamma$-rays from neutron capture reactions and positive rest gas ions from secondary ionization processes in Penning-like traps of the $a$SPECT spectrometer.

Independent of its origin, the background can be categorized into a component that depends on the retardation voltage and one that does not. The latter can be readily tolerated since it simply represents a count rate offset in the integral proton spectrum which can be considered as free fit parameter in the fit function of the $\chi^2$ minimization. Thus, this background (if small) may only slightly worsen the purely statistical sensitivity in the determination of $a$.

On the other hand, an $U_{\text{AP}}$-dependent background changes the shape of the spectrum and therefore the value of $a$ extracted from the fit, unless a quantitative description of its functional dependence is given and taken into account accordingly. In previous beam times, the origin of the $U_{\text{AP}}$-dependent background was investigated and measures to reduce or even to get rid of it were implemented. 

The main source of the retardation voltage-dependent background is residual gas ionization due to electrons from neutron decay and field electron emission in combination with Penning-like traps inside $a$SPECT which amplify this kind of background. Field emission often originates from microprotrusions and particulate contamination on the surface of the electrode, which would enhance the local electric field. With the consequent and sustainable measures to improve the electrode surface quality (cf.~section \ref{sec:meas}), these particular sources of ionization could be largely eliminated. Beam off measurements during the 2013 run have shown that the field emission induced ion count rate in the proton region is $\approx$ $5\times{}10^{-3}$ cps and its impact on $a$ is negligibly small ($\ll$ 0.1 \%) \cite{mai2014}.

Looking at the composition of the rest gas inside $a$SPECT at low pressure and low temperature, hydrogen (H$_2$) accounts for the largest fraction\footnote{Measurements were performed with a mass spectrometer Pfeiffer Vacuum QMG-220 mounted at one of the $a$SPECT side ports. We identified the ratios H$_2$ : H$_2$0 : N$_2$ as 1 : 0.16 : 0.17 \cite{mai2014}.}. The small bump in the proton region of the pulse height spectrum at 780 V (cf.~Fig.~\ref{fig:detspec}) stems from collisions of trapped low-energy electrons in the AP region with hydrogen molecules. These secondary electrons are mainly produced by the $\beta$-electrons from neutron decay whose trajectories along the magnetic flux tube hit the AP electrode \cite{kite2003, reimer1977}.  The ionization cross section for electron impact on H$_2$ is highest for energies around 50 eV \cite{pa2009, yoon2008}, the energy range of secondary electrons which can be easily stored in the Penning-like trap around the AP electrode (cf.~Fig.~\ref{fig:aspectb}).

\begin{figure}
	\includegraphics[width=\linewidth]{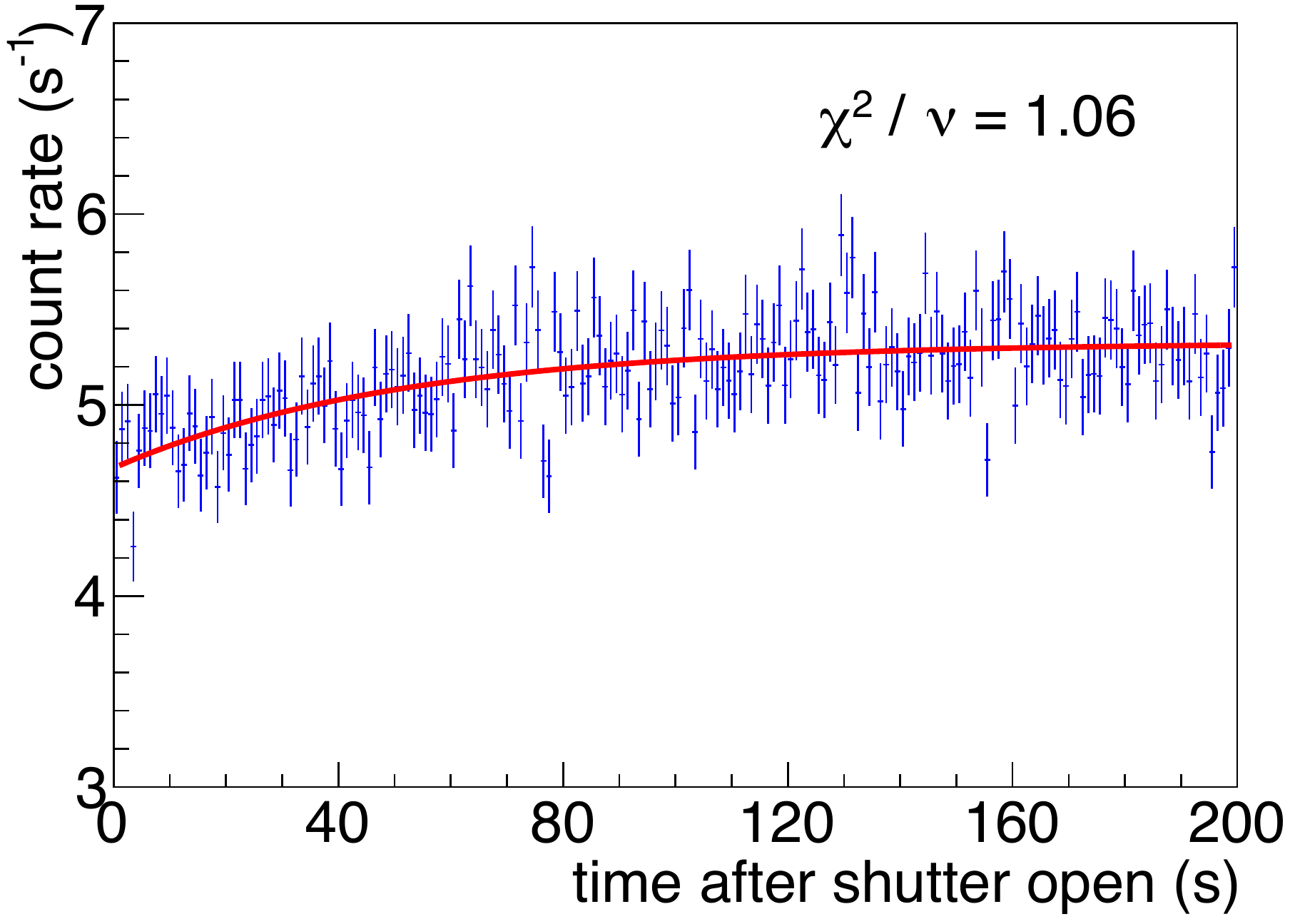}
	\caption{Evolution of the background count rate in the proton region as a function of the time after opening the fast neutron shutter for $U_{\text{AP}} =$ 780 V. The red solid curve is a fit to the data: $y=p_0+p_1\cdot{}\left(1-\exp{}(-t/\tau_s) \right)$. The constant part of the background, $p_0$, is calculated to be (4.67 $\pm$ 0.06) cps, the non-constant part shows an exponential saturation behaviour with $\tau_s = (51$ $\pm$ $10)$ s being the characteristic time constant and $p_1 = $ (0.66 $\pm$ 0.06) cps the resulting count rate after saturation is reached. These background investigations were carried out in the commissioning phase before the runs config 1-config 7 used in the analysis. During commissioning, the somewhat higher residual gas pressure produced a higher non-constant background ($\approx$ factor of 2) as compared to config 1 (cf.~Fig.~\ref{fig:pulseheight}).}
	\label{fig:evobg}
\end{figure}

H$_2^+$ and H$^+$ ions that are produced above the AP (or have sufficient energy to pass the AP) are accelerated towards the detector electrode (ions produced below the AP are stored and removed by the E$\times$B electrode E8). If they hit the detector, they are a potential cause of background events. Depending on the applied AP voltage the trap depth for those low energy electrons changes and with it the yield of hydrogen ions, leading to the retardation voltage dependent background. This background component cannot be measured directly during normal data taking due to the presence of protons from neutron decay, which result in a signal much larger than the background. Only for the 780 V measurement the background is directly accessible. Figure~\ref{fig:evobg} shows the evolution of the background count rate in the proton integration window after opening the fast neutron shutter (cf. section \ref{sec:proced} for the measuring sequence). The retardation voltage-dependent background represents the non-constant part, the time-evolution of which reflects the filling of the trap, where saturation is reached after a characteristic time constant of about 50 s. Note that the data in Fig. \ref{fig:evobg} were taken during commissioning at a higher pressure than during data taking.

\begin{figure}
	\includegraphics[width=\linewidth]{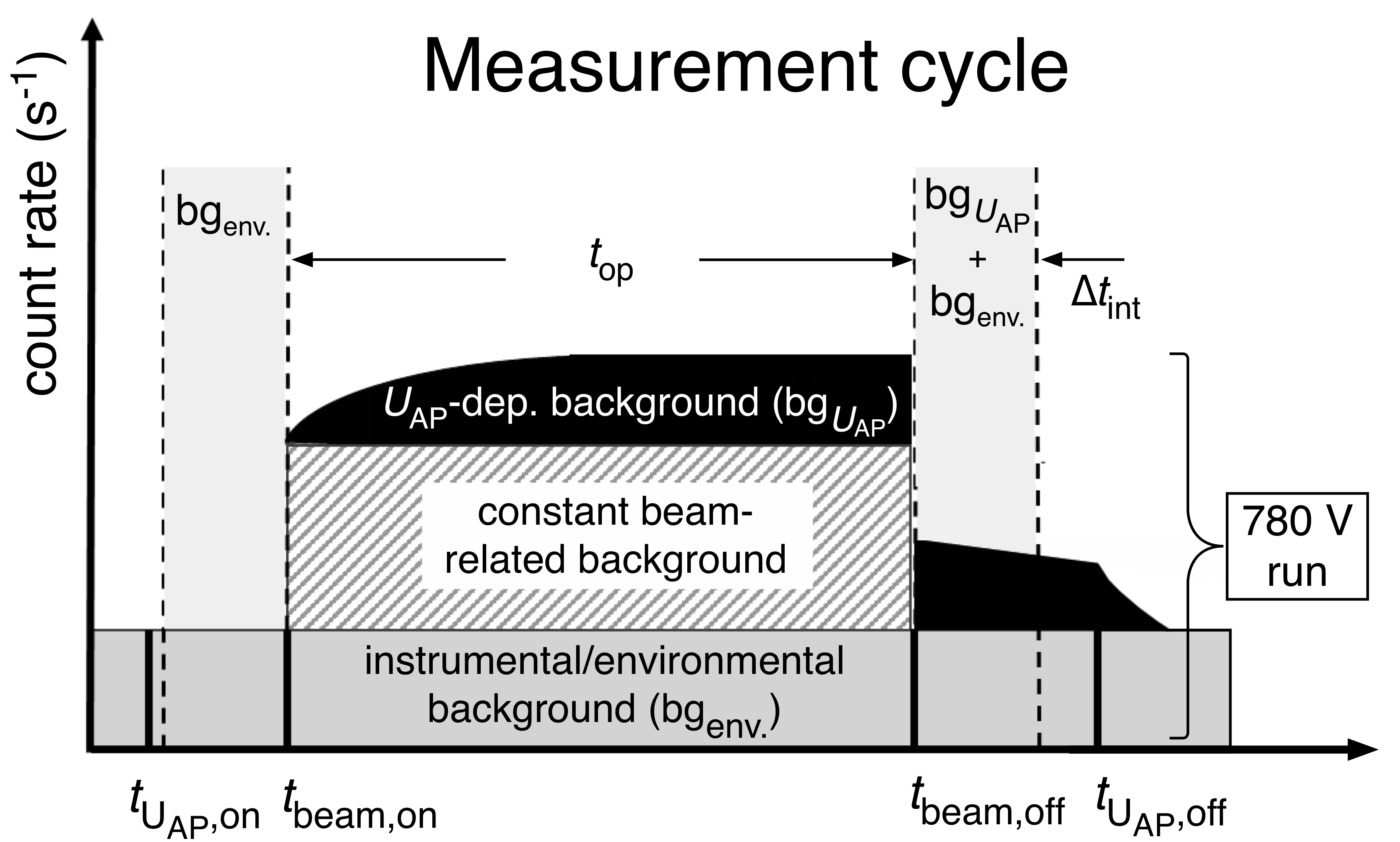}
	\caption{Temporal sequence of a measurement cycle showing the different background contributions (not to scale). The $U_{\text{AP}}$-dependent background ($\text{bg}_{U_{\text{AP}}}$) can be extracted from the counting rate difference of two measurement intervals ($\Delta{}t_{\text{int}}$) displaced in time, one before shutter opening ($t_{\text{beam, on}}$), the other immediately after closing the shutter ($t_{\text{beam, off}}$), \textit{i.e.,} the regions enclosed by vertical dashed lines. For $t\ge{} t_{\text{beam, off}}$, the trap empties again with the time constant $\tau_{\text{s}}$ allowing to monitor the yield of the rest gas ions ($\text{bg}_{U_{\text{AP}}}$).}
	\label{fig:tempcycle}
\end{figure}

For all other voltage settings, this background component must be extracted from the measured count rates in two distinguished time windows of the measurement cycle, the temporal sequence of which is depicted in Fig.~\ref{fig:tempcycle}.

\begin{figure}
	\includegraphics[width=\linewidth]{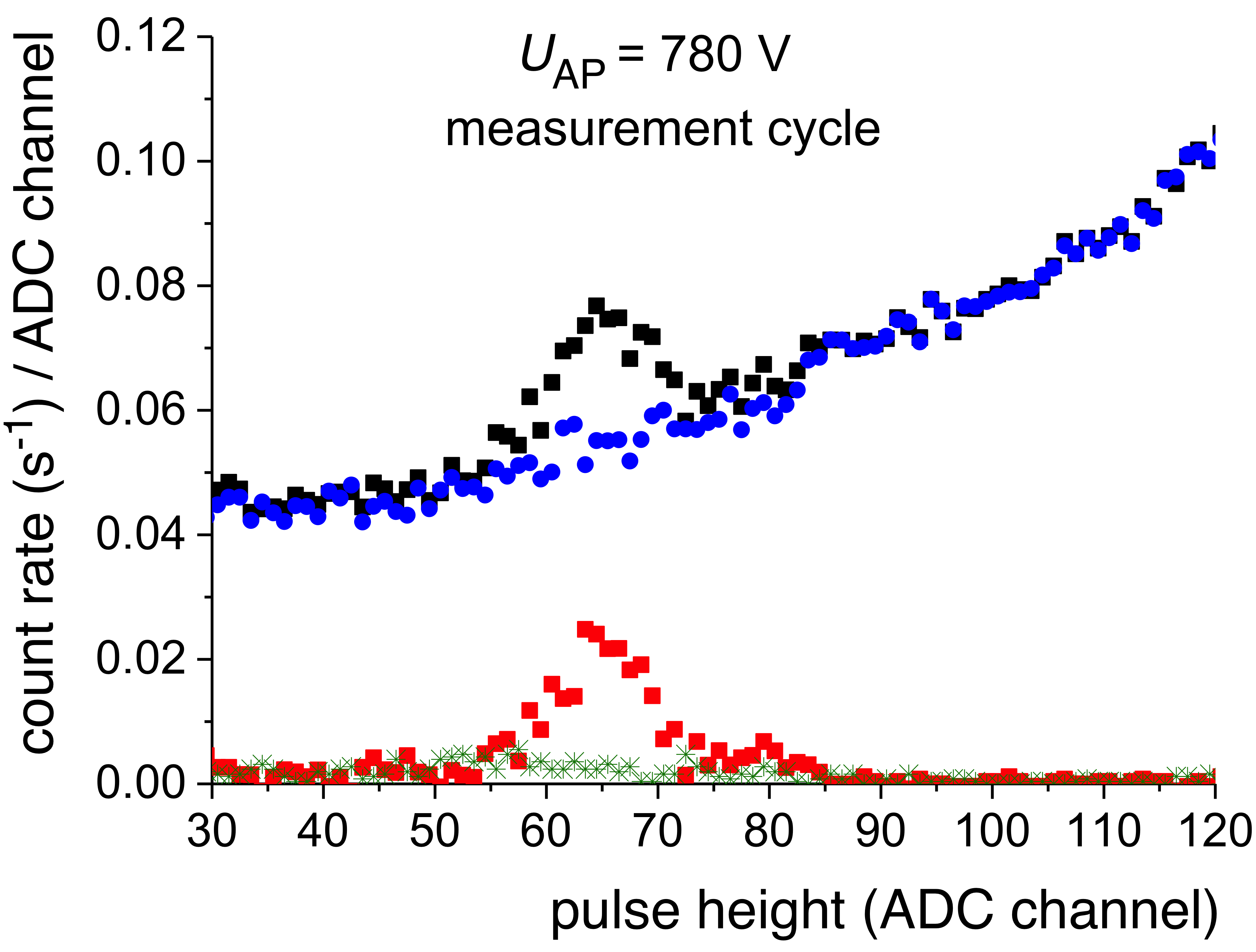}
	\caption{Pulse height spectra measured at $U_{\text{AP}}$ = 780 V within the proton integration window for config 1. Red squares: Spectrum of $\text{bg}^{II}_{\text{780 V}}$ extracted from the difference measurement (cf.~Fig.~\ref{fig:tempcycle}) with $\langle \text{bg}^{II}_{\text{780 V}}\rangle \approx$ 0.34 cps. Black squares: Spectrum measured with beam on (shutter opened) with instrumental/environmental background (green stars) already subtracted. Subtraction of  $\text{bg}^{I}_{\text{780 V}} = \text{bg}^{II}_{\text{780 V}} \cdot{}R$ according to the background model yields the blue data points (essentially electrons from neutron decay) with an integral count rate of $\approx$ 5.75 cps.}
	\label{fig:pulseheight}
\end{figure}

As consistency check, the 780 V measurement cycle apart from a known conversion factor should give the same values for the retardation voltage dependent background rate, once directly extracted from the integral value of the proton-like peak in the pulse height spectrum of Fig.~\ref{fig:detspec} (I) and then from the measurement procedure depicted in Fig.~\ref{fig:tempcycle} (II). A simple background model to describe the build up of ($\text{bg}_{U_{\text{AP}}}$) and its relaxation after shutter closed predicts for the ratio $R$ of the time-averaged background rates with shutter open and after closing the shutter:
\begin{equation}
R = \frac{1-\tau_{\text{s}}/\tau_{\text{op}}}{\tau_{\text{s}}/\Delta{}t_{\text{int}}\left(1-\exp(-\Delta{}t_{\text{int}}/\tau_{\text{s}}) \right)} = (0.9 \pm 0.1) \; .
\label{eq:ratioshutter}
\end{equation}
Eq. (\ref{eq:ratioshutter}) holds for $t_{\text{op}} \gg \tau_{\text{s}}$ which is valid for $t_{\text{op}}$ = 200 s. The chosen time interval is $\Delta{}t_{\text{int}}$ = 20 s (cf. Fig. \ref{fig:tempcycle}). The error bar reflects the uncertainty in $\tau_{\text{s}}$. The direct comparison $\langle \text{bg}^{I}_{780 \; \text{V}} \rangle / \langle \text{bg}^{II}_{\text{780 V}} \rangle \approx$ $0.9$ confirms the expected ratio (cf.~Fig.~\ref{fig:pulseheight}). 

\begin{figure}
	\includegraphics[width=\linewidth]{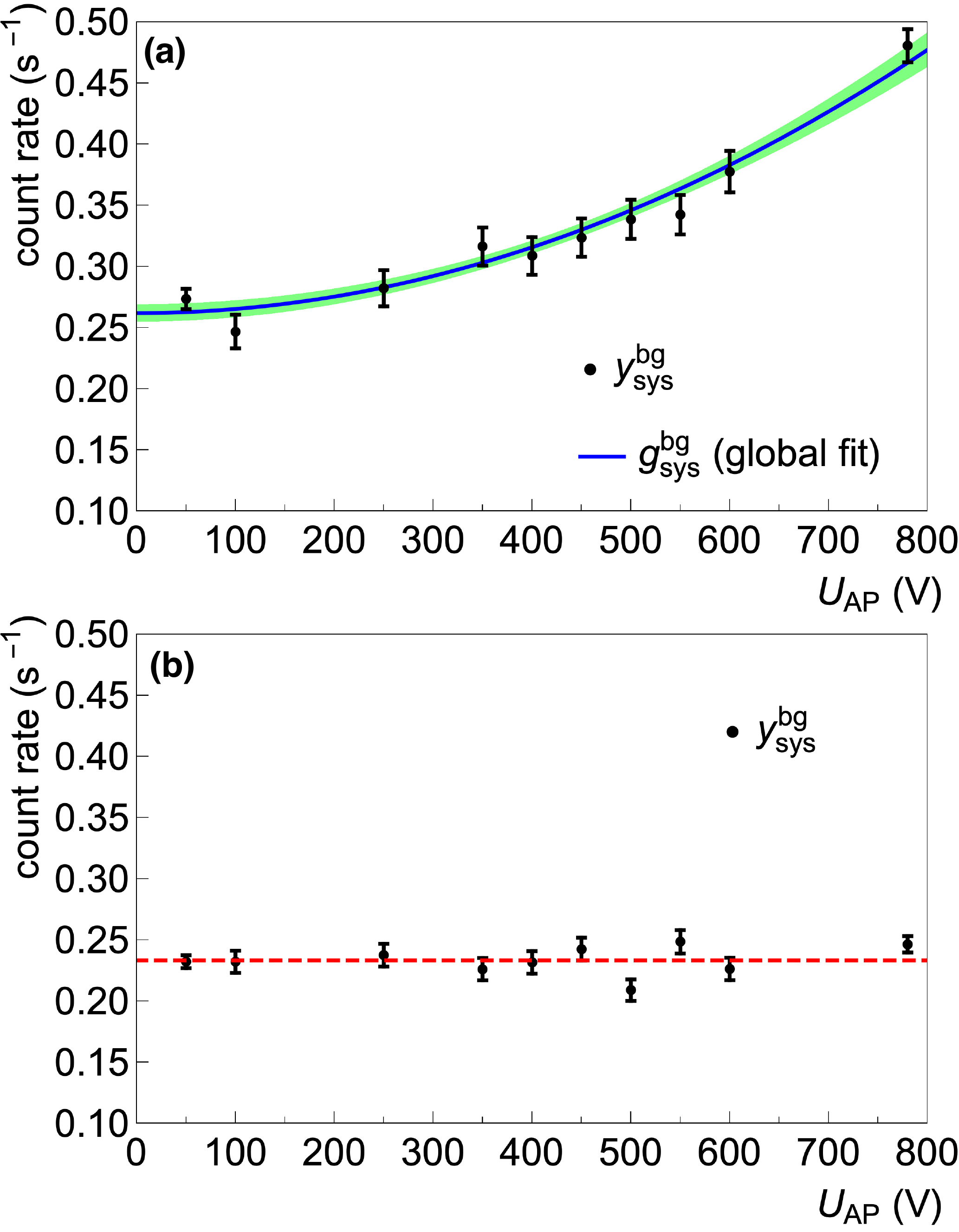}
	\caption{Measured retardation voltage-dependent background in the proton region with pad 2. Shown is the count rate in the measurement interval immediately after $t_{\text{beam, off}}$ (cf.~Fig.~\ref{fig:tempcycle}) in dependence of the applied retardation voltage $U_{\text{AP}}$ for config 1 (a) and config 3 (b). A clear increase of the count-rate is the finding for config 1, whereas a voltage dependency is no longer observed (dashed horizontal line to guide the eyes) for config 3 (as well as for the subsequent configuration runs). The constant instrumental /environmental background contributes with $\langle \text{bg}_{\text{env}} \rangle \approx$ 0.14 cps. Further drawn is the fit result of $g^{\text{bg}}_{\text{sys}}$ for the global fit to the config 1 dataset.}
	\label{fig:backgroundretvolt}
\end{figure}

Since the vacuum conditions inside $a$SPECT continuously improved during the 2013 measurement run, the background from ionized rest gas atoms was steadily decreasing. In addition, the electrode E15 was used as a dipole electrode (E$\times$B drift electrode) which considerably reduced the number density of secondary electrons trapped in the AP region. Therefore, from config 3 on no AP voltage-dependent background could be identified anymore. Figure~\ref{fig:backgroundretvolt} shows the extracted background component $\left(\text{bg}_{\text{env}} + \text{bg}^{II}_{U_{\text{AP}}} \right)_{\text{pad 2}}$ immediately after $t_{\text{beam, off}}$ at the different $U_{\text{AP}}$ voltage settings for config 1 and config 3. To incorporate the retardation voltage dependent background in the fitting procedure, the data have to be added as $y^{\text{bg}}_{\text{sys}, k}$ to the overall dataset with their statistical errors $\Delta y^{\text{bg}}_{\text{sys}, k}$. To these data the following function has been fitted\footnote{The function $g_{\text{sys}}^{\text{bg}}$ was orthogonalized in a way to reduce correlations between other fit parameters below 0.1. Similarly, it was done for $g_{\text{sys}}^{\langle U_{\text{A}} \rangle}$ (cf.~Eq.~(\ref{eq:fitproc})) and $g_{\text{sys}}^{\text{ee}}$ (cf.~Eq.~(\ref{eq:edge})).}
\begin{footnotesize}
\begin{equation}
g_{\text{sys}}^{\text{bg}}\left(U_{\text{AP}}; \{ c_0^{\text{bg}}, c_2^{\text{bg}} \} \right) = c_0^{\text{bg}} + c_2^{\text{bg}} \cdot{}\left(\left(\frac{U_{\text{AP}}}{700~\mathrm{V}} \right)^2 - \frac{1}{3}\right)
\end{equation}
\end{footnotesize}
From config 3 on, the constant fit function $g_{\text{sys}}^{\text{bg}} = c_0^{\text{bg}}$ was sufficient to describe the data. The retardation voltage-dependent term is then included in the fit function of Eq.~(\ref{eq:chisquare}) according to
\begin{footnotesize}
\begin{equation}
f_{\text{sys}}^{\text{bg}}\left(U_{\text{AP}}; \{ c_0^{\text{bg}}, c_2^{\text{bg}} \} \right) = R\cdot{}c_2^{\text{bg}} \cdot{}\left(\left(\frac{U_{\text{AP}}}{700~\mathrm{V}} \right)^2 - \frac{1}{3}\right) + c_{\text{bg}}
\label{eq:termretvoltage}
\end{equation}
\end{footnotesize}
The first term on the RHS has been multiplied by the conversion factor $R$ to adapt it to the real voltage dependent background during `beam on'. The second term represented by the free fit parameter $c_{\text{bg}}$ includes all constant background components, so also $c_0^{\text{bg}}$. After the first two config runs, $f_{\text{sys}}^{\text{bg}}$ of Eq.~(\ref{eq:termretvoltage}) could be replaced by $f_{\text{sys}}^{\text{bg}} = c_{\text{bg}}$.

\subsection{Edge effect}
\label{sec:ee}

The so-called edge effect originates from the gyration of the protons in the magnetic field. The radius of gyration, $r_{\text{g}}$, is the radius of the circular motion of a charged particle ($q$) of mass $m$ in the presence of a uniform magnetic field given by
\begin{equation}
r_{\text{g}} = \frac{m\cdot{}v_{\perp}}{|q|B}
\end{equation}
where $v_{\perp}$ is the component of the velocity perpendicular to the direction of the magnetic field $B$. Hence transmitted protons which arrive close to the edges of the detector\footnote{The detector reaches its full response at a distance $<$ 0.1 mm from its edges \cite{sim2010}. This was measured at PAFF at Technische Universit\"at M\"unchen \cite{mueller2007}.} have a certain probability to be either detected or not, due to their gyration\footnote{The gyration radius of the protons at the height of the detector ($B_{\text{DV}} = $ 4.4 T) is $r_{\text{g}} <$ 1.3 mm.}. The probability to be detected depends on the initial transverse energy $T_{\perp} = T\sin^2\theta$ of the proton and thus via the transmission function on the retardation voltage. Given a homogeneous spatial distribution of the incident neutron beam in the DV, the gain and loss of protons at the edges of the detector cancel. Fig.~\ref{fig:beamprofile} shows our measured neutron beam capture flux profiles along the y-axis. We find an almost linear drop of intensity $-|dI/dy|$ at the site of the detector edges.

\begin{figure}
	\includegraphics[width=\linewidth]{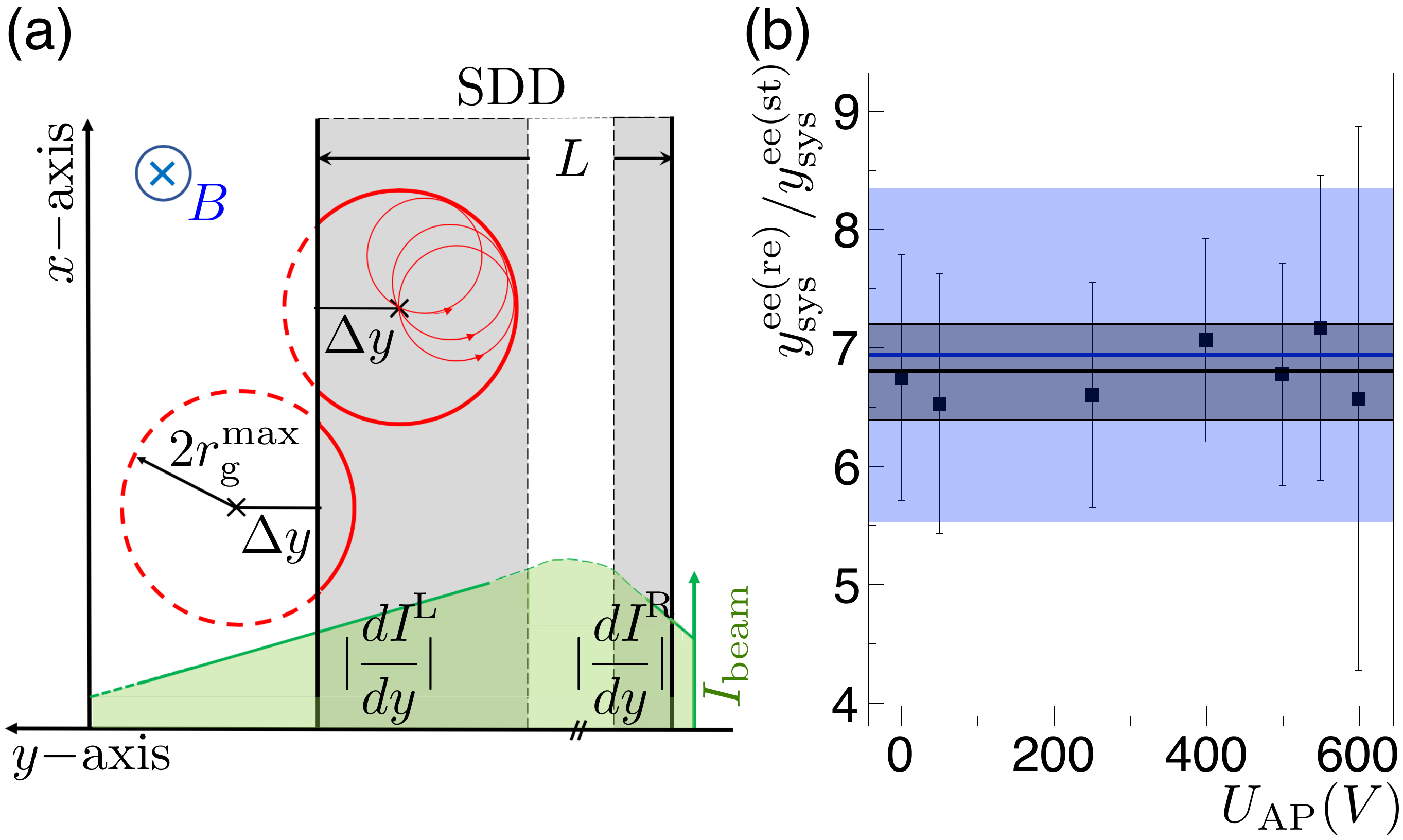}
	\caption{(a) Radial spread $r \le 2\cdot{}r_{\text{g}}^{\text{max}}$ of decay protons at the height of the detector plane emitted from a point source in the DV. The relative count rate losses due to the edge effect are illustrated for two radial probability distributions of gyrating protons at mean distance $\Delta{}y < 2\cdot{}r_{\text{g}}^{\text{max}}$ left and right from the detector edge in case of $|dI/dy| > 0$. (b) Ratio $\left(y_{\text{sys}}^{\text{ee(re)}}/y_{\text{sys}}^{\text{ee(st)}} \right)_{U_{\text{AP}}}$ of relative count rate losses for the reduced and standard beam profile from particle tracking simulations. Within the error bars, no dependence on the chosen retardation voltage settings (7 in total) can be observed (black horizontal line and grey shaded area represent the mean and its standard error). This finding coincides with the simple expression $\langle \varepsilon_{\text{re}} \rangle / \langle \varepsilon_{\text{st}} \rangle$ from Eq.~(\ref{eq:ratiolosses}) which gives (6.9 $\pm$ 1.4) (mean and $\sigma$-error shown in blue).}
	\label{fig:relcountlosses}
\end{figure}

The density of monoenergetic particles per unit area, $dP/dA$, which are isotropically emitted from a point source in a magnetic field is given by \cite{sjue2015}
\begin{equation}
\frac{dP}{dA} = \frac{1}{4\pi\cdot{}r\cdot{}r_{\text{g}}^{\text{max}}}
\end{equation}
with $P = \int_0^{2\cdot{}r_{\text{g}}^{\text{max}}} \; \left(\frac{dP}{dA}\right) \; dA = 1$, $r_{\text{g}}^{\text{max}} = r_{\text{g}}(v_\bot = v)$, and $r \le 2\cdot{}r_{\text{g}}^{\text{max}}$. From that the fraction of particles, $P(\alpha)$, can be derived which hit the detector at distance $\Delta{}y \le 2\cdot{}r_{\text{g}}^{\text{max}}$ left ($P(\alpha)$) and right ($1-P(\alpha)$) from the detector edge as illustrated in Fig.~\ref{fig:relcountlosses}~(a):
\begin{equation}
P(\alpha) = \frac{1}{\pi} \cdot{}\left(\pi/2 - \cos(\alpha)\cdot{}\text{asinh}(\tan(\alpha)) - \text{asin}(\cos(\alpha))\right)
\end{equation}
with $\alpha = \text{acos}(\Delta{}y / (2\cdot{}r_{\text{g}}^{\text{max}}))$.

For the average relative loss across the width $L$ of the detector pad, one finally obtains:
\begin{eqnarray}
\langle \varepsilon \rangle & = & 2\cdot{}\sqrt{\frac{B_{\text{DC}}}{B_0}} \cdot{}\frac{|dI/dy|}{L\cdot{}\langle I_{\text{beam}} \rangle} \nonumber \\
& \cdot{} & 2\cdot{}\int_0^{2\cdot{}r_{\text{g}}^{\text{max}}} \Delta{}y \cdot{} P(\Delta{}y) \cdot{} d(\Delta{}y) \; . 
\end{eqnarray}
Here we assumed $|dI/dy|^{\text{L}} = |dI/dy|^{\text{R}} = |dI/dy|$ and $\langle I_{\text{beam}} \rangle$ being the average beam intensity across the detector acceptance (shaded area in Fig.~\ref{fig:beamprofile}). The factor $\sqrt{\frac{B_{\text{DC}}}{B_0}} \approx \sqrt{2}$ compensates for the reduced slope (cf. Eq.~(\ref{eq:radialdisp})) of $|dI/dy|$ if this quantity is extracted from Fig.~\ref{fig:beamprofile}. From that it results:
\begin{equation}
\label{eq:losspad}
\langle \varepsilon \rangle \approx 0.94\cdot{}\frac{|dI/dy|}{L \cdot{} \langle I \rangle} \cdot{} \left(r_{\text{g}}^{\text{max}} \right)^2 \; .
\end{equation}
For a given retardation voltage $U_{\text{AP}}$ one can formally introduce an effective gyration radius (squared), $\left(r_{\text{g}}^{\text{eff}}(U_{\text{AP}})\right)^2$, which comprises the spectrum of gyration radii for transmitted protons which hit the detector. The latter number must be determined by particle tracking simulations to give precise numbers for the average relative loss rates, in particular their dependence on $U_{\text{AP}}$.

However, for the standard (st) and reduced (re) beam profile the \textit{ratio} of the relative count rate losses results in a simple expression
\begin{equation}
\frac{\langle \varepsilon_{\text{re}} \rangle}{\langle \varepsilon_{\text{st}} \rangle}\approx{}\frac{|dI/dy|_{\text{re}}}{|dI/dy|_{\text{st}}}\cdot{}\frac{\langle I_{\text{beam}}^{\text{st}} \rangle}{\langle I_{\text{beam}}^{\text{re}} \rangle}
\label{eq:ratiolosses}
\end{equation}
which directly can be calculated from Fig.~\ref{fig:beamprofile} (or Table \ref{tab:cr50V}) giving: $\langle \varepsilon_{\text{re}} \rangle / \langle \varepsilon_{\text{st}} \rangle = \left(6.9 \pm 1.4\right)$. The error bar mainly results from the uncertainties in determining the actual slopes $|dI/dy|$ at the detector edges.

\begin{figure}[h]
	\includegraphics[width=\linewidth]{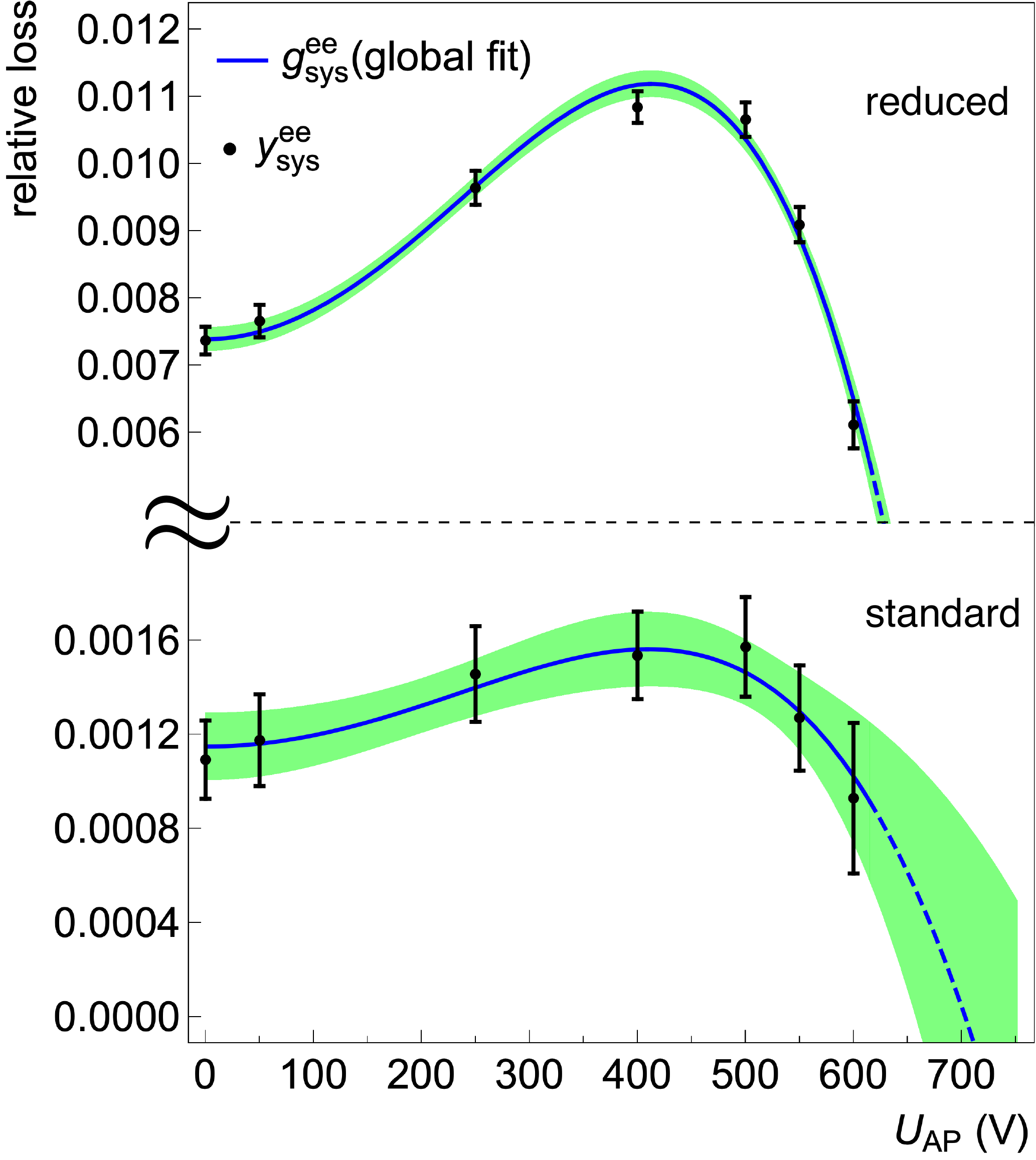}
	\caption{Simulation of the retardation voltage dependence of the relative edge-effect losses $y_{\text{sys}}^{\text{ee(st)}}$ and $y_{\text{sys}}^{\text{ee(re)}}$ for the standard and reduced beam profile, respectively. Further drawn is the fit result of $g_{\text{sys}}^{\text{ee(re)}}(U_{\text{AP}})$ and $g_{\text{sys}}^{\text{ee(rst)}}(U_{\text{AP}})$ for the global fit to the overall data set.}
	\label{fig:simretvoltedge}
\end{figure}

More precise numbers for the edge effect, particularly its dependence on the retardation voltage, are obtained from particle tracking simulations. In these simulations, a homogeneous profile in the DV has been simulated. The actual profiles were then implemented by weighing the simulated homogeneous start distribution with the measured profile distributions. The relative loss rate $y_{\text{sys}}^{\text{ee}}=1-\Re$ then results from the ratio $\Re$ of the simulated hits at the detector with the actual beam profile and the homogeneous one. This procedure easily allows to vary the position of the beam in the DV region relative to the detector to determine the uncertainty due to an overall position uncertainty of $\pm$ 1 mm. The simulations have been performed for each detector pad and measurement configuration separately. It turned out that the differences in the edge effect for pad 2 and 3 are marginal. The same is true for the differences between configurations measured with the same beam profile. Therefore, results of the different pads and configurations have been combined. The resulting relative edge-effect losses at different retardation voltages are shown in Fig.~\ref{fig:simretvoltedge}. The uncertainty $\Delta{}y^{\text{ee}}_{sys}$ incorporates the MC statistics and the uncertainty in the beam position ($\pm$ 1 mm). For the ratio $\left( y_{\text{sys}}^{\text{ee(re)}}/y_{\text{sys}}^{\text{ee(st)}}\right)_{U_{\text{AP}}}$ we obtain the data points depicted in Fig.~\ref{fig:relcountlosses} (b). Within error bars, these ratios show no dependence on the retardation voltage with their mean given by $\langle y_{\text{sys}}^{\text{ee(re)}}/y_{\text{sys}}^{\text{ee(st)}} \rangle$ = (6.8 $\pm$ 0.4). This result is in very good quantitative agreement with the ratio $\langle \varepsilon_{\text{re}} \rangle / \langle \varepsilon_{\text{st}} \rangle$ (cf.~Eq.~(\ref{eq:ratiolosses})) in which only the characteristics of the respective beam profile\footnote{This comparison serves as a consistency test between a simple estimation and a complex simulation of the edge effect, which of course increases the confidence in the results.} enter. The data depicted in Fig.~\ref{fig:simretvoltedge} can be described by the function
\begin{eqnarray}
\label{eq:edge}
& & {} g_{\text{sys}}^{\text{ee}}\left(U_{\text{AP}}; \{ c_0^{\text{ee}}, c_2^{\text{ee}}, c_4^{\text{ee}} \} \right) \nonumber \\
& & {} = c_0^{\text{ee}} + c_2^{\text{ee}} \cdot{}\left( \frac{3\cdot{}\left(\frac{U_{\text{AP}}}{700~\mathrm{V}}\right)^2 - 5\cdot{}\left( \frac{U_{\text{AP}}}{700~\mathrm{V}} \right)^4}{8} \right) \nonumber \\
& & {} + c_4^{\text{ee}} \cdot{}\left( \frac{3\cdot{}\left(\frac{U_{\text{AP}}}{700~\mathrm{V}}\right)^2 + 5\cdot{}\left( \frac{U_{\text{AP}}}{700~\mathrm{V}} \right)^4 - 2}{8} \right) \; .
\end{eqnarray} 
The relative edge-effect losses are then included in the fit function of Eq.~(\ref{eq:chisquare}) according to
\begin{footnotesize}
\begin{equation}
\label{eq:rellosses}
f_{\text{sys}}^{\text{ee}} = \left(-g_{\text{sys}}^{\text{ee}}\left(U_{\text{AP}}; \{ c_0^{\text{ee}}, c_2^{\text{ee}}, c_4^{\text{ee}} \} \right)\right)\cdot{}y_{\text{theo(n)}}
\end{equation}
\end{footnotesize}
with $y_{\text{theo(n)}}$ from Eq.~(\ref{eq:fitn}).

\subsection{Backscattering and below-threshold losses}
\label{sec:lld}

Protons reaching the detector can get backscattered due to scattering off the nuclei of the detector material (silicon). Consequently, these protons deposit only a fraction of their kinetic energy inside the active detector volume and the resulting pulse height may fall below the threshold of the DAQ system. Backscattering depends on the energy of the proton, $15 \; \text{keV} < (15 \; \text{keV} + T) < 15.75 \; \text{keV}$, and its impact angle. The distribution of both quantities is affected by the applied retardation voltage, $U_{\text{AP}}$. The $U_{\text{AP}}$-dependence of the detection efficiency may change the value $a$ extracted from the integral proton spectrum.

The protons relevant for $a$SPECT have a very short range in the detector (about 200~nm), thus the efficiency for proton detection is extremely sensitive to the detector properties near the surface. A proton penetrating the detector first needs to penetrate the entrance window, which is comprised of 30~nm of aluminium. Free charge carriers produced by the proton in this region will not be detected. Even after the entrance window, not all charge carriers will be collected in the central anode of the SDD. Close to the surface, a large fraction of the created electron-hole pairs will recombine. The charge-collection efficiency at the border ($z' = 0$) between the entrance window and active silicon bulk is approximately 50~\% and rises with increasing depth ($z'>0$) according to the following equation \cite{pop2000}:

\begin{math}
  f_{\text{CCE}}(z) = \left\{
    \begin{array}{l}
      0 \hspace{3.22cm} \text{for} \; z' = z-\Delta z < 0 \\
      S + B\left(\frac{z}{L}\right)^c  \hspace{1.58cm} \text{for} \; 0 \le z' \le L \\
      1 - A\cdot{}\exp{}\left(-\frac{z-L}{\tau}\right) \hspace{0.25cm} \text{for} \; L < z' \le D
    \end{array}
  \right. \\
\end{math}
\begin{eqnarray}
\text{with} \; A & = & (1-S)\frac{\tau\cdot{}c}{L+\tau\cdot{}c} \nonumber \\
B & = & (1-S)\left(1-\frac{\tau\cdot{}c}{L+\tau\cdot{}c}\right)
\end{eqnarray}
With $\Delta z$ we introduced an additional parameter which characterizes the effective thickness, $d_\mathrm{eff}$, of the SDD deadlayer with $d_{\mathrm{eff}} =30~\mathrm{nm}+ \Delta z$.
The total thickness $D$ of the detector is about 450~$\mu$m. As this is much thicker than the maximum penetration depth of low energy protons, the exact thickness of the detector is of minor importance. In order to determine the four remaining parameters, $S, c, \tau,$ and $L$ of the charge-collection efficiency function (including $\Delta z$), one has to calculate the effective deposited ionization energy of each proton
\begin{equation}
E_{\text{eff ion.}} = \sum_i E_{\text{ion.}}(z_i)\cdot{}f_{\text{CCE}}(z_i) \; ,
\end{equation}
a quantity which is proportional to the measured pulse height. Hence, the histogram of $E_{\text{eff ion.}}$ from many simulated proton events reproduces our pulse height spectra, if the parameters of $f_{\text{CCE}}$ are correct. The ionization depth profile $E_{\text{ion.}}(z_i)$ is determined by the SRIM code \cite{srim2008} (version 2012.3). SRIM is a collection of software packages which calculate many features of the transport of ions in matter, here in particular the amount of ionization, \textit{i.e.,} the amount of electron-hole pair production in silicon caused by a penetrating proton. For this purpose, the depth of 300~nm was partitioned into 100 bins ($z_i$) of 3~nm depth. The first 10 bins account for the aluminium cover layer, the following 90 for the active silicon bulk of the detector.

\begin{figure}
	\includegraphics[width=\linewidth]{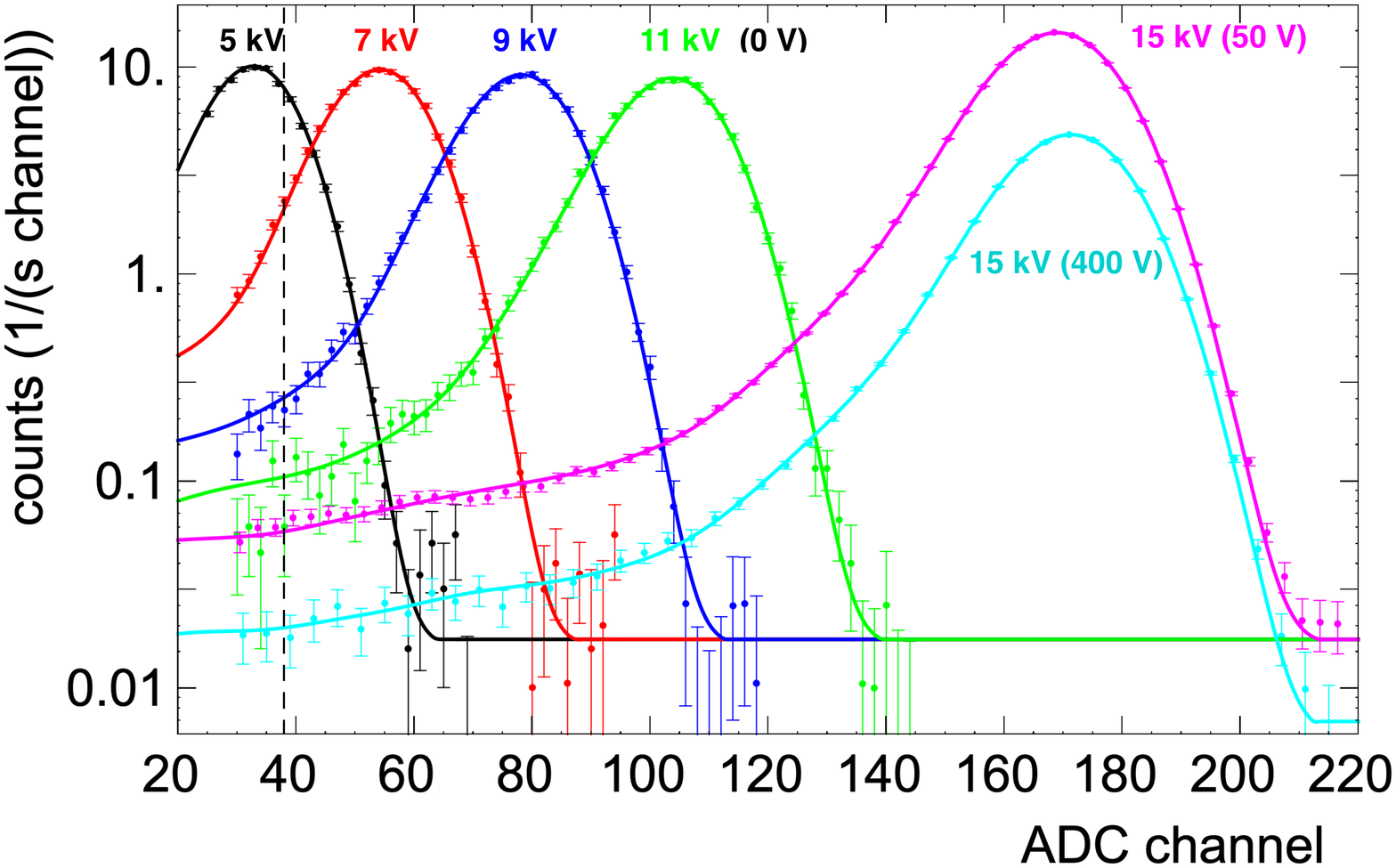}
	\caption{Measured pulse height spectra (pad 2) at different acceleration voltages, $U_{\text{acc}}$, and retardation voltages, ($U_{\text{AP}}$). The data were taken with a linear shaper at the end of the 2013 beam time to avoid spectral distortions, which would otherwise make the combined fitting cumbersome. The corresponding histograms of the calculated pulse heights $\propto{}E_{\text{eff ion.}}$ are shown as continuous lines to improve readability. From the fit, the parameters of the charge-collection efficiency could be deduced: $S = 0.5562(10)$, $L = 38.01(27)$~nm, $c = 3.171(23)$, and $\tau =  84.39(37)$~nm. For $\Delta z$ we obtained $\Delta z= 6.54(4)~\mathrm{nm}$ resulting in $d_\mathrm{eff}= 36.54(4)~\mathrm{nm}$. From the comparison with pulse height data taken with the logarithmic shaper (see text), we determined the proper threshold (ADC channel: 37.6; dashed line) which was needed to calculate the below-threshold losses including the events with no energy deposition inside the detector.}
	\label{fig:phsaccel}
\end{figure}

Figure~\ref{fig:phsaccel} shows a combined fit to different pulse height spectra (measured with the linear shaper) with one common parameter set of $f_{\text{CCE}}$ together with the result. From the fits, the calibration constant to convert ADC channel to ionization energy was also extracted. Because the two detector pads (2, 3) have different gains and differences also in their charge-collection efficiency, each pad has to be treated separately.

For the SRIM simulation, the physical $(E, \alpha)$ distributions of protons impinging on the detector ($U_{\text{acc}} := -U_{\text{DC}} = +15$~kV and $U_{\text{AP}} = 0$~V) were extracted from the particle tracking simulations, \textit{i.e.,} a data set of $\approx{}3\times{}10^{6}$ protons which hit pad 2 and about the same amount which hit pad 3. The corresponding distributions of energy $E'$ and the impinging angle $\alpha'$ for acceleration voltages $U_{\text{acc}}$ less than 15~kV could be deduced from those by using:
\begin{eqnarray}
E' & = & E- e\cdot{}\left(15 \; \text{kV} - U_{\text{acc}} \right) \; \text{and} \nonumber \\
\alpha' & = & \arctan{}\sqrt{\frac{E\cdot{}\sin^2(\alpha)}{E\cdot{}\cos^2(\alpha) - e\cdot{}\left(15 \; \text{kV} - U_{\text{acc}} \right)}}
\end{eqnarray}
For retardation voltages $U_{\text{AP}} > 0$~V, protons for which the following inequality holds were filtered out from the data set
\begin{equation}
T\left(1-r_{\text{B}}\cdot{}\sin^2(\theta) \right) > e\cdot{}U_{\text{AP}}
\end{equation}
(truncation of the simulated parameter space $(T, \theta)$ at the decay point in the DV).

Backscattered protons may return to the detector after motion reversal due to the electrostatic potential of the AP electrode. Those protons hit the detector again with the energy and angle to the normal they had when leaving the dead layer. In the simulation, all possible hits of a proton due to backscattering were taken into account by adding the collected charge from all hits in the active region of the detector.

In order to extract the detection efficiency from the simulated pulse height spectra, we analyzed the pulse height spectra at different acceleration voltages measured with the logarithmic shaper. The empirically-found functional relationship between peak position (ADC channel) and $U_{\text{acc}}$ gives us the respective acceleration voltages at the experimentally-set lower integration limits, \textit{i.e.,} $U_{\text{acc}}=5.75$~kV @ ADC channel 29 (pad 2) and 4.97~kV @ ADC channel 28 (pad 3). Transferred to the $U_{\text{acc}}$-dependent course of the peak position in case of the linear shaper, this method determines the relevant lower threshold in the respective region, \textit{i.e.,} pad 2: 37.6 and pad 3: 34.5.

\begin{figure}
	\includegraphics[width=\linewidth]{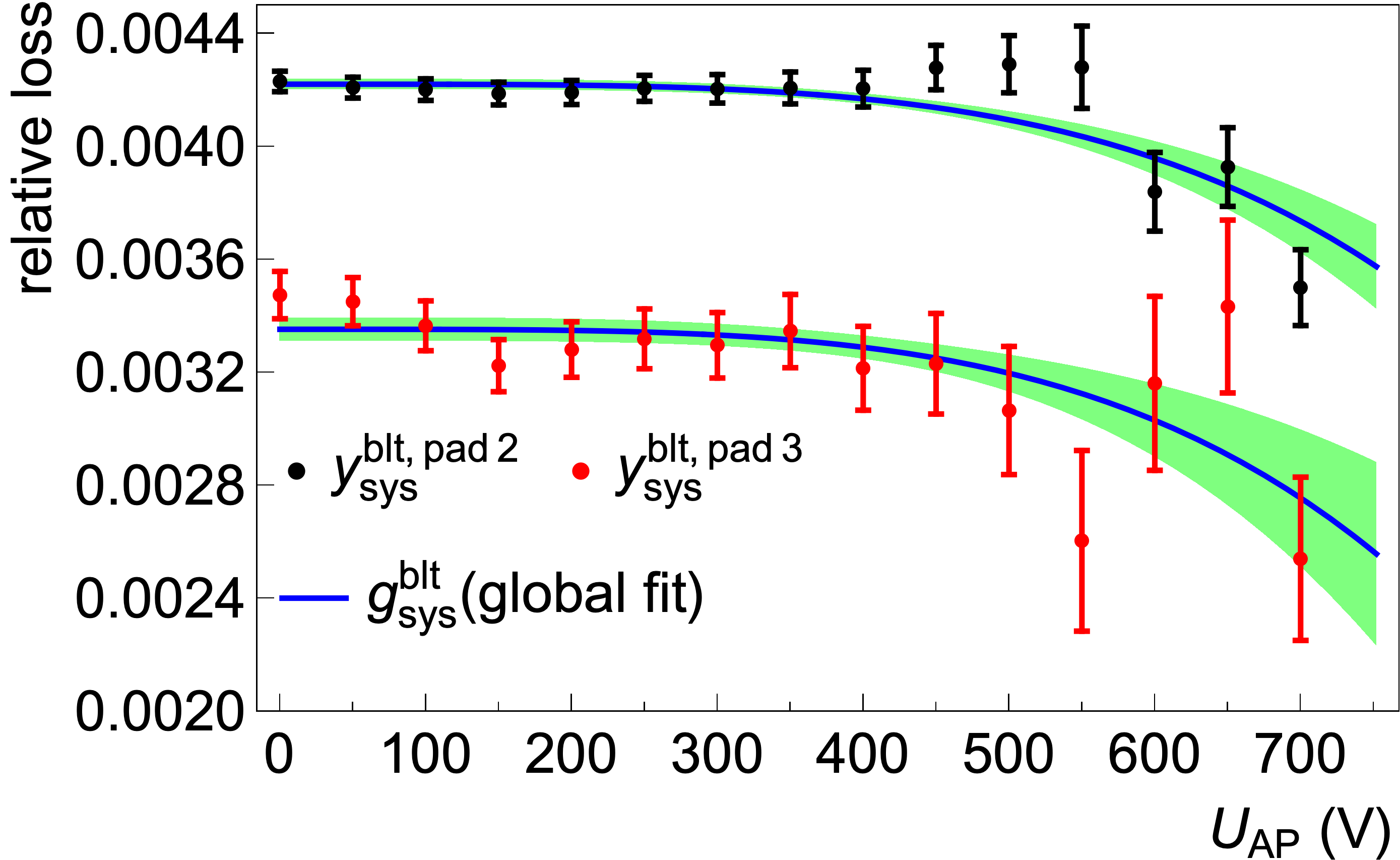}
	\caption{Fraction of undetected protons of the integral proton spectrum whose corresponding pulse heights fall below the threshold of the DAQ system. The $U_{\text{AP}}$-dependence of the fractional losses for pad 2 and pad 3 can be described by Eq. (\ref{eq:lossesretvolt}). The solid lines are the fit functions $g_{\text{sys}}^{\text{blt}}(U_{\text{AP}})$ from the global fit to the overall data set.}
	\label{fig:fraundetected}
\end{figure}

Finally, the number of simulated events below these thresholds includes the events with no energy deposition inside the detector. Figure \ref{fig:fraundetected} shows the fractional loss obtained in those calculations.

Equation (\ref{eq:lossesretvolt}) describes the dependency of these losses on the retardation voltage, $U_{\text{AP}}$:
\begin{equation}
\label{eq:lossesretvolt}
g_{\text{sys}}^{\text{blt}}\left(U_{\text{AP}}; \{c_0^{\text{blt}}, c_4^{\text{blt}} \} \right) = c_0^{\text{blt}} + c_4^{\text{blt}}\cdot{}\left(U_{\text{AP}}\right)^4
\end{equation}
The factor, which has to be included in the fit function of Eq. (\ref{eq:chisquare}) to account for these losses with $y_{\text{theo(n)}}$ from Eq. (\ref{eq:fitn}) is given by
\begin{equation}
f_{\text{sys}}^{\text{blt}}=\left(-g_{\text{sys}}^{\text{blt}}\left(U_{\text{AP}}; \{ c_0^{\text{blt}}, c_4^{\text{blt}} \} \right) \right)\cdot{}y_{\text{theo(n)}} \; .
\end{equation}
In this context, we also investigated how a change of the threshold of the DAQ system affects the integral 
proton spectra. A change of the threshold of $\pm$~5~\% changes the $c_4^{\text{blt}}$ coefficient also by 5~\% which corresponds to a fraction of 0.3 of its standard error, a small effect we could include in the error bars of the simulation results shown in Fig. \ref{fig:fraundetected}. The change of the $c_0^{\text{blt}}$ coefficient is 5~\% of its value and much bigger than its standard error. But because the $c_0^{\text{blt}}$ coefficient is just a constant completely independent of the spectral shape of the integral proton spectra, it does not contribute to our error budget.

\subsection{Dead time and pile-up}
\label{sec:uld}

\begin{figure}[h]
\includegraphics[width=\linewidth]{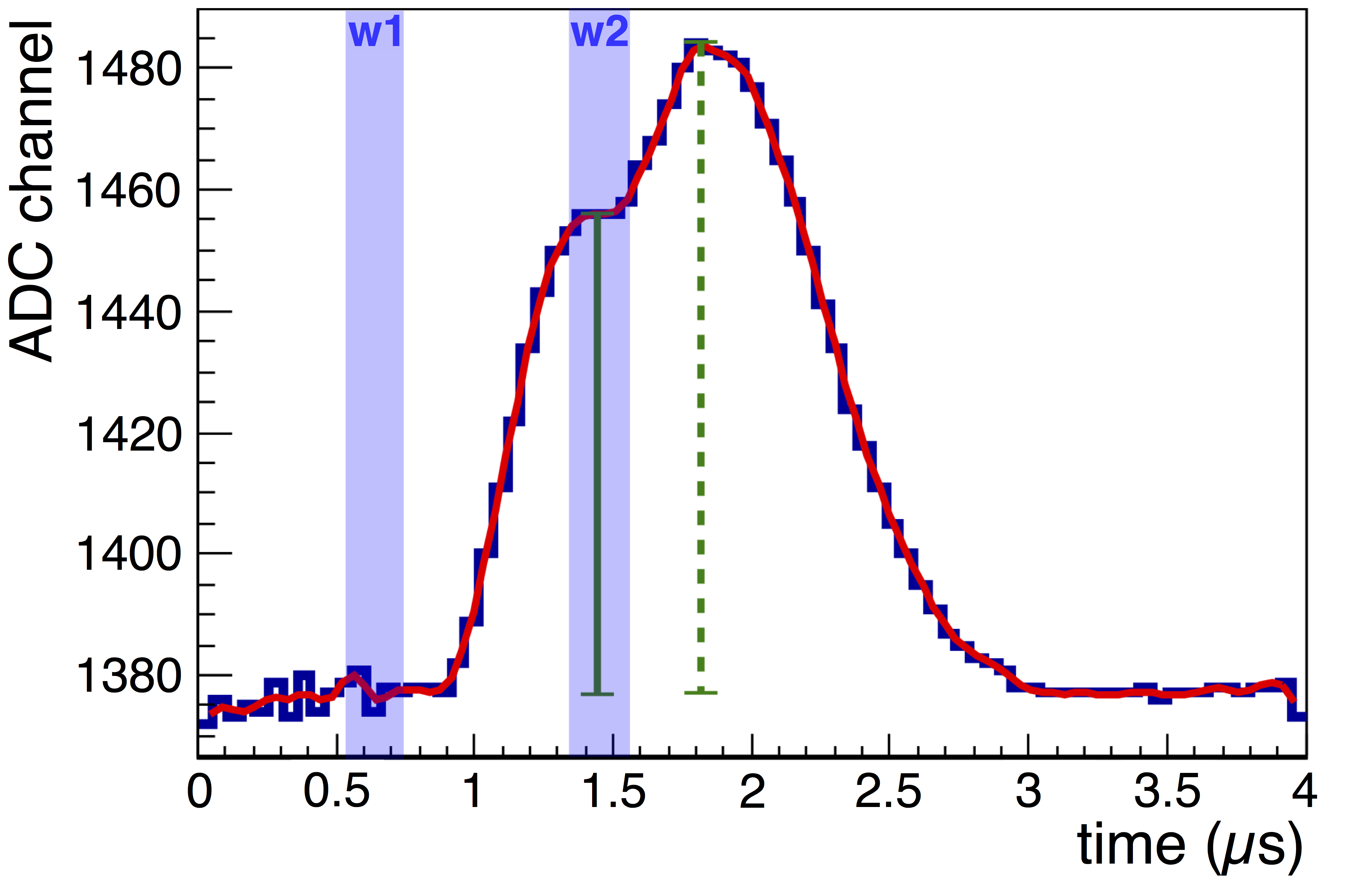}
\caption{Example of two individual proton events within one event window. The signal as recorded by the ADC is shown in blue, the spline interpolation in red. The position of the two events is indicated by the vertical green lines. The trigger algorithm is based on the comparison of two windows (w1, w2) within the shift register the data from the ADC is continuously shifted through. If the mean values of those two windows differ by more than an externally set threshold, the trigger condition is fulfilled. Window w1 is used to determine the baseline (first 15 time bins of 0.75 $\mu$s), whereas w2 is separated from w2 by 0.8 $\mu$s. For the subsequent signal analysis, the baseline is subtracted in each case (cf.~Fig.~\ref{fig:detspec}).\label{fig:twoevents}}
\end{figure}

The dead time of the DAQ as well as the pile up both depend on the total count rate. This rate in turn depends, a.o., on the retardation voltage. Hence, both effects introduce a retardation voltage-dependent effect. As described in \cite{sim2010}, $a$SPECT uses a sampling ADC\footnote{Sampling frequency is 20 MHz, resulting in time bins with a width of 50 ns.}. If a trigger has occured, the ADC values for a time window of 4 $\mu$s (event window) are stored in a memory buffer (cf. Fig.~\ref{fig:twoevents}). A second event arriving within this time will be recorded in the same event window. Due to the nature of the trigger (the DAQ system processes an event in 0.2 $\mu$s) the next event window has a minimum time difference of $T_{\text{dead}}$ = 4.2 $\mu$s. As per event window only one event, namely the first one, is counted in the analysis, a non-extendable dead time correction \cite{leo1994} has been performed in the following way
\begin{equation}
y_{\text{exp}} = \frac{y_{\text{exp}}^{\text{meas}}}{1-y_{\text{tot}}^{\text{meas}}\cdot{}T_{\text{dead}}} \; .
\label{eq:deadtimecorr}
\end{equation}
$y_{\text{tot}}^{\text{meas}}$ is the total count rate detected\footnote{In config 1 (pad 2), the total count rate at 50 V was $y_{\text{tot}}^{\text{meas}}\approx{}$ 530 cps with the following partial count rates in the respective integration regions: 439 : 74 : 17 for $N_{\text{p}}(50 \; V)$ : $N_{\text{el}}$ : $N_{\text{noise}}$.}, whereas $y_{\text{exp}}^{\text{meas}}$ is the measured integral count rate in the proton region. This correction for the dead time has been applied to the pulse height spectra of each pad, retardation voltage and configuration separately, resulting in $y_{\text{exp}, i}^{c, p}$ used for the analysis (cf.~section \ref{sec:fit}). It is important to know $T_{\text{dead}}$ precisely in order to apply a good correction. With $\Delta{}T_{\text{dead}}$ unknown by $\pm$ 50 ns, the uncertainty results in a negligible systematic error of $\Delta{}a/a\approx{}\pm{}$ 0.04 \%. This has been extracted from Eq.~(\ref{eq:countrate}) using the reference value $a_{\text{ref}}$ \cite{kon2011}. In the dead time correction of Eq.~(\ref{eq:deadtimecorr}) it is assumed that the events are occurring randomly, \textit{i.e.,} obey Poisson statistics. This, however, is not fulfilled since in $a$SPECT a maximum of 13.1 \% of the decay electrons (electron count rate: $N_{\text{el}}$) can be detected in coincidence with their correlated proton (proton count rate: $N_{\text{p}}$) \cite{kon2011}. In the experiment in the limit $U_{\text{AP}}\rightarrow{}$ 0 V (Fig.~\ref{fig:spectrum} (b)) we observe a slightly larger number of $N_{\text{el}}/N_{\text{p}} \; (0 V) \approx$ 16 \%, due to electron backscattering \cite{kon2011}. The influence of these correlated events on the deadtime correction (Eq.~(\ref{eq:deadtimecorr})) has been investigated by MC simulations. The total count rate can be decomposed according to
\begin{eqnarray}
y_{\text{tot}}^{\text{meas}}(U_{\text{AP}}) & = & N_{\text{p}}(U_{\text{AP}}) + N_{\text{el}} + N_{\text{noise}} \nonumber \\ & \equiv & \left(0.68\cdot{}N_{\text{p}}(U_{\text{AP}}) + N_{\text{el}} + N_{\text{noise}}\right) \nonumber \\ & + & 2\cdot{} \left(0.16\cdot{}N_{\text{p}}(U_{\text{AP}})\right) \; , 
\label{eq:totalcr}
\end{eqnarray}
where $N_{\text{noise}}$ denotes the rate of the electronic noise. The first term on the RHS represents the uncorrelated count-rate events, which are randomly distributed. The second term gives the rate of correlated electron/proton pairs. The time difference between correlated pairs (TOF spectrum) can be parametrized by a log-normal distribution
\begin{equation}
y=U_{\text{peak}}\cdot{}\exp{}\left(-\frac{1}{2}\left(\frac{\ln{}((t-t_0)/\tau{})}{\sigma}\right)^2 \right)
\end{equation}
where the minimum TOF of decay protons detected with their correlated electrons is $t_0 =$ 7.2 $\mu$s for $U_{\text{AP}}=50$ V up to $t_0$ = 10.0 $\mu$s for $U_{\text{AP}}$ = 600 V with $\tau{}\approx$ 2.8 $\mu$s and $\sigma{}\approx$ 0.7, typically \cite{kon2011}. In the MC simulation, the count rate events from $0.16\cdot{}N_{\text{p}}(U_{\text{AP}})$ are again randomly distributed over the unit time interval of 1 s and the associated proton events are added with a time offset that reflects the TOF spectrum. Finally, dead time losses are determined by the query: $t_{i+1} - t_i \le 4.2 \; \mu$s; in chronological order of the simulated events which differ due to the retardation voltage dependence of the total count rate (cf.~Eq.~(\ref{eq:totalcr})). The simulation showed that the inclusion of correlated events in the dead time correction shifts the $a$ coefficient by $|(a_{\text{corr}} - a_{\text{uncorr}}) / a_{\text{uncorr}}| = 0.1 \%$ compared to Eq.~(\ref{eq:deadtimecorr}) which assumes a purely statistically distributed event rate. Therefore, in our dead time correction this effect was taken into account.

\begin{figure}[h]
\includegraphics[width=\linewidth]{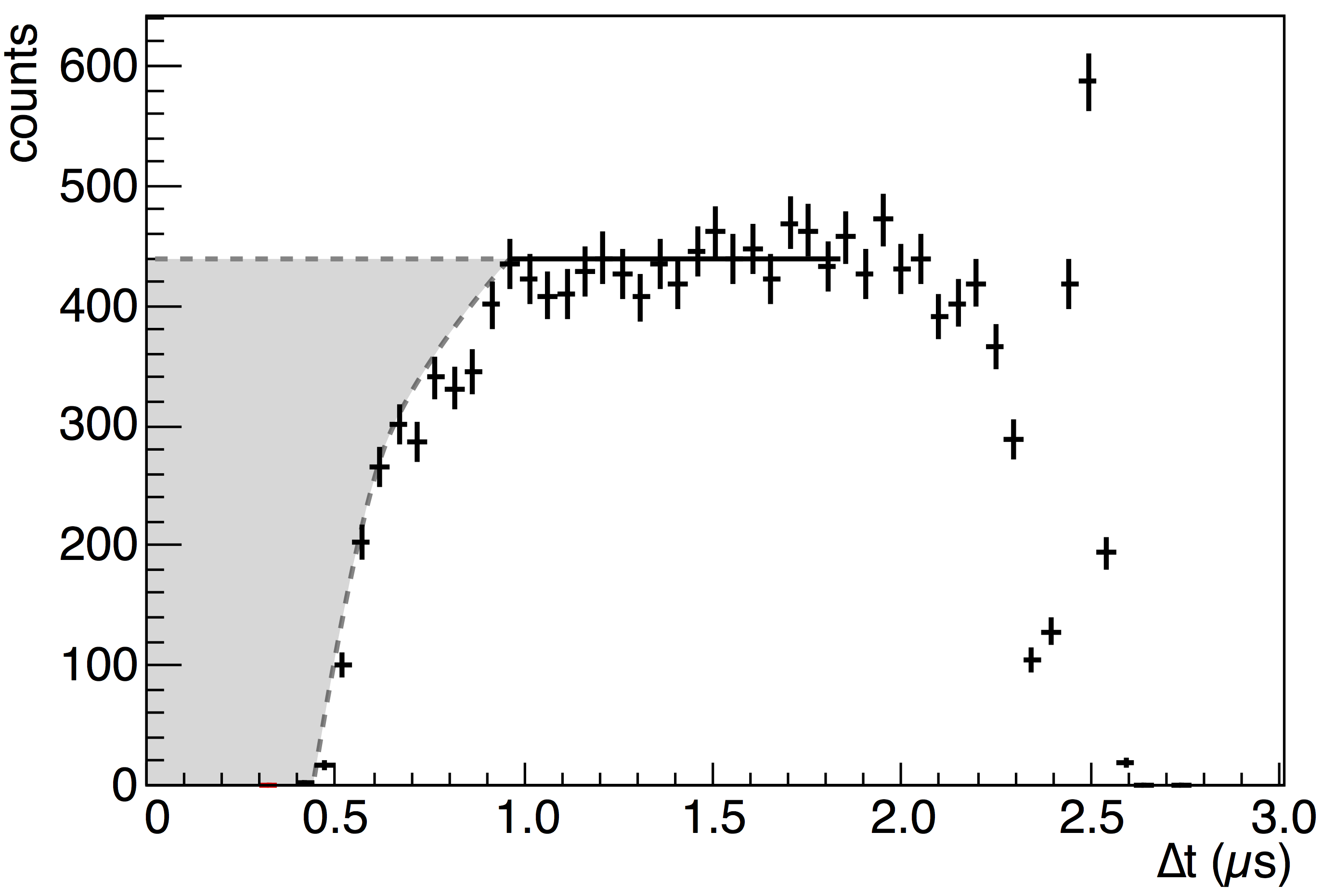}
\caption{Quantitative determination of pile up events (blue area) within the event window. The counts per time bin (50 ns) of the separated events (crosses) which reach a plateau at 1 $\mu$s $\le{}\Delta{}t_{\text{s}} \le{}2 \; \mu$s are extrapolated to $\Delta{}t_{\text{s}}\rightarrow{}0$. The integral number of pile up events divided by the measurement time is then the rate of pile up events. The data shown are from config 1 (pad 2) at $U_{\text{AP}}$ = 50 V with the total measuring time of 38600 s. The reduction of the distribution above 2 $\mu$s is caused by the finite length of the event window in which the maximum of the 2$^{\text{nd}}$ pulse no longer falls.\label{fig:pileupquantitative}}
\end{figure}

For a proper pulse height determination, possible multiple pulses within the same event window have to be separable. In Fig.~\ref{fig:twoevents}, two pulses are shown occurring within one event window. To determine the correct pulse height of the first pulse, a spline interpolation $f_{\text{spline}}(t)$ has been performed. Using a simple curve sketching, pulse maxima, inflection points, etc. can be identified which allows to reconstruct the true pulse height (\textit{i.e.,} that of the first, triggering event) even in case of overlapping pulses. This method works down to a separation time between the two pulses of $\Delta{}t_{\text{s}} \approx$ 0.5 $\mu$s (cf. Fig.~\ref{fig:pileupquantitative}). Two pulses with closer separation can only be partially differentiated or not at all which will lead to pile-up events. This effect is rate- and thus $U_{\text{AP}}$-dependent and has to be accounted for. In a first step, all events with two clearly separated peaks and the first peak having a pulse height in the proton region are identified. To this pulse height of the first peak, the pulse height of the second is added. If the sum of both pulse heights is higher than the upper integration limit, the event is selected. This selection ensures that only events from the proton region are taken, in which a pile up would push the first peak out of the proton region\footnote{Pile up events which would still be within the proton region are not considered, as they are counted anyway.}.

\begin{figure}[h]
\includegraphics[width=\linewidth]{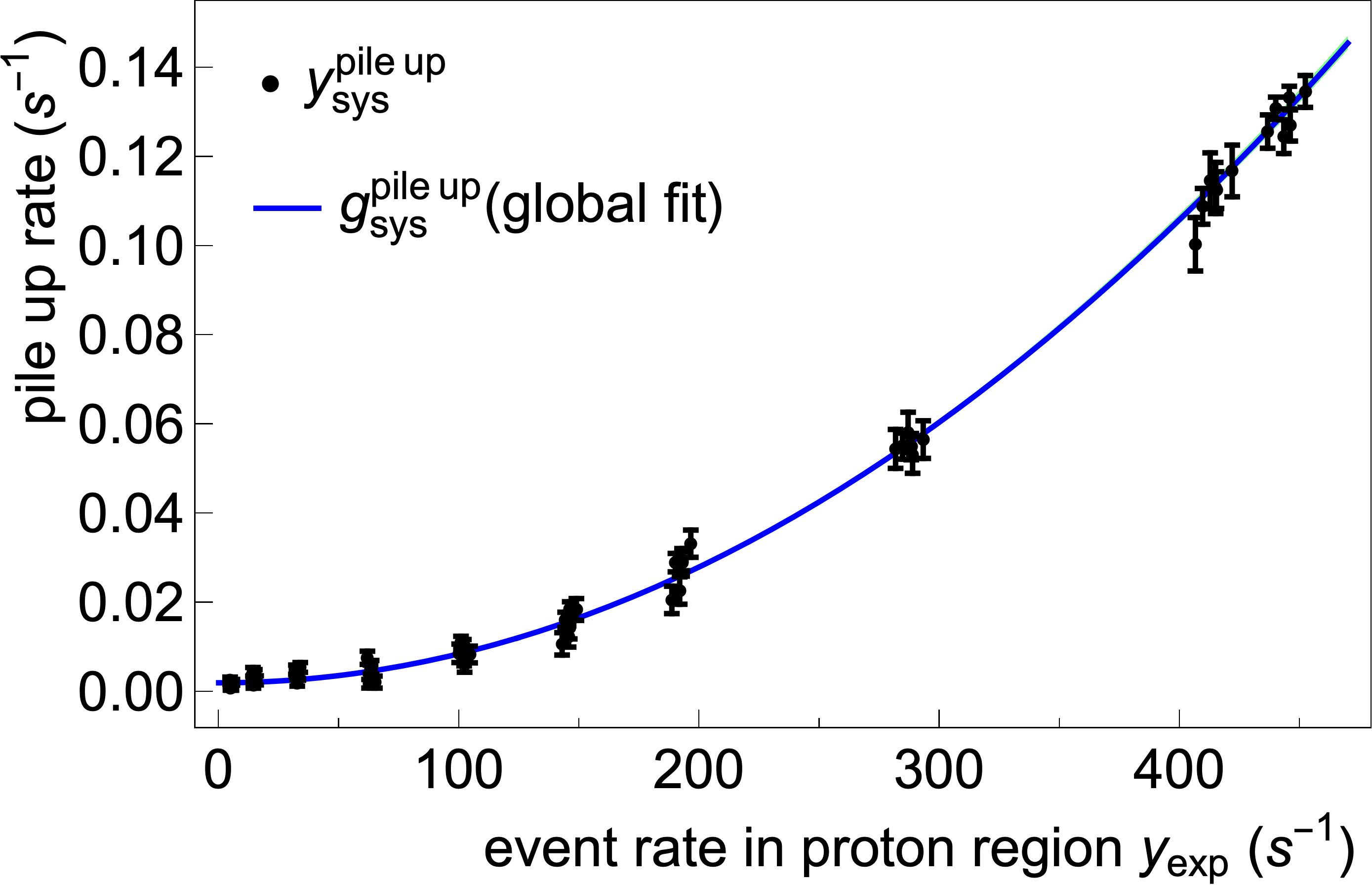}
\caption{Pile up rate $y_{\text{sys}}^{\text{pile up}}$ as function of the event rate $y_{\text{exp}}$ in the proton region. Since $y_{\text{exp}}$ depends, a.o., on the retardation voltage, the pile up results in a retardation voltage-dependent loss. Data taken from config 1, 3, 7, pad 2 and 3 are bundled in clusters for a given retardation voltage setting. Further drawn is the global fit result of $g_{\text{sys}}^{\text{pile up}}$ to the overall data set.\label{fig:pileuprate}}
\end{figure}

By counting all separable double pulses in the event window and creating their distribution as a function of their respective separation times $\Delta{}t_{\text{s}}$, the fraction of pile up events can be determined (cf.~Fig.~\ref{fig:pileupquantitative}). An almost constant number $C_0$ of counts per time bin is observed in the range 1 $\mu$s $< \Delta{}t_{\text{s}} < 2 \mu$s. For $\Delta{}t_{\text{s}} \le{}1 \; \mu$s the number of separable pulses starts to decrease due to pile up, for $\Delta{}t_{\text{s}} > 2 \; \mu$s the second pulse maximum starts to move out of the event window. The number of pile up events is then extracted by extrapolating the constant $C_0$ to $\Delta{}t_s = 0$ and integrating the missing counts represented by the grey shaded area in Fig.~\ref{fig:pileupquantitative}. The integral number of missing events divided by the measurement time is then the rate of pile up events used as correction. This procedure was performed for the high statistics runs config 1, 3, 7, for each pad and retardation voltage separately. The resulting count rate loss $y_{\text{sys}}^{\text{pile up}}$ as a function of the actual count rate in the proton region, $y_{\text{exp}}$, is shown in Fig.~\ref{fig:pileuprate} with the statistical uncertainties $\Delta{}y_{\text{sys}}^{\text{pile up}}$. The resulting functional dependence can be used as a correction for all configurations, as it originates from the DAQ being independent of the individual configurations. Hence, the pile up correction $y_{\text{sys}}^{\text{pile up}}$ shown in Fig.~\ref{fig:pileuprate} has been included in the fit as common correction with
\begin{eqnarray}
g_{\text{sys}}^{\text{pile up}}(y_{\text{exp}}; \{c_0^p, c_2^p \}) & = & c_0^p + c_2^p\cdot{}\left(y_{\text{exp}}(U_{\text{AP}})\right)^2 \nonumber \\
\text{and} \; f_{\text{sys}}^{\text{pile up}} & = & - \left(c_0^p + c_2^{p}\cdot{}(y_{\text{theo(n)}})^2 \right) \; .
\end{eqnarray}
As the pile up is a loss of count rate, it has to enter with a negative sign in the fit function of Eq.~(\ref{eq:chisquare}).

\subsection{Proton traps in the DV region}
\label{sec:mirror}

Protons with low kinetic energy $T$ and emission angle close to 90$^\circ$ with respect to the magnetic field can be trapped in the DV region. For example, the applied axial magnetic field gradient ($dB/dz < 0$) across the DV (cf.~Fig.~\ref{fig:nmrDVAP} (a)) was a targeted measure to prevent such protons to be trapped between the DV and EM by the magnetic mirror effect, if they have been emitted into the rear hemisphere. In combination with an inhomogeneous electric potential $\phi_0$, Penning-like traps can easily be created inside the DV region. Therefore, great care has been taken in the design of the electrode system of the $a$SPECT spectrometer to avoid these traps. In axial direction, the beneficial effect of field leakages from the positively-charged EM electrode (+ 860 V) and the negatively charged (E$\times$B) electrode E8 (- 200 V) to some extent prevents protons from being stored in the DV region. WF inhomogeneities of the various electrode segments, however, lead to the actual potential inside the DV region as shown in Fig.~\ref{fig:simpotentialDV}.

Protons with low longitudinal energy can be trapped by this potential and thus are lost for the measurement. Such a loss would bias the measured $a$ value. To investigate traps inside the DV and their effect on $a$ we performed
\begin{itemize}
	\item[1.] particle tracking simulations including the measured work function distributions in the DV along with an analytical approach to quantify the retardation voltage-dependent losses due to stored protons in the DV region,
	\item[2.] measurements of $a$ with an additional extraction field in the DV.
\end{itemize}

\subsubsection{Particle tracking simulations}

\begin{figure}
\includegraphics[width=\linewidth]{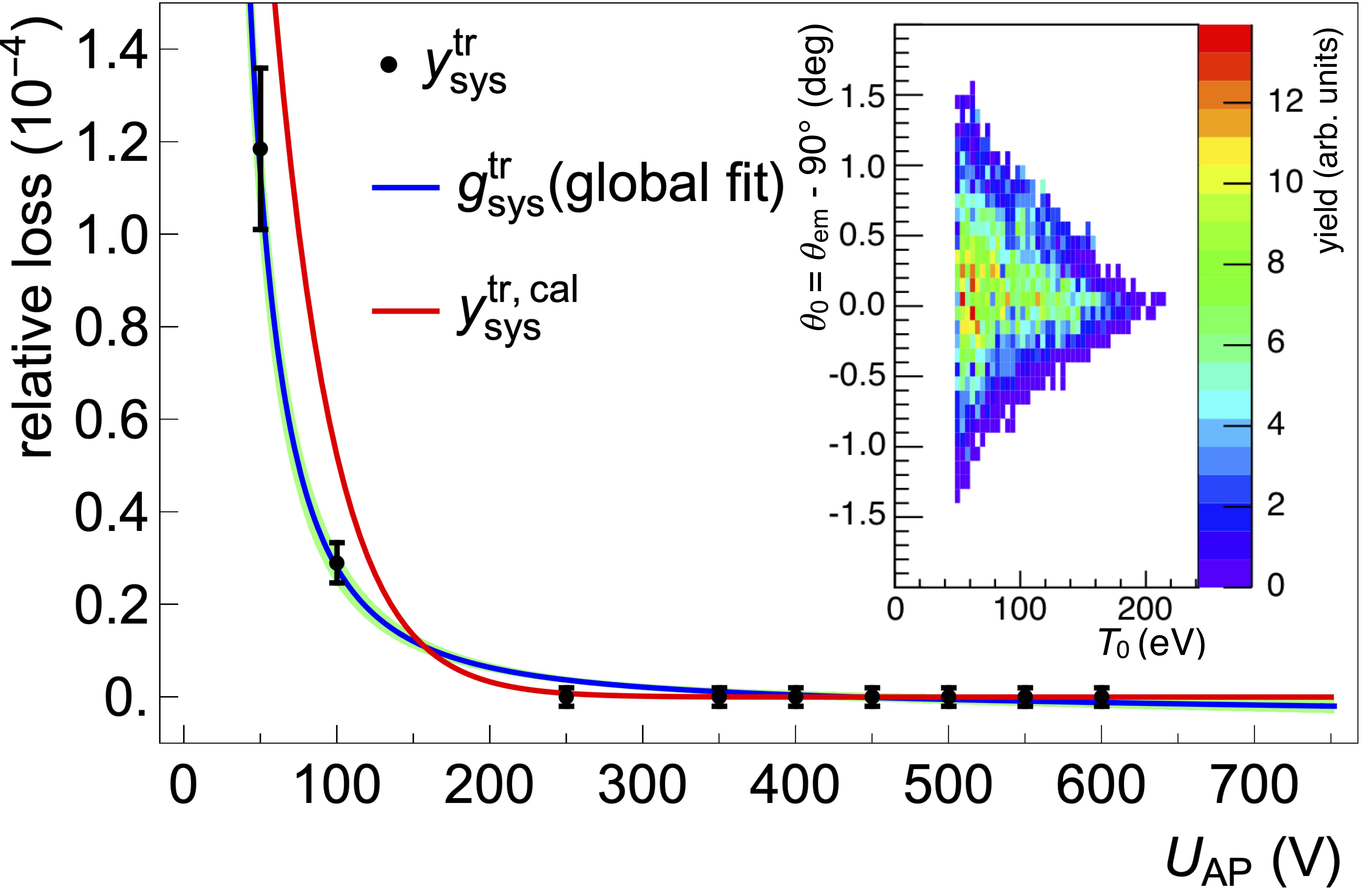}
\caption{Relative loss $y_{\text{sys}}^{\text{tr}}$ due to trapped protons in the DV region as a function of the retardation voltage. The black data points are from particle tracking simulations whereas the red solid line is the result of an analytical calculation $(y_{\text{sys}}^{\text{tr, cal}})$ under simplified assumptions (proton trajectories: on axis). Further drawn is the global fit result of $g_{\text{sys}}^{\text{tr}}$. Inset: Conditions for protons to be stored in the DV region. Shown is the color-coded yield (arb.~units)  for the parameter space $\theta_0$ and $T_0$, the proton emission angle $\theta_0=\theta_{\text{em}} - 90^{\circ}$ and its kinetic energy $T_0$ at the decay point.\label{fig:rellosstrapped}}
\end{figure}

In the simulation, protons are generated throughout the DV weighted with the measured neutron beam profile. Here, we only consider protons from the fiducial decay volume, which under optimal conditions would be losslessly guided along the magnetic flux tube onto the two detector pads (2, 3). In the actual B- and E-field configuration, their motion is tracked and if a proton is trapped or can leave the flux tube in radial direction by E$\times$B drift, this proton is counted as lost. We note that the E$\times$B drift is fast enough so that scattering on residual gas can be neglected. Figure~\ref{fig:rellosstrapped} (inset) shows the yield of trapped protons as a function of the kinetic energy of the proton at its decay point inside the fiducial volume and the emission angle. The corresponding relative loss of protons $y_{\text{sys}, k}^{\text{tr}}$ as a result from particle tracking simulations is shown in the same figure as a function of the retardation voltage ($k = 1, \cdots{}, 9$). The uncertainties $\Delta{}y_{\text{sys}, k}^{\text{tr}}$ shown in Fig.~\ref{fig:rellosstrapped} include statistical uncertainties from the Monte Carlo simulations as well as the uncertainties of the WF and field leakages and the uncertainty of the neutron beam profile. This loss is implemented in the fit function by
\begin{eqnarray}
g_{\text{sys}}^{\text{tr}}\left( U_{\text{AP}}; \{c_{-2}^{\text{tr}}, c_1^{\text{tr}}\} \right) & = & c_{-2}^{\text{tr}}\cdot{}U_{\text{AP}}^{-2}+c_1^{\text{tr}}\cdot{}U_{\text{AP}} \nonumber \\
\text{and} \; f_{\text{sys}}^{\text{tr}} & = & -g_{\text{sys}}^{\text{tr}}\cdot{}y_{\text{theo(n)}}
\end{eqnarray}

For an analytical calculation of the expected relative proton losses in the DV region, one can use Eq. (3) from \cite{glu2005}, which describes the longitudinal energy $T(P)$ of the proton at any trajectory point $P$ and which after some manipulation using $T(z)=0$ (on-axis trajectories are only considered) can be written as:
\begin{equation}
\theta_0=\sqrt{\frac{B_{z_0}}{B_z}\left(\frac{e\phi_z - e\phi_0}{T_0}\right) - \frac{B_{z_0} - B_z}{B_z}} \; .
\end{equation}
$T_0$ is the proton kinetic energy at decay point $z_0$ with $\phi_0, B_{z_0}$ the respective local electric potential and magnetic field. Correspondingly we have $\phi_z, B_z$ along the z-axis. $\theta_0$ is the proton emission angle at $z_0$ related to the direction perpendicular to $B_{z_0}$ which causes proton reflection at position $z$. Using the distribution of the electric potential and the magnetic field along the z-axis in the DV region (Fig.~\ref{fig:simpotentialDV}), the maximum emission angle $\theta_0^{\text{max}}$ referred to 90$^\circ$ can be determined: $\theta_0^{\text{max}}(T_0, z_0) = \text{max}|\theta_0(T_0, z_0; V_z, B_z)|$. This angle also represents the relative number of stored protons of energy $T_0$ at $z_0$ for isotropically emitted protons, since we have: $2\cdot{}\{\int_0^{\theta_0^{\text{max}}}2\pi{}\cos{}\theta{}d\theta / 4\pi \} \approx \theta_0^{\text{max}}$.

The weighting with the normalized beam profile $I(z_0)$ (cf.~Fig.~\ref{fig:simpotentialDV}) along the z-axis gives $\langle \theta_0^{\text{max}}(T_0) \rangle = \int_{-5 \; \text{cm}}^{+5 \; \text{cm}}I(z_0) \cdot{}\theta_0^{\text{max}}(T_0, z_0)dz_0$ which to a good approximation can be parametrized by $\langle{}\theta_0^{\text{max}}(T_0) \rangle = 0.00485\cdot{}\exp{}\left(-(T_0 - 50 \; \text{eV})/39 \; \text{eV} \right) \; \text{rad}$. Finally, the relative count rate loss due to trapped protons can be determined by including the differential proton spectrum and the transmission function:
\begin{footnotesize}
\begin{equation}
y_{\text{sys}}^{\text{tr, cal}} = \frac{\int{}\langle \theta_0^{\text{max}}(T_0) \rangle \cdot{}\omega_{\text{p}}(T_0, a)\cdot{}F_{\text{tr}}(U_{\text{AP}}, r_{\text{B}}; a, T_0)dT_0}{\int{}\omega_{\text{p}}(T_0, a)\cdot{}F_{\text{tr}}(U_{\text{AP}}, r_{\text{B}}; a, T_0)dT_0}
\end{equation}
\end{footnotesize}
The result is shown in Fig.~\ref{fig:rellosstrapped} where $y_{\text{sys}}^{\text{tr, cal}}$ is plotted versus $U_{\text{AP}}$ for $a_{\text{ref}}$ = -0.103. The relative loss rate is about 30 \% higher than the one derived from particle tracking simulations. This is reasonable since the simplifications made, \textit{i.e.,} x, y-dependence of the electric potential (off-axis) were not taken into account, slightly overestimate the actual losses.

\subsubsection{Measurement with additional extraction field}

In order to quantify the effect of trapped protons on $a$, an E-field was applied along the z-axis of the DV electrode, strong enough to extract any trapped proton. To generate such a field, the connecting electrodes below and above the DV electrode have been set to +4 V and -4 V, respectively. This does not change the mean potential in the DV, but generates an electric field of the order of $\mathcal{O}(6 \; \text{V/m})$ along the z-axis, see Fig.~\ref{fig:simpotentialDV}. This field prevents protons from being stored in the DV region. A measurement of $a$ with this field (config 7) coincides with $a$ derived from config 1\footnote{In config 1 to 6 these electrodes like the DV electrode are at ground potential.} within their respective uncertainties (cf.~section \ref{sec:globalfit}). In the fit routine of Eq.~(\ref{eq:chisquare}), $f_{\text{sys}}^{\text{tr}}$ was not used for config 7.

\subsection{Miscellaneous effects}
\label{sec:misc}

\subsubsection{Proton scattering off residual gas}
\label{sec:protonscattres}

The transmission of protons through $a$SPECT may be modified by scattering of the protons off residual gas atoms. Three different kinds of collision can be distinguished: The protons may be neutralized by charge exchange processes, or change their energy and direction due to elastic or inelastic scattering. This problem has already been taken into account in the design phase of $a$SPECT: In order to be negligible for an experiment at the 0.3 \% level, the residual gas pressure between the DV and the AP has to be below $10^{-8}$ mbar \cite{glu2005}. With all the vacuum improvements in place since the offline beam time in 2012, we measured a pressure of $\approx$ $5\times{}10^{-10}$ mbar with a pressure sensor outside the magnet at a port that reaches directly into the decay volume. This indicates that the pressure in the spectrometer bore tube is well below the critical pressures given in \cite{glu2005}.

\subsubsection{Adiabaticity}

The calculation of the integral proton spectrum in Fig. \ref{fig:spectrum} (b) is based on exact adiabatic proton motion from DV to AP. The adiabaticity of the protons in the $a$SPECT spectrometer was tested in \cite{glu2005} by high-precision tracking simulations for various magnetic fields and for $U_{\text{E8}}=-3$ kV dipole potential of the lower E$\times$B electrode E8. According to Table I of \cite{glu2005}, the relative change of $a$ due to non-adiabaticity at $B_0\approx2.2$ T is smaller than $4\times{}10^{-4}$. The proton motion adiabaticity improves with smaller absolute values of $U_{\text{E8}}$ (due to the smaller kinetic energy of the protons in the E8 region), and we used $U_{\text{E8}} = -200$ V in our measurements (cf. Table \ref{tab:potentials}). Therefore, the systematic relative change of the $a$ value due to non-adiabaticity is far below $4\times{}10^{-4}$ in our measurements.

\subsubsection{Doppler effect due to neutron motion}

The motion of the decaying particle also changes the observed energies of the outgoing particles relative to the energies in the center-of-mass system (CMS) of the decaying particle according to:
\begin{eqnarray}
T_{\text{LAB}} & = & T_{\text{CMS}} + \frac{m_p}{m_n}T_{\text{n}} \nonumber \\ & + & 2 \sqrt{\frac{m_p}{m_n}}\sqrt{T_{\text{CMS}}\cdot{}T_n}\cos{}\theta_{\text{CMS}}
\label{eq:energylab}
\end{eqnarray}
where $\theta_{\text{CMS}}$ is the polar angle in the CMS and $T_{\text{n}}\approx$ 4 meV is the average energy of the cold neutron beam at PF1B. In $a$SPECT, the magnetic field is transverse to the neutron beam and protons are detected with 4$\pi$ acceptance. We find therefore a large cancellation of Doppler effects. After averaging over all $\theta_{\text{CMS}}$ angles (cf.~Eq.~(\ref{eq:energylab})), the lab energies of the protons are systematically higher by $\Delta{}T\approx$ 4 meV than their corresponding CMS energies. From section \ref{sec:transmissionfunc} one can estimate this effect on $a$ by $\Delta{}a/a \approx$ 0.05 \% if $\Delta{}T$ is attributed to a corresponding uncertainty in the retardation voltage of $\Delta{}U_{\text{AP}}=\Delta{}T/e$. A more refined analysis done by \cite{glu2005} predicts even smaller relative changes. Hence, we do not expect any essential systematic uncertainty from the Doppler effect at our current level of accuracy.

\section{Fit results and extraction of $a$}
\label{sec:globalfit}

\begin{figure}
\includegraphics[width=\linewidth]{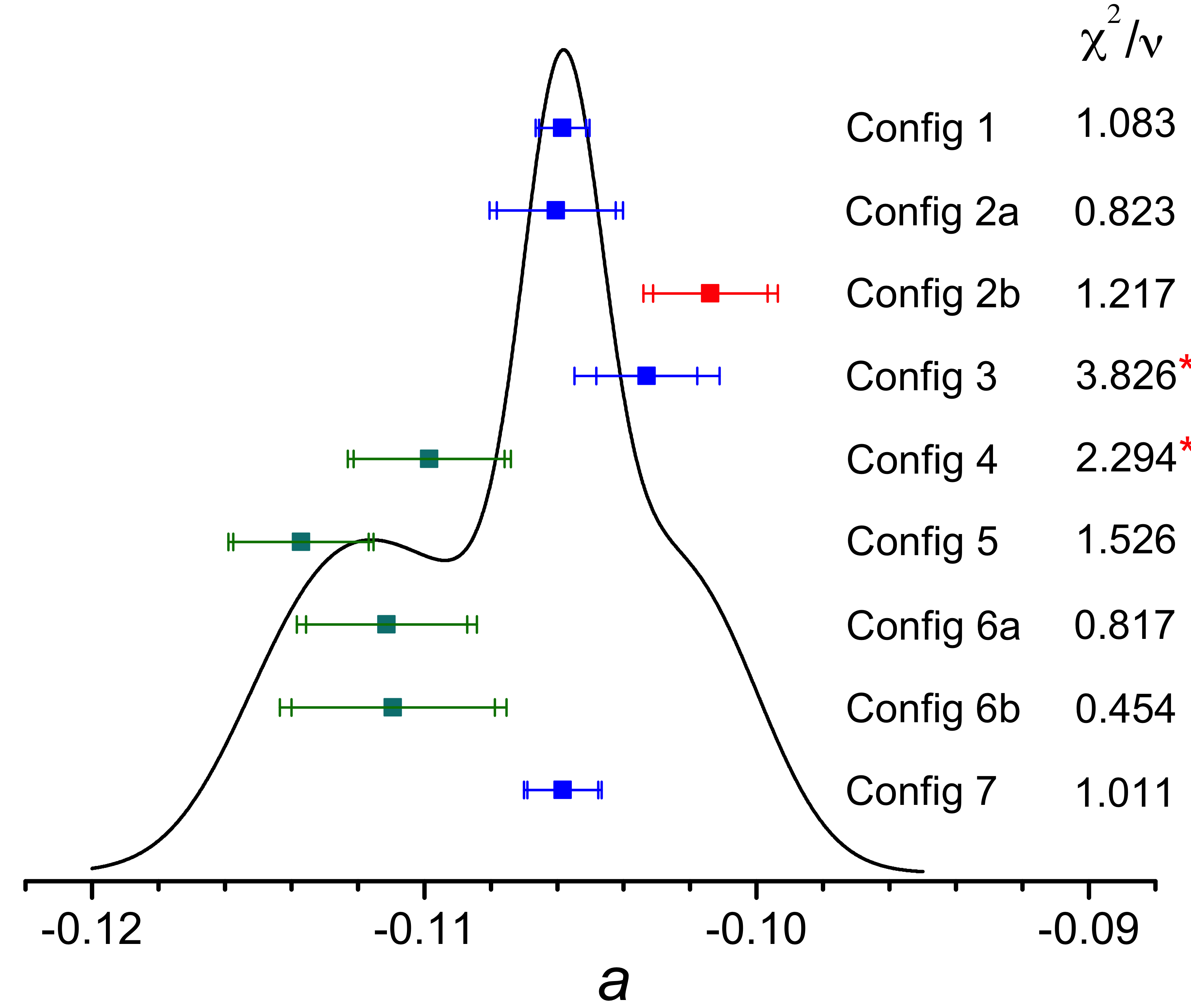}
\caption{Ideogram of $a$ values for each configuration fitted separately without any systematic correction. The blue data points are derived from configuration runs with standard parameter settings. Forced enhancement of the edge effect by a reduced beam profile leads to more negative $a$ values (green data points), whereas config 2b (EM off) shows the trend from systematic shifts to positive $a$ values. The total error from the fit is shown together with the uncorrelated error which reflects the count rate statistics and which essentially constitutes the total error. The error bars of the extracted $a$ values are scaled with $\sqrt{\chi^2/\nu}$ whenever the $p$-value is less than the conventionally accepted significance level of $\alpha=0.05$ (cf. Table~\ref{tab:cr50V}) which is indicated by an asterisk (*).  The degrees of freedom are $\nu=15$ for each configuration. 
\label{fig:ideogram}}
\end{figure}

In order to get a first impression of the quality of the raw data, the individual configurations are fitted separately without any systematic correction. For that we use the (normalized) theoretical integral proton spectrum (cf.~Eq.~(\ref{eq:fitn})) and consider the background signal by a constant term, $c_{\text{bg}}$, which besides $N_0$ and $a$ is another free fit parameter of the fit function given by:
\begin{eqnarray}
f_{\text{fit}}(U_{\text{AP}}, r_{\text{B}}; a, N_0, c_{\text{bg}}) & = & y_{\text{theo(n)}}(U_{\text{AP}}, r_{\text{B}}; a, N_0) \nonumber \\
& + & c_{\text{bg}} \; .
\end{eqnarray}
Figure~\ref{fig:ideogram} shows the ideogram of $a$ values for each configuration ($j$). The ideogram was built in the same manner of the PDG review \cite{beringer2012} to convey information about possibly inconsistent measurements. Each data point is represented by a Gaussian with a central value $a_j$, error $\sigma_{a_j}$, and area proportional to $1/\sigma_{a_j}$. The error bars shown in Fig.~\ref{fig:ideogram} include correlations between fit parameters. The inner tick marks at the error bars denote what the statistical uncertainties would be if these correlations were not present. The uncorrelated error from the fit can be deduced from $\sigma_a^{\text{stat}} = \left(\sum_i (1/\Delta{}a_i)^2 \right)^{-1/2}$ with $\Delta{}a_i = \Delta{}y_{\text{exp}, i} / \left(dy/da\right)_i$. The $\Delta{}y_{\text{exp}, i}$ are the statistical uncertainties of the measured count rates $y_{\text{exp}, i}$ at the respective retardation voltage settings\footnote{For $U_{\text{AP}}$ = 50 V, the $\Delta{}y_{\text{exp}}$ values for the different configurations are listed in Table \ref{tab:cr50V}.} ($i$). The derivative $(dy/da)_i$ expresses the sensitivity of $y_{\text{exp}, i}$ to changes in $a$ at measurement point ($i$) of the integral proton spectrum (cf.~Fig.~\ref{fig:spectrum} (b)).

The central peak of the ideogram which culminates at $a \approx$ -0.106 comprises the configuration runs (blue) with the standard parameter settings. At its wings a shoulder towards positive $a$ values and a bump structure on the opposite side can be identified. For configuration runs 4, 5, 6a, and 6b (green data points) with the reduced beam profile, the enhanced edge effect leads to a shift in $a$ towards negative values with the common mean at $a \approx$ -0.111. On the other hand, the weakly prominent shoulder can be attributed to config 2b (red data point), where the electrostatic mirror was switched off.

The reduced $\chi^2/\nu$ values to test the goodness of the fit are listed in Fig.~\ref{fig:ideogram} for the individual configuration runs.

In order to investigate the effect of the individual systematic corrections onto $a$ and its uncertainty it seems obvious to take the $\chi^2$ function of Eq.~(\ref{eq:chisquareoverall}) which includes all systematic corrections and to extract the $a$ value for the data set of the particular configuration run. Then in a second step, the fit procedure is repeated but now with the systematic effect of interest turned off. Any change in $a$ should therefore be attributable to the influence of the systematic effect under investigation. This procedure, however, does not lead to unambiguous quantitative results on the influence of the respective systematic effect on $a$. This is due to the fact that with the removal and addition of a systematic effect, the correlations between the fit parameters also change, which in turn influences the value on $a$ as result of the $\chi^2$ minimization. This is particularly the case when it comes to small systematic shifts which lie within the statistical (uncorrelated) error. In order to get an overview of the contribution of the systematic effects discussed in section \ref{sec:systematics}, a different procedure is chosen:
Starting point is the integral proton spectrum $y_{\text{theo(n)}}^{a_{\text{ref}}}\left(U_{\text{AP}}, r_{\text{B}}; a_{\text{ref}}, N_0 \right) + c_{\text{bg}}$ to which the reference value $a_{\text{ref}} = -0.103$ of the $\beta$-$\overline{\nu}_e$ angular correlation coefficient is assigned. The count rate amplitude $N_0$ is adapted to the measured count rates in the respective configuration runs (cf. Table \ref{tab:cr50V}), as well as the constant background of $c_{\text{bg}}\approx$ 6 cps measured at 780 V. In the next step, this spectrum is modified with the contributions of the systematic effect under investigation. For instance in case of the edge effect, the function $f_{\text{sys}}^{\text{ee}} = \left(-g_{\text{sys}}^{\text{ee}}\left(U_{\text{AP}}; \{c_0^{\text{ee}}, c_2^{\text{ee}}, c_4^{\text{ee}} \} \right) \right)\cdot{}y_{\text{theo(n)}}^{a_{\text{ref}}}$ is added which describes the relative count rate losses due to this effect (cf.~Eqs. (\ref{eq:edge}) and (\ref{eq:rellosses})). The coefficients were determined from the global fit to the overall data set. Finally, a $\chi^2$ fit yields the potential change of $a$ according to
\begin{eqnarray}
\chi^2 & = & \sum_{i=1}^{10} \frac{1}{\left(\Delta{}y_{\text{theo(n)}, i} \right)^2} \nonumber \\
& & {} \cdot{}(y_{\text{theo(n)}, i}^{a_{\text{ref}}}\cdot{}(1- g_{\text{sys}}^{\text{ee}}\left(U_{\text{AP}}; \{ c_0^{\text{ee}}, c_2^{\text{ee}}, c_4^{\text{ee}} \})\right) \nonumber \\
& & {} + c_{\text{bg}} - f_{\text{fit}}^a)^2
\end{eqnarray}
with the fit function given by $f_{\text{fit}}^a=y_{\text{theo(n)}}^a(U_{\text{AP}}, r_{\text{B}}; a, \widetilde{N}_0) + \widetilde{c}_{\text{bg}}$. Prior to a $\chi^2$ fit, the respective count rate at measurement point ($i$) (cf.~Fig.~\ref{fig:spectrum} b) was modified by an offset count rate which is Gaussian distributed around zero mean with standard deviation $\Delta{}y_{\text{theo(n)}, i}$. For $\Delta{}y_{\text{theo(n)}, i}$ we take a statistical error $\approx$ 5 times smaller in total than the actual count rate error for config 1. This measure is a compromise between measurement sensitivity to trace tiny systematic shifts and the goodness of fit testing with a reduced $\chi^2$ of $\chi^2/ \nu \le{}2$.

Table \ref{tab:relchanges} shows the influence of systematics discussed in section \ref{sec:systematics} on the extracted value $a_{\text{fit}}$ from the fit.

\begin{table}
  \flushright
  \caption{Relative changes of $a$ values as result of a $\chi^2$ fit in which a reference integral proton spectrum ($a_{\text{ref}} = -0.103$) was modified by systematic effects as discussed in section \ref{sec:systematics}. The relative uncertainty of the extracted $a_{\text{fit}}$ values is $\approx$ 0.2 \%, essentially determined by the chosen statistical errors $\Delta{}y_{\text{theo(n), i}}$ at the data points ($i$) of the integral proton spectrum. The respective $U_{\text{AP}}$ and $r_{\text{B}}$ offset error does not change the input reference value $a_{\text{ref}}$, but only increases its uncertainty as result of the fit\footnote{Relative change of $a$ using the fit result of the retardation voltage-dependent background in config 1 (cf. Fig. \ref{fig:backgroundretvolt} (a)). For config 2, this value is already reduced by a factor of $\approx$~2 and there will be no shift in $a$ for the subsequent configuration runs.}.}
  \begin{center}
    \begin{tabular}{lccc}
    \hline
    \hline
    & Section & $a_{\text{fit}}$ & $\left(a_{\text{fit}} - a_{\text{ref}}\right) / |a_{\text{ref}}|$ \\
    & & & (\%) \\
    \hline
    No systematic & --- & -0.1031 & -0.1 \\
    $U_{\text{AP}}$-dep.~background$^{\text{a}}$ & \ref{sec:bg} & -0.1044 & -1.4 \\
    Trapped protons in DV & \ref{sec:mirror} & -0.1028 & +0.3 \\
    Edge effect (standard) & \ref{sec:ee} & -0.1041 & -1.1 \\
    Edge effect (reduced) & \ref{sec:ee} & -0.1121 & -8.8 \\
    Backscattering/threshold & \ref{sec:lld} & -0.1031 & $-0.1$ \\
    Pile up & \ref{sec:uld} & -0.1029 & +0.1 \\
    $\langle U_{\text{AP}} \rangle$ & \ref{sec:ua} & -0.1025 & +0.5 \\
    $U_{\text{AP}}$ offset & \ref{sec:ua} & $a_{\text{ref}}$ & 0.0 $\pm$ 0.3 \\
    $\langle r_{\text{B}} \rangle$ & \ref{sec:rb} & -0.1030 & $< |0.1|$ \\
    $r_{\text{B}}$ offset & \ref{sec:rb} & $a_{\text{ref}}$ & 0.0 $\pm$ 0.1 \\
    \hline
    \hline
    \end{tabular}%
  \label{tab:relchanges}%
  \end{center}
\end{table}%

The expected finding here is the dominant shift of the $a$ value by the edge effect with reduced beam profile, which was already observed in the raw data fits without systematic corrections (cf.~Fig.~\ref{fig:ideogram}). From the ratio of the relative count rate losses for the standard (st) and reduced (re) beam profile, see Eq.~(\ref{eq:ratiolosses}), we further expect $(a_{\text{fit}}^{\text{re}} - a_{\text{ref}}) / (a_{\text{fit}}^{\text{st}} - a_{\text{ref}}) \approx{} \langle \varepsilon_{\text{re}} \rangle / \langle \varepsilon_{\text{st}} \rangle$. This functional relationship matches well within the specified error bars of $\langle \varepsilon_{\text{re}} \rangle / \langle \varepsilon_{\text{st}} \rangle = (6.9 \pm 1.4)$, see section \ref{sec:systematics}, and $(a_{\text{fit}}^{\text{re}} - a_{\text{ref}}) / (a_{\text{fit}}^{\text{st}} - a_{\text{ref}}) = (8.8 \pm 1.8)$. In the latter case, the relative uncertainty of the $a_{\text{fit}}$ values with $\delta a_{\text{fit}}/a_{\text{ref}}\approx$ 0.2 \% determines this error.

Among the configuration runs with the standard parameter settings the listed systematic effects may add up to a relative shift in $a$ of $\delta a_{\text{sys}}/a \approx$ 1 \%. All in all, this is a relatively small effect for the systematic corrections on the measurement values. The error on the individual systematic corrections ($j$) listed in Table \ref{tab:relchanges} can be estimated from the corresponding error band on $g_{\text{sys}}^{j}$ as a result of the global fit. Taking, for example, the edge effect (standard beam profile) as one of the major systematic corrections, $\Delta{}g_{\text{sys}}^{\text{ee,st}} / g_{\text{sys}}^{\text{ee,st}} \approx 15$ \% can be inferred from Fig. \ref{fig:simretvoltedge}. Thus, the relative uncertainty on the extracted $a$ value due to the edge effect correction ($\text{st}$) is $\Delta{}a/a^{\text{ee,st}} \le |-0.011\cdot{}0.15| \approx{}$ 0.15 \%. In a similar way, this can be done for the other systematic corrections shown in Table \ref{tab:relchanges} in order to get an estimate on their relative contributions to the overall uncertainty in $a$. To derive the total error on $a$ (including the correlated error) correctly, the global fit needs to be performed in which we minimize $\chi^2$ as defined in Eq. (\ref{eq:finalfit}).

In fact, two global fits have been performed: \textit{Global-Config-}$a_c$ and \textit{Global-}$a$. In both cases, all systematic errors and their uncertainties are included. The difference between the two was only in the parameter space of the $a$ values to be fitted. In \textit{Global-Config-}$a_c$, independent fit parameters $a_c$ for the $\beta-\overline{\nu}_e$ angular correlation coefficient have been assigned to each configuration run ($c$). This approach leads to equal corrections of systematic effects as far as they are relevant for the respective configuration runs. Additionally, it indicates if the distribution of the $a_c$ values does scatter statistically or not. 

\begin{figure}
\includegraphics[width=\linewidth]{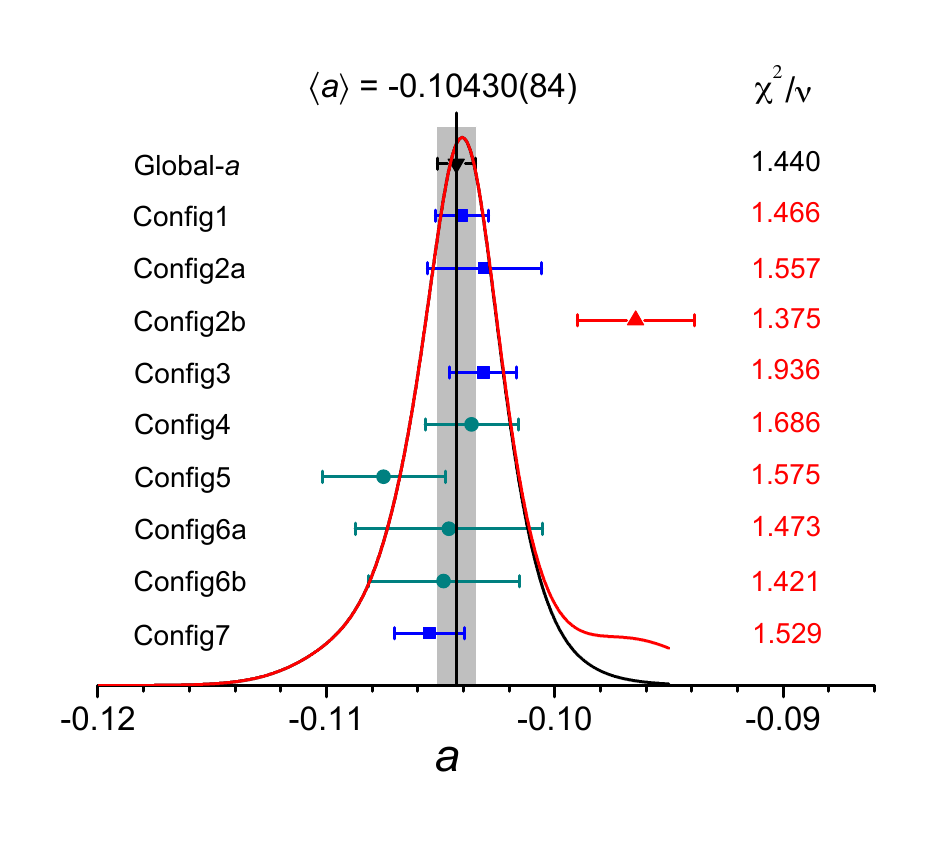}
\caption{Global fit results on $a$ with systematic corrections included. Ideograms (black/red) of extracted $a$ values where independent fit parameters $a_c$ for the $\beta-\overline{\nu}_e$ angular correlation coefficient have been assigned to each configuration run ($c$) (\textit{Global-Config-}$a_c$). The red one includes the outlier value of $a$ from config 2b (red data point), where $a$SPECT operated as a $2\pi$ spectrometer (EM off). The global fit (\textit{Global-}$a$) to the overall dataset (except config 2b) with only one common fit parameter for $a$ yields $\langle a \rangle = -0.10430(84)$ for the production beam time in 2013 (black data point with vertical line and error band indicated by gray bar). The respective error bars are scaled with $\sqrt{\chi^2 / \nu}$, \textit{i.e.,} the $\chi^2 / \nu$ values listed on the right hand side, since the associated $p$ values are in all cases less than the significance level (cf. Table~\ref{tab:cr50V}) for $\nu > 100$. For details, see text. \label{fig:ideogramcorr}}
\end{figure}

From the \textit{Global-Config-}$a_c$ fit we get a reduced $\chi^2$ of $\chi^2_{GC} / \nu = 1.399$ $(\nu = 292)$. Since the data statistics (weighting) as well as the contribution of systematic errors differ significantly for the different configurations (cf. Fig. ~\ref{fig:ideogram}), $\chi^2_{a_c} / \nu$ values were calculated for each configuration. They are displayed (in red) on the right side of Fig.~\ref{fig:ideogramcorr}. Hereby only the data set of the respective configuration and the extracted fit-parameters which enter the corresponding model function $f_{\text{fit(n)}}^{c, p}$ (cf. Eq.~(\ref{eq:chisquare})) are taken into account. Related to the $\chi^2_{GC}/\nu$ value from the \textit{Global-Config-}$a_c$ fit, no outliers can be identified in the listed $\chi^2_{a_c} / \nu$ values which in turn does not provide any hints to additional systematics for a particular configuration.

Except for the $a$ value extracted from config 2b (EM off), all other values behave as expected, which manifests in the depicted Gaussian ideogram (black) of Fig.~\ref{fig:ideogramcorr}. On the other hand, an ideogram (red curve in Fig.~\ref{fig:ideogramcorr}) which includes the config 2b value shows a pronounced tail towards positive $a$ values. This value deviates by $\approx 3$ standard deviations from the peak position of the ideogram(s). The latter, in turn, matches almost perfectly with the extracted $\langle a \rangle$ value from \textit{Global-}$a$, in which the overall data set (except config 2b) was fitted with only one common fit parameter for $a$ (cf. Fig.~\ref{fig:ideogramcorr}). 

We can identify two reasons why this non-standard measurement of config 2b does not allow us to extract a precise value on $a$ via the $\chi^2$ fit:
\begin{enumerate}
\item{Protons which are emitted into the rear hemisphere are guided along the magnetic field lines onto the bottom flange (stainless steel) of the $a$SPECT spectrometer (B $\approx$~0.11 T) if the electrostatic mirror (EM) is off. The angle- and energy-resolved intensity distributions of reflected H$^+$ ions were measured, \textit{e.g.}, in \cite{sas2017} for incident proton beams in the energy range $<$ 1 keV. So a fraction of them is backscattered and may pass the AP if they can overcome the magnetic mirror below the DV and if their energy is higher than the applied retardation potential. Accordingly, one may expect a change of the integral proton spectrum which, however, cannot be quantitatively determined with sufficient accuracy. In case of EM `on' (for all other configurations), there is no backscattering off materials, but rather it is a reversal of motion without energy loss.}
\item{In config 2b, $a$SPECT operated as a 2$\pi$ spectrometer (EM off). In that case the differential proton recoil spectrum $\omega_{\text{p}}(T, a)$ (cf.~Fig.~\ref{fig:spectrum} a) must be supplemented by a $\cos{}\vartheta$ term (see Appendix \ref{sec:prrec}) according to
\begin{equation}
\label{eq:polarized}
W(T,a,c) = \omega_{\text{p}}(T,a)+P\cdot{}\omega_{\text{ps}}(T,a)\cdot{}\cos{}\vartheta
\end{equation}
where $\vartheta$ is the angle between neutron spin and proton momentum and $c$ denotes the product $c=P\cdot{}\cos{}\vartheta$. The second term vanishes for $P=0$ and/or in case of a 4$\pi$ detection of the decay protons (for the latter reason all other configurations are insensitive to a residual polarization). The H113 beam is nominally unpolarized, but the neutron guide wall of the ballistic $^{58}$Ni/Ti supermirror guide \cite{abele2006} could cause a slight unwanted neutron polarization as observed on the NG-6 beam ($P~\approx0.6\%$) of the aCORN experiment \cite{darius2017}. Moreover, one must assume that the polarization is not homogeneously distributed over the beam profile. The lack of knowledge about the finite beam polarization and its spatial distribution in the decay volume does not allow to determine the model function of the integral proton spectrum from Eq.~(\ref{eq:polarized}) good enough.}
\end{enumerate}

Both aspects, therefore, suggest not to take config 2b into account in the final data analysis. Hence, the value for the $\beta-\overline{\nu}_e$ angular correlation coefficient $a$ obtained from the \textit{Global-}$a$ fit ($\chi^2_G / \nu = 1.440$ ($\nu = 268$); $p$ value: $3.0 \times 10^{-6}$) is
\begin{equation}
\langle a\rangle = -0.10430 \pm 0.00084 \; ,
\label{eq:finalavalue}
\end{equation}
which results in a relative uncertainty of $\Delta{}a/a$ = 0.8 \% in the determination of this quantity.

\begin{figure}
\includegraphics[width=\linewidth]{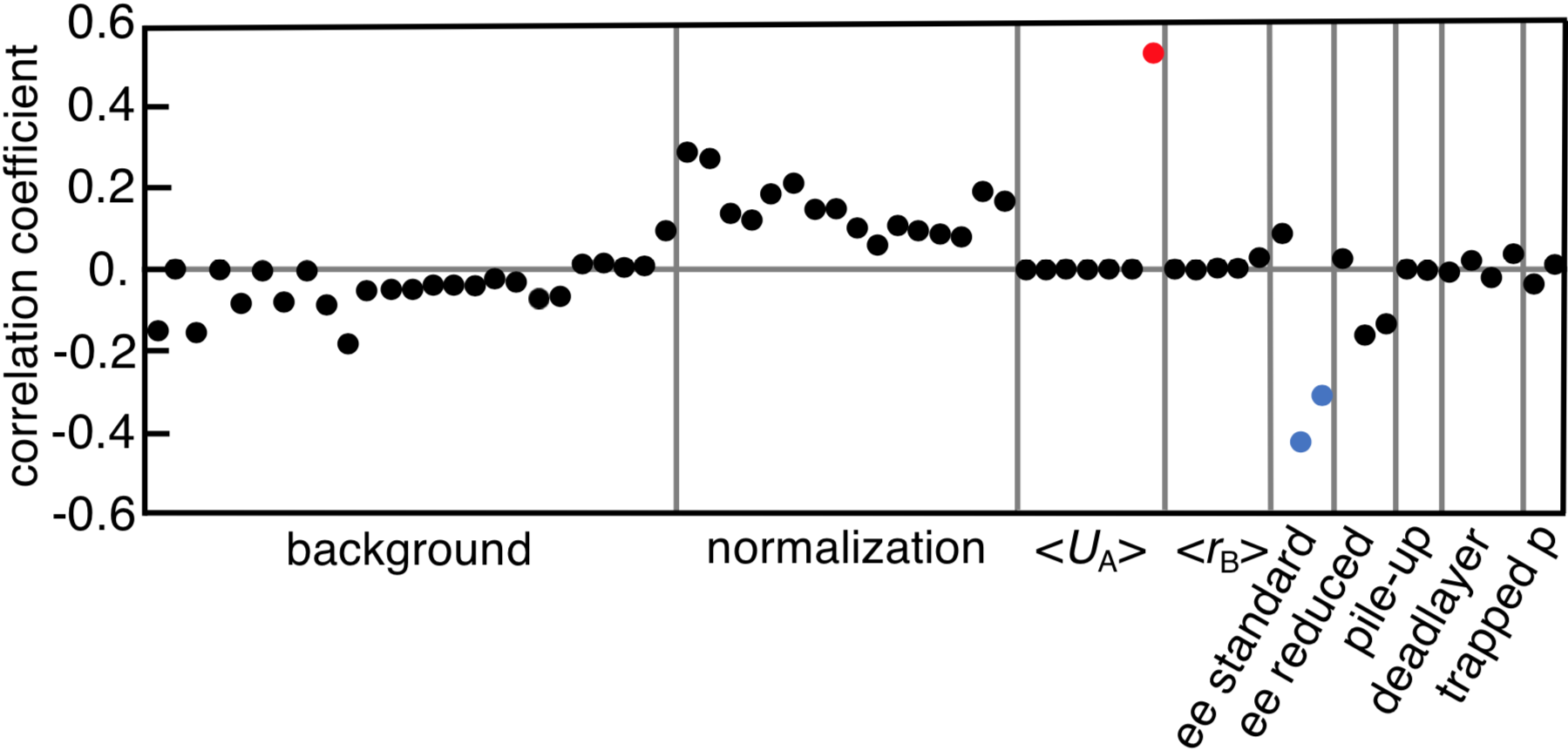}
\caption{Linear relationship (correlation) coefficient of $a$ with various fit parameters (67 in total) as result of the global fit to the overall data set. Most of the parameters show correlation coefficients $<|0.2|$ which is weak and likely insignificant. Strong correlations, \textit{i.e.,} values that surpass $|0.8|$ are not found. Moderate correlations (+ 0.5) are found for $c_{\text{AP, offset}}^{\langle U_{\text{A}} \rangle}$ (data point highlighted in red) to describe the offset error common to all $\langle U_{\text{A}} \rangle$ values and for the parameters $c_{2,4}^{\text{ee(st)}}$ ($\approx |0.4|$; data points highlighted in blue) to correct for the edge effect (standard beam profile). A further breakdown according to detector pads ($p$) and configuration run ($c$) was not made in this correlation plot.\label{fig:fitpar}}
\end{figure}

The error on $a$ is the total error scaled with $\sqrt{\chi^2_G / \nu}$ (cf. Fig.~\ref{fig:ideogramcorr}). Besides the statistical error, it contains the uncertainties of the systematic corrections and the correlations among the fit parameters which enter the variance-covariance matrix to calculate the error on the derived quantity from the fit. Figure~\ref{fig:fitpar} shows the correlation coefficients between $a$ and the various fit parameters which in most cases are $< \left| 0.2 \right|$.

In our global fits, we used low-order polynomials to describe all possible modifications on the spectrum's shape by the investigated systematic effects listed in section~\ref{sec:systematics}. This approach greatly facilitated the convergence of the fit, but partly introduced some minor discrepancies between the data from supplementary measurements and simulations of the various systematic effects ($j$) and their functional descriptions $g_{\text{sys}}^j$ (cf. section~\ref{sec:fit}).
In the corresponding figures of section~\ref{sec:systematics}, the error bands of the $g_{\text{sys}}^j$ values as a result of the \textit{Global-}$a$ fit have already been scaled with $\sqrt{\chi_G^2/\nu}=\sqrt{1.44}=1.20$.

In order to get a hint that the elevated $\sqrt{\chi^2/\nu}$ values of the global fits are due to the simplifications made to model systematic effects or more likely arise due to the non-white reactor power noise (cf.~section~\ref{sec:tempstab}), i.e., non-statistical count rate fluctuations of the data points of the integral proton spectra, \textit{Global-a} fits with different error scalings, $f_{\text{scal}}$, have been performed as listed in Table~\ref{tab:scalederrors}. The factor $f_{\text{scal}}$ was set $f_{\text{scal}}=1.20$ to reach a reduced $\chi^2$ of $\chi^2_G/\nu=1$ (last row in Table~\ref{tab:scalederrors}).
%The elevated $\sqrt{\chi^2 / \nu}$ values of the global fits are likely to be the result of this fact. In the corresponding figures of section~\ref{sec:systematics}, the error bands of the $g_{\text{sys}}^j$ values as a result of the \textit{Global-}$a$ fit have already been scaled with $\sqrt{\chi^2_G / \nu}$.
%\begin{table}
%	\begin{center}
%		\begin{tabular}{cccccc}
%		\hline
%		\hline
%		Error scaling&
%		Error scaling&
%		$\chi^2_G/\nu$&
%		$p$-value&
%		$a$&
%		Error\\
%		data of integral&
%		systematic&
%		$(\nu = 268)$&
%		&
%		&
%		on $a$\\
%		proton spectra&
%		corrections&
%		&
%		&
%		&
%		on $a$\\
%		\hline
%		1.00&	1.00&	1.44&	3.1$\cdot 10^{-6}$&	-0.10430&	0.00084*\\
%		1.20&	1.00&	1.17&	0.029&	-0.10430&	0.00084*\\
%		1.00&	1.20&	1.27&	0.0018&	-0.10433&	0.00082*\\
%		1.20&	1.20&	1.00&	0.49&	-0.10432&	0.00080*\\
%		\hline
%		\hline
%		\end{tabular}%
%		\label{tab:scalederrors}%
%	\end{center}
%\end{table}%

\begin{table}
	\flushright
	  \caption{\textit{Global-$a$} fit results on the central value of $a$ and its error using different error scalings for the data of the integral proton spectra and/or the systematic corrections. Note: $\nu = 268$.}
	\begin{center}
		\begin{tabular}{ccccc}
			\hline
			\hline
			\footnotesize{Error scaling}&
			\footnotesize{Error scaling}&
			\footnotesize{$\chi^2_G/\nu$}&
			\footnotesize{$p$-value}&
			\footnotesize{$a$}\\
			\footnotesize{data of integral}&
			\footnotesize{systematic}&
%			\footnotesize{$(\nu = 268)$}
			&
			&
			\\
			\footnotesize{proton spectra}&
			\footnotesize{corrections}&
			&
			&\\
			\hline
			1.00&	1.00&	1.44&	3.1$\times 10^{-6}$ &	-0.10430(84)*\\
			1.20&	1.00&	1.17&	0.029&	-0.10430(84)*\\
			1.00&	1.20&	1.27&	0.0018&	-0.10433(82)*\\
			1.20&	1.20&	1.00&	0.49&	-0.10432(80)\\
			\hline
			\hline
		\end{tabular}%
		\label{tab:scalederrors}%
	\end{center}
\end{table}%

The error scaling of the data of the integral proton spectra has a bigger effect on the reduced $\chi^2$ value than error scaling of the systematic corrections (column 3 in Table~\ref{tab:scalederrors}). This is an indication that our measurements are more limited by the non-white noise of the integral proton spectra. Furthermore, error scaling of the systematic corrections has a negligible effect on the value of $a$. This verifies that our method of systematic corrections has no pull, i.e., engenders no bias to the extracted central values of $a$. The error of $a$ is almost independent of error scaling because the respective $\chi^2_G /\nu$ value is already taken into account by the fit procedure while calculating the fit parameter errors whenever the $p$-value is less than the significance level of $p=0.05$. The asterisk (*) indicates these cases (last column in Table~\ref{tab:scalederrors}).

%A better functional description of those data requires more fit parameters (higher-order polynomials), which may reduce the $\chi^2 / \nu$ values, \textit{i.e.,} the data are statistically consistent with the model function. The extracted error on $a$, however, will not necessarily be reduced due to the extended fit parameter space (correlations). Besides, there is also the risk of an overfitted model function, a problem encountered in the statistical modeling of data.

Our new value (cf.~Eq.~(\ref{eq:finalavalue})) is in good agreement with the present PDG value of (-0.1059 $\pm$ 0.0028) \cite{tanabashi2018} for the $\beta$-$\overline{\nu}_e$ angular correlation coefficient of the free neutron but with the overall accuracy improved by a factor of 3.3. Using Eq.~(\ref{eq:lambda}) one can deduce a value for the ratio of the weak axial-vector and vector coupling constant $\lambda = g_{\text{A}}/g_{\text{V}}$ given by
\begin{equation}
\label{eq:res}
\lambda = (-1.2677 \pm 0.0028)
\end{equation}

\begin{figure}
\includegraphics[width=\linewidth]{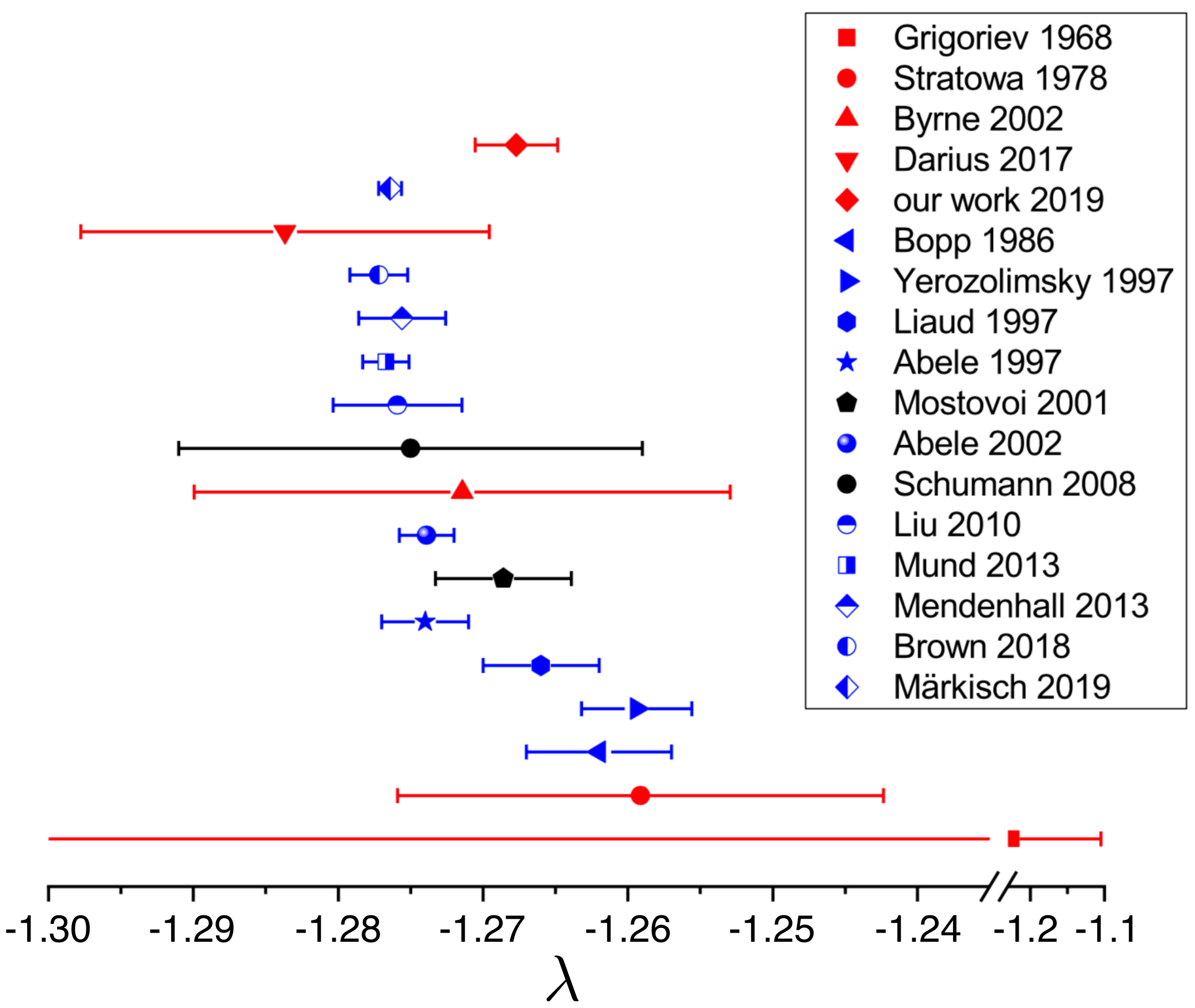}
\caption{Published results on $\lambda$ derived from $\beta-\bar{\nu}_e$ angular correlation measurements (red data points: Grigoriev \cite{grig1968}, Stratowa \cite{stratowa1978}, Byrne \cite{byrne2002}, Darius \cite{darius2017}, our work), $\beta$-asymmetry measurements (blue data points: 
	Bopp \cite{bopp1986}, Yerozolimsky \cite{yero1997}, Liaud \cite{liaud1997}, Abele \cite{abele1997, abele2002}, Liu \cite{liu2010}, Mund \cite{perkeo2013}, Mendenhall \cite{menden2013}, Brown \cite{brown2018}, and Märkisch \cite{maerk2018}), and other measurements (black data points: Mostovoi \cite{mostov2001} and Schumann \cite{schuma2008}).\label{fig:lambda}
}
%	), $\beta$-asymmetry measurements (blue data points), and other measurements (black data points) \cite{abele1997, abele2002, liu2010, menden2013, maerk2018, brown2018, perkeo2013, schuma2008, mostov2001, liaud1997, yero1997, bopp1986, darius2017, byrne2002, stratowa1978, grig1968}.\label{fig:lambda}
\end{figure}

Figure \ref{fig:lambda} shows the status of $\lambda$ measurements (including our result) in which the distinction is made between measurements which determine the $\lambda$ value from the beta-asymmetry $A$ (blue data points), from the $\beta$-$\overline{\nu}_e$ angular correlation coefficient $a$ (red data points) and from other observables (black data points).

An overall systematic difference at the 1-2 $\sigma$ level cannot be identified between the different measures of $\lambda$ extraction, although comparable accuracies are obtained with the most recent results.

Under the assumption of the conserved vector current (CVC) hypothesis, experimentally determined values for $\lambda$ directly determine $g_{\text{A}}$. This serves as a benchmark for lattice QCD calculations and determines the relationship among parameters of the weak hadronic current. Recent improvements in lattice QCD calculations which approach the per-cent-level determination in the physical point \cite{alex2017, chang2018, liang2018, gupta2018, ottnad2018, capitani2019} show promising agreement between theory and experiment. A comparison of experimental values for $g_{\text{A}}$ with lattice values by itself constitutes a new physics test of nonstandard couplings \cite{gonz2016}.

\section{Conclusion and outlook}

In summary, we have measured the $\beta$-$\overline{\nu}_e$ angular correlation coefficient $a$ with $a$SPECT resulting in a fractional precision of $\approx$ 0.8 \%. This result is in good agreement with the present PDG value but with the overall accuracy improved by a factor of 3.3. Within the SM, the correlation coefficients in neutron $\beta$-decay can be expressed in terms of one parameter, $\lambda$, which is the ratio of the weak coupling constants: $\lambda = g_{\text{A}}/g_{\text{V}}$. With $a = -0.10430(84)$ we obtain $\lambda = -1.2677(28)$. This value deviates by 2.8~$\sigma$ from the most recent  $\lambda$ measurement of the PERKEO III collaboration \cite{maerk2018}, which was determined via the $\beta$-asymmetry parameter $A$. This experimental situation calls for further improvements in the measurement accuracy; in particular being on par with the Perkeo result in terms of accuracy presents a major challenge. 

The 4$\pi$ detection of the decay protons with the $a$SPECT spectrometer which is based on the electrostatic MAC-E filter principle helps to a great deal to suppress unwanted systematics. From the analysis of the systematic effects we are confident that with an upgrade of the present spectrometer, a relative accuracy of $\Delta{}a/a \; \approx$~0.2 \% can be reached.

The essential improvements in the order of their importance are:
\begin{enumerate}
\item{WF differences of polycrystalline gold surfaces as well as their temporal fluctuations result in the current uncertainty of $da/a \; \approx$~0.3 \%. For this reason, electrode surfaces with better uniformity of the work function (\textit{e.g.}, as for gold single crystal layers) have to be used. Surface dipoles caused by adsorption of contaminants if exposed to ambient conditions may lead to potential changes of the electrode, but those are spatially uniform for Au surfaces in a defined crystallographic orientation \cite{leu2003}. As only the potential difference between the DV and AP electrode is of relevance, this WF offset (and its possible temporal drift) drops out. In this context, the current accuracy ($\approx$ 13 mV) in the voltage mesurement must be improved accordingly.}
\item{The electrode system has to be redesigned. In particular, the use of a broader magnetic flux tube onto the enlarged SDD detector area of 3$\times$3 pads should be realized. The uncorrelated statistical error at present contributes with $\approx$ 0.4 \% to the total error and was obtained within $\approx$ 20 days effective data taking time at the PF1b beam line at ILL with two detector pads in operation. This measure will allow to reach a statistical limit of $\Delta{}a/a_{\text{stat}}\approx$ 0.1 \% within 100 hours of effective data taking.}
\item{The major remaining systematic correction (after having eliminated the retardation voltage-dependent background by improved vacuum conditions) is the edge effect and proton backscattering at the SDD detector. At present, the edge effect corrections (standard beam profile) are under control to a level of $\Delta{}a/a \approx$ 0.1 \%. A better adapted collimation of the incoming neutron beam will reduce the slope $dI/dy$ of the beam profile in the DV and along with it the edge effect correction (cf. Eq. (\ref{eq:losspad})). Proton backscattering at the SDD has been thoroughly investigated (cf. section \ref{sec:lld}) and is under control at the level of $\Delta{}a/a <$ 0.1 \%.}
\end{enumerate}

The envisaged relative accuracy in the determination of $a$ in turn will result in a determination of $\lambda$ of $\Delta{}\lambda/\lambda \approx \; 4\times{}10^{-4}$. This is the sensitivity range which was recently achieved by the PERKEO III collaboration. From neutron decay data, not only a precise V-A SM value of $\lambda$ can be extracted. Of particular interest is the search for right-handed currents and for S and T interactions where the measurement of $\tau_n, A$ and $a$, \textit{e.g.}, exhibit different dependencies \cite{glu1995, gonz2019, pattie2013, pattie2015}. A common fit to the neutron decay data is all the more predictive on Beyond the Standard Model contributions if comparable accuracies are achieved.

\section{Acknowledgements}

This project was funded by the DFG priority program 1491 under grants HE 2308/9-1 and -2 as well as SPP\_ZI 816/1-1 and SPP\_ZI 816/4-1. At an earlier stage it was supported by the German Federal Ministry for Research and Education under Contract No. 06MZ989I, 06MZ170, and 06MT196, by the European Commission under Contract No. 506065, and by the Internal University Research Funding of the Johannes Gutenberg University (JGU) Mainz.

The authors gratefully acknowledge the strong support by the mechanical and electronic  workshops at the Institute of Physics and the computing time granted on the supercomputer MOGON at JGU Mainz.

We appreciate the generous ILL support in the setup phase  of the experiment and during the 100-day beam time. Further, we would like to thank for the access to the UHV Kelvin probe at the Karlsruhe Institute of Technology.
 
In particular we thank the following persons for their help and valuable contributions:  D. Berruyer (ILL technician), P. Bl\"umler (NMR detection), and  H. Lenk (detector electronics).

\begin{appendix}

\section{Work function measurements using a Kelvin probe}
\label{app:wf}

\begin{figure}
\includegraphics[width=\linewidth]{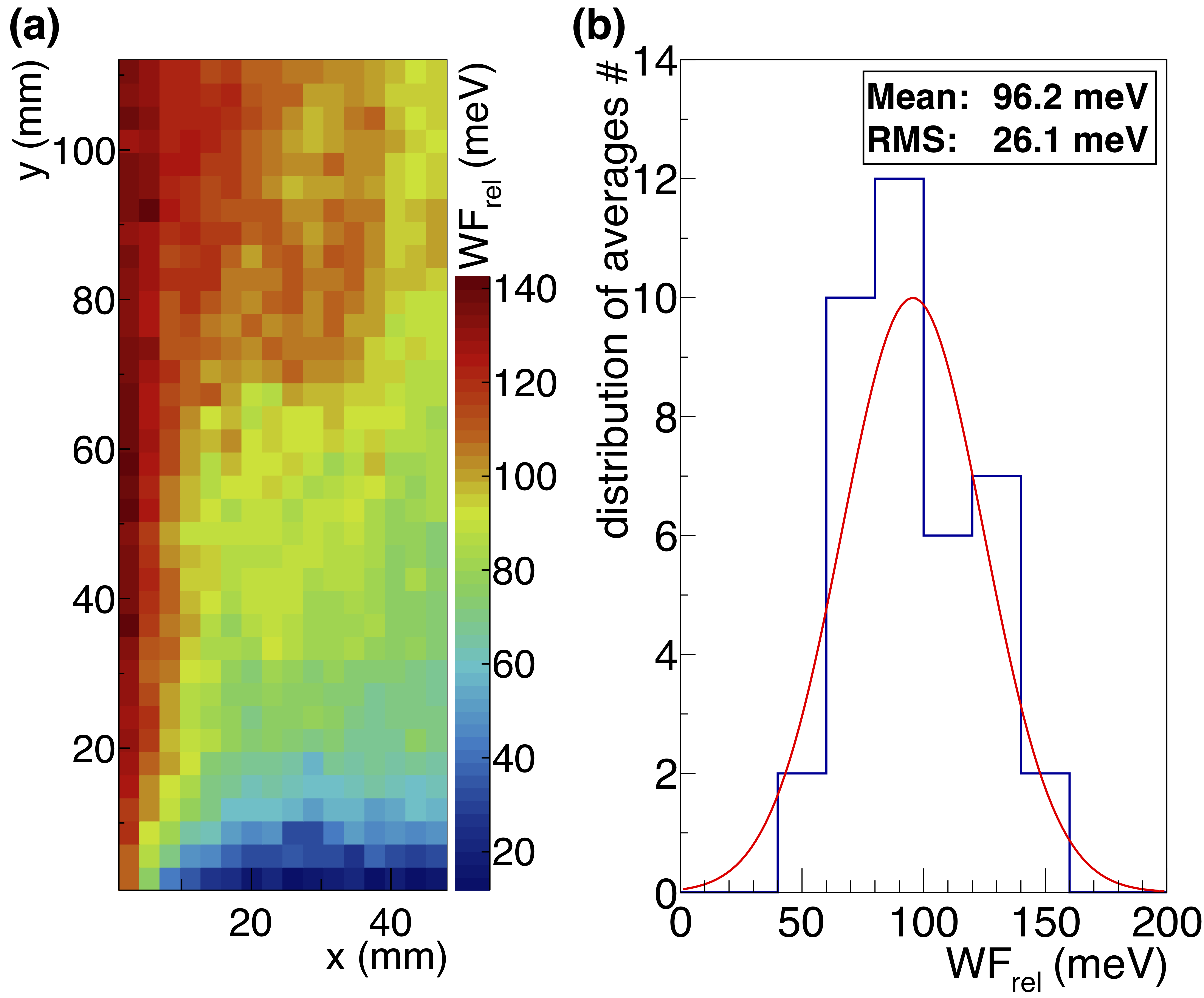}
\caption{WF scan of a flat electrode segment (AP-83) of size 48 x 108 mm$^2$. The false color map shows the $\text{WF}_{\text{rel}}$ distribution (meV) across the surface with  $\overline{\text{WF}}_{\text{rel}} = 92.76$ meV and RMS fluctuation of 24.73 meV.  Scan time: $\approx$ 1.5 h. b) Repeated $\overline{\text{WF}}_{\text{rel}}$ measurements for segment AP-83 using a time span overnight ($\approx$ 12 h) for the system to stabilize. To a good approximation the $\overline{\text{WF}}_{\text{rel}}$ distribution can be described by a Gaussian function (solid curve) with mean $\langle \overline{\text{WF}}_{\text{rel}} \rangle$ = 96.2 meV , $\sigma =$ 26.1 meV, and $\delta \langle \overline{\text{WF}}_{\text{rel}} \rangle \approx$ 4.2 meV (error on mean value determined on a statistical basis).\label{fig:WFAP83}}
\end{figure}

The work function (WF) of different electrodes and its variation across the surface of each electrode was measured after the 2013 beam time by means of a Kelvin probe. Kelvin probe systems are vibrating capacitor systems and are based on the experimental approaches of \cite{Kelvin1898} and \cite{Zisman1932}. The capacitor is formed by the sample electrode and the gold-plated tip ($\O$ = 2 mm) of the Kelvin probe. The term relative work function ($\text{WF}_{\text{rel}}$) in the context of Kelvin probe measurements tells that the WF of the electrode is measured relative to the WF of the probe tip, \textit{i.e.,} $\text{WF}_{\text{rel}} := \text{WF}_{\text{tip}} - \text{WF}_{\text{sample}}$. The Kelvin probe used at ambient conditions (KP Technology SKP150150) is a scanning Kelvin probe system with scan size 15 cm $\times$15 cm specified to have a precision of 1 - 3 meV. It is contained in an enclosure for reproducible and stable results. The enclosure contains an open container of saturated MgCl$_2\cdot$6H$_2$O solution to stabilize the air humidity at (33.2 $\pm$ 0.1) \%. The latter is monitored by means of a humidity sensor. Prior to each WF scan the electrodes were wiped with isopropyl using fine-grade wipers (BEMCOT M-3) in order to receive similar initial conditions with lowest levels of lint and particles. After having put an electrode under the Kelvin probe for a scan measurement the environmental conditions have to stabilize. From the WF scans (step sizes of 1 mm or 3 mm) the average $\overline{\text{WF}}_{\text{rel}}$ of each electrode was computed as well as the RMS fluctuation across its surface. Figure~\ref{fig:WFAP83}~(a) gives an example of a WF scan for segment AP-83 as part of the AP electrode.

\begin{figure}
\includegraphics[width=\linewidth]{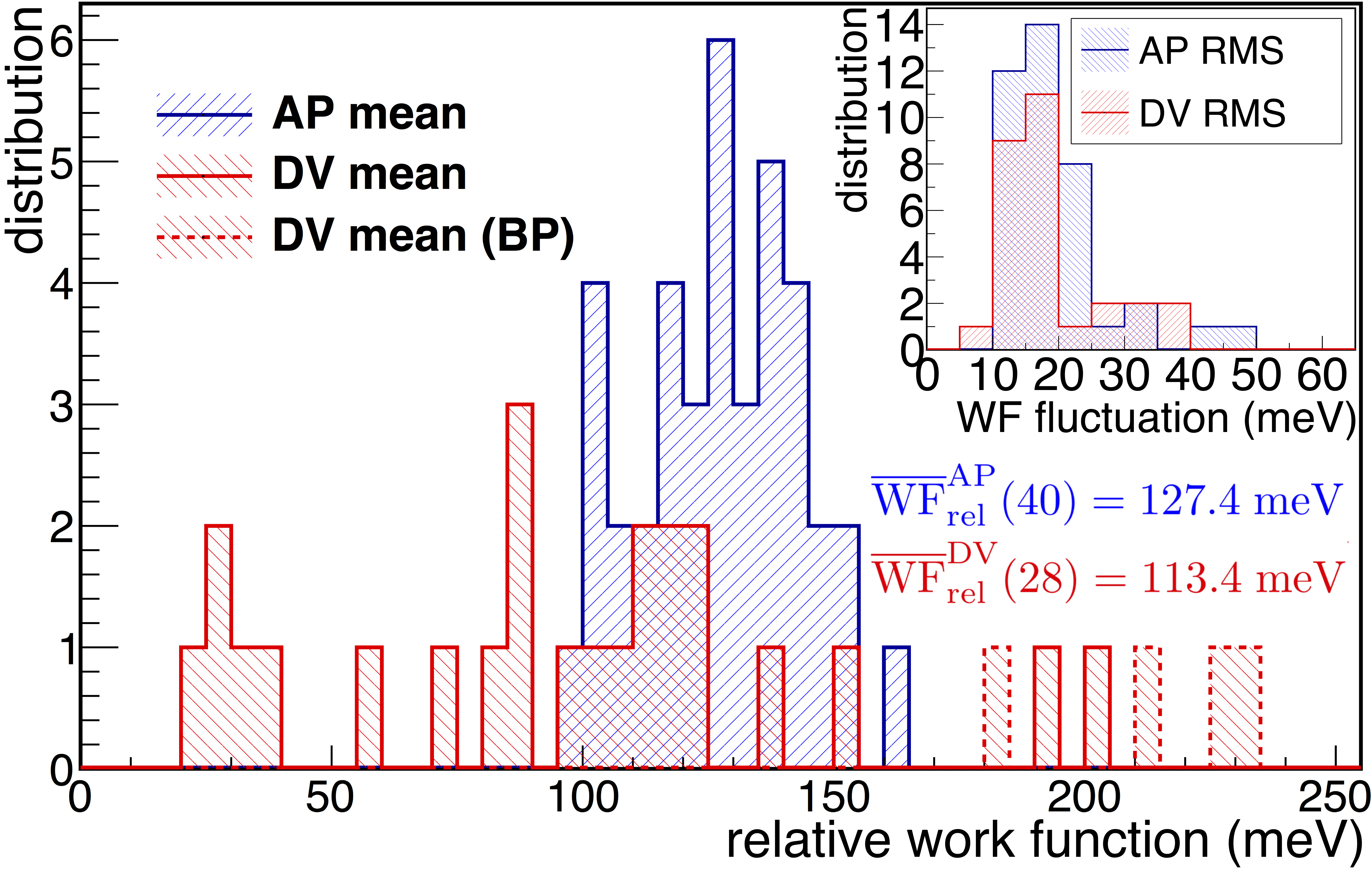}
\caption{Distribution of $\overline{\text{WF}}_{\text{rel}}$ for each segment of the DV (red) and the AP electrode (blue). The bars indicated in dashed red are from the bottom part (BP) of the DV electrode which was re-machined and led to slightly different surface properties. Otherwise the $\text{WF}_{\text{rel}}$ averages of both electrodes overlap to quite some extent. Inset:  Distribution of the RMS fluctuation of $\text{WF}_{\text{rel}}$ across each electrode. All electrodes show a remarkably similar behavior of their WF fluctuations. \label{fig:WFdist}}
\end{figure}

The time for the system to stabilize for reproducible results on the level of $\approx$ 3 meV strongly depends on the air humidity at the time of placing the sample into the enclosure. WF changes accompanying the adsorption/desorption of water on gold surfaces have been investigated in \cite{wells1972}. Repeated scans after lock-up showed that it takes several days up to one week to reach stable conditions. This characteristic stabilization time is too long to perform all necessary scans of the DV and AP electrode segments which had to be cut into smaller pieces (28 DV and 40 AP segments) to fit into the scanning area. As a good compromise we took the time span overnight ($\approx$ 12 h) for the system to stabilize, which allowed us to scan two electrode sample pieces per day. The choice of a shorter time span than the one required to equilibrate the sample electrode with the environment causes a larger uncertainty in the measured WF averages. This is shown in Fig.~\ref{fig:WFAP83} b for repeated $\overline{\text{WF}}_{\text{rel}}$ measurements of a single AP electrode segment (AP-83). The RMS fluctuations are $\approx$ 26 meV around the common mean of $\langle \overline{\text{WF}}_{\text{rel}} \rangle \approx$ 92 meV. Therefore, for all the single-unit $\overline{\text{WF}}_{\text{rel}}$ measurements of electrode segments we take the somewhat higher value $\pm$ 30 meV as common uncertainty. Figure~\ref{fig:WFdist} shows the distribution of the $\overline{\text{WF}}_{\text{rel}}$ values for the DV and the AP electrode segments and the distribution of the RMS fluctuation across the individual surfaces.

The quantity of interest for $a$SPECT is the difference of the potentials of AP and DV. Any common change or drift of the WF will drop out in this difference. Since the Au-coated electrodes were manufactured and treated in the same way, most of the changes due to adsorbates, temperature, as well as temporal drifts are expected to cancel in the difference. The challenge lies in the quantification of residual changes. The issue of the quantification of the residual change of the WF difference between AP and DV electrodes can be subdivided into four categories: 
\begin{itemize}
\item{Accuracy of the WF measurements}
\item{Temporal changes of the WF}
\item{Transferability of WF measurements to UHV conditions inside $a$SPECT, and}
\item{Influence of the temperature differences on WF}
\end{itemize}

\subsubsection{Accuracy of the WF measurements}
The exact electrode geometry with the segments’ associated relative WF is implemented and used as input in KEMField for electromagnetic field and potential calculations in the DV and the AP region.  For the AP electrode segments, the complete WF pattern (see Fig.~\ref{fig:WFAP83}~(a) was taken into account, while only the mean value was used to characterize the WF across the surface of a DV electrode segment. This is due to the fact that the flux tube inside the AP electrode passes closer to the surfaces of the segments than in the DV, where WF differences (patch sizes of $\approx$ 3$\times$3 cm$^2$) are smeared out by the appropriate distance\footnote{ The minimal distance of the effective decay volume to one of the DV electrodes is $>$ 40 mm (cf.~Fig.~\ref{fig:APDVpicture}), \textit{i.e.,} larger than the patch sizes . Therefore, we can use for each segment its surface-averaged WF \cite{bundaleski2013}.}. The impact of the measurement uncertainty ($\pm$ 30 meV) on the extracted values $\langle U_{\text{A}} \rangle$ is considered in the particle tracking simulation by modifying the measured WF of the individual electrode segments statistically with WF offsets generated from a Gaussian distribution with mean zero and $\sigma$ = 30 meV. The resulting RMS fluctuations in $\langle U_{\text{A}} \rangle$ are then taken as error on the common mean. 

\subsubsection{Temporal changes of the WF}
The measurements of the work function took place in 2014 and 2015 whereas the production beam time for the measurement of the beta-neutrino angular correlation was in summer 2013 and the gold plating of the electrodes was performed in early spring 2013. On these time scales one has to consider the issue of a changing WF over time.

\begin{figure}
\includegraphics[width=\linewidth]{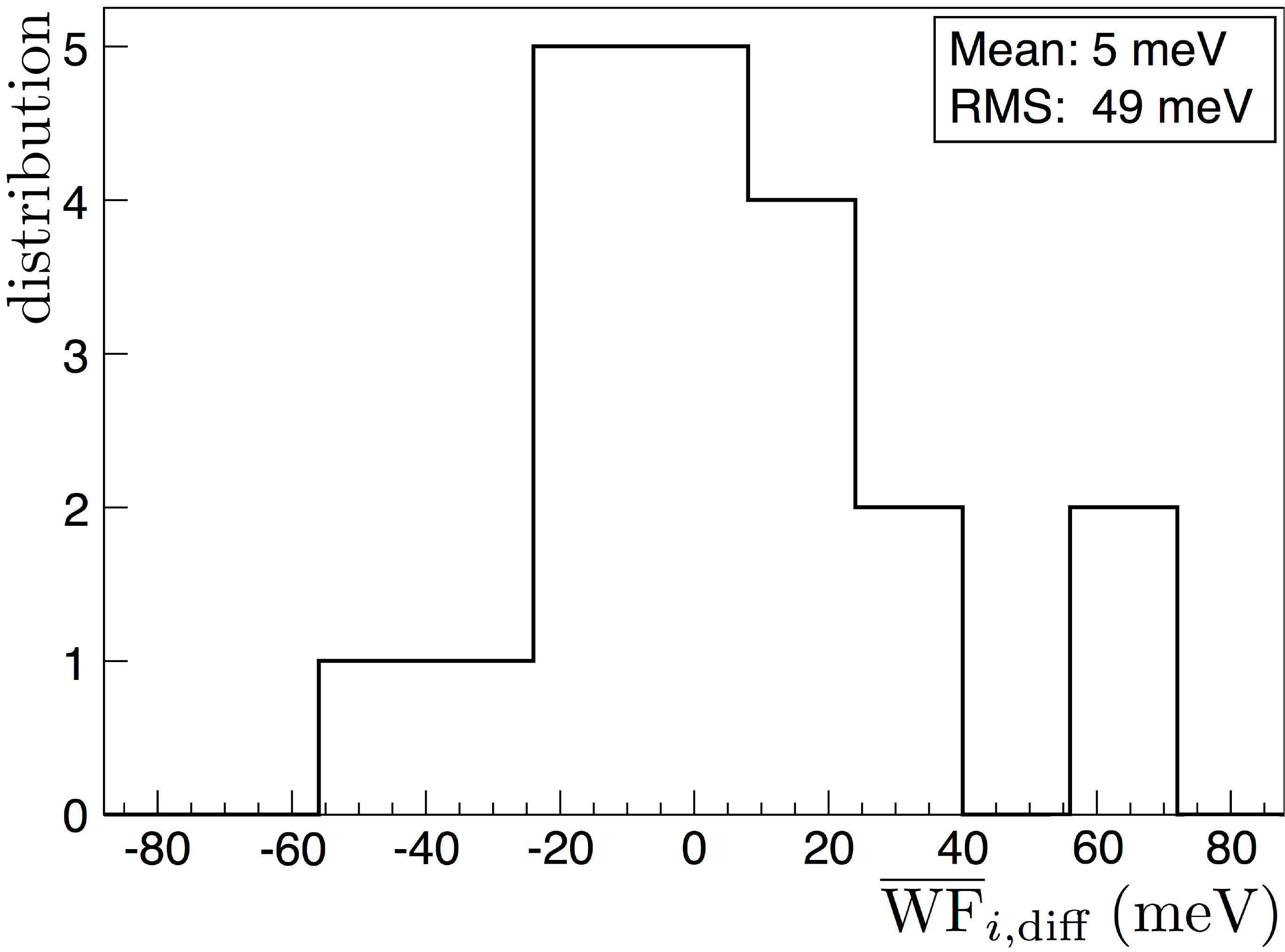}
\caption{Distribution of the WF differences $\overline{\text{WF}}_{i, \text{diff}}$ used to extract temporal WF changes within a time span of about one year.\label{fig:WFdiffdist}}
\end{figure}

Since only the difference between the WF of the AP and the DV electrode is of interest at $a$SPECT, the relatively large number of electrode segments can be used to investigate possible WF changes over time. A total of 40 electrode segments is used and divided into two subsets ($i,j$) of 20 pieces each. The pairwise WF difference $\Delta{}\overline{\text{WF}}_{i, j = i + 20} := \overline{\text{WF}}_i - \overline{\text{WF}}_{j = i + 20}$ of two segments, one from each subset, is calculated for the chosen division and the segment numbering used. This procedure was performed in a measuring campaign in 2014 and then approximately one year later in 2015. It has the advantage that on a statistical basis the measurement uncertainty of $\pm$ 30 meV largely drops out and a potential temporal effect can be observed. Figure~\ref{fig:WFdiffdist} shows the distribution of the differences $\overline{\text{WF}}_{i, \text{diff}} = \Delta{}\overline{\text{WF}}_{i, j = i + 20} \; \text{(2014)} - \Delta{}\overline{\text{WF}}_{i, j = i + 20} \; \text{(2015)}$. From this we can deduce the average change of $\langle \overline{\text{WF}}_{\text{diff}} \rangle = (5 \pm 6)$ meV. This is compatible with zero. The uncertainty yields the limit on the temporal stability of 11 meV.

For the 2013 measurement run we take 20 meV as a conservative upper limit for possible WF differences between the DV and AP electrode due to aging effects. 

\subsubsection{Transferability of WF measurements to UHV conditions inside $a$SPECT}

\begin{figure}
\includegraphics[width=\linewidth]{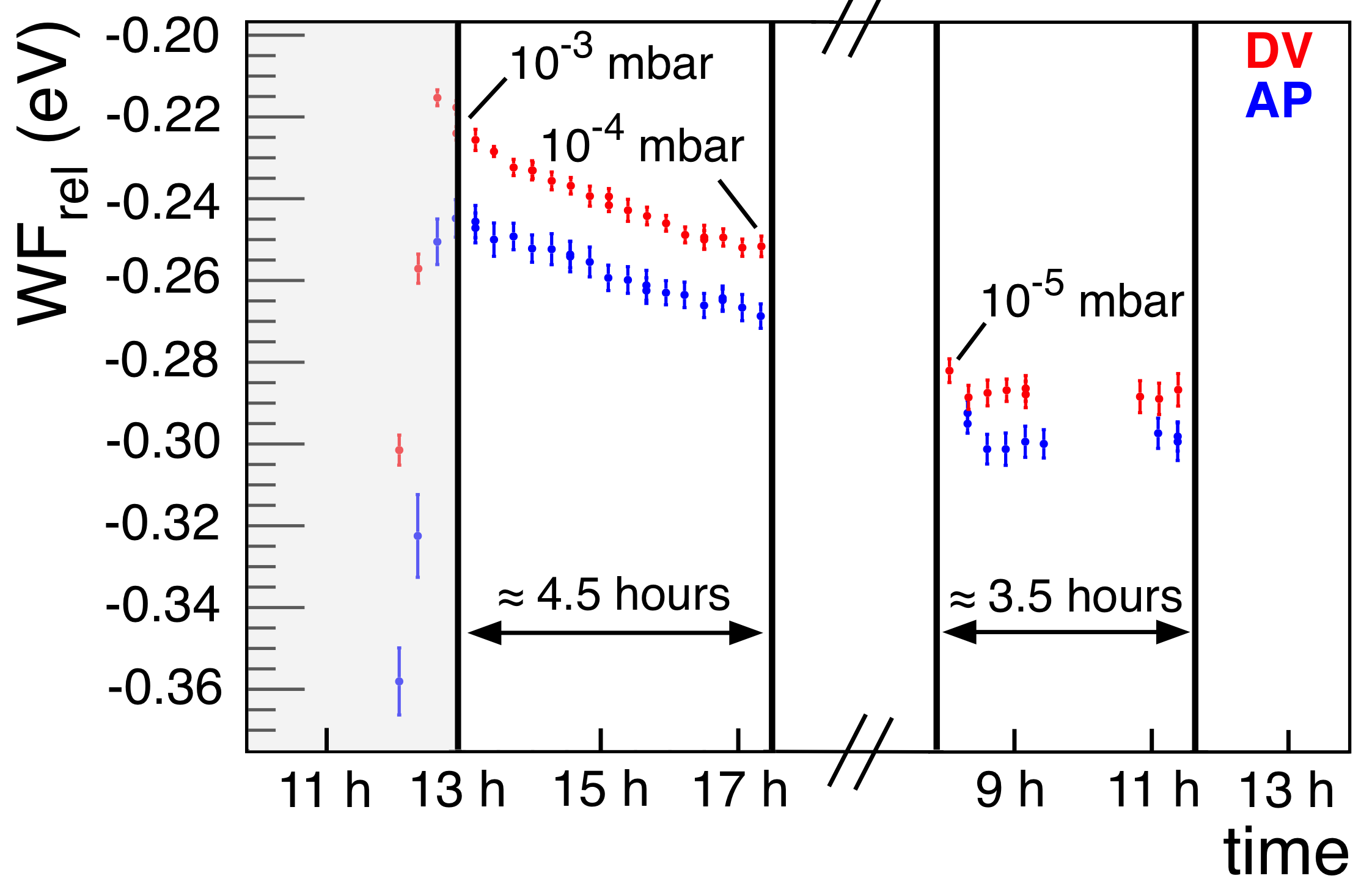}
\caption{Time sequence of line scans showing the extracted means of the relative WF during evacuation of a UHV Kelvin probe. The two electrode samples were measured alternately. \label{fig:timeWFscan}}
\end{figure}

In the 2013 beam time, the electrodes were mounted in $a$SPECT with the surface adsorbates still present. Since the electrodes are located in the cold bore of the superconducting magnet, they cannot be baked out. Therefore, adsorbates like water are not fully removed under vacuum and the modification of the WF or what is more relevant to our case: the change in WF differences had to be investigated. To experimentally check WF changes, we put two electrode segments of $a$SPECT in a Kelvin probe at vacuum (end pressure $\approx 2 \times{}10^{-5}$~mbar) which had been set-up for WF measurements for the KATRIN experiment. To get reliable and stable values the work functions were measured after the system had been stabilized. In order to have an almost simultaneous WF comparison, only line scans ($\approx$ 15 min) across the surfaces were performed, one immediately after the other with the samples in alternation. Figure~\ref{fig:timeWFscan} shows the sequence of the average relative WF extracted from such line scans for both electrode samples. During the initial phase of pumping down relatively large $\overline{\text{WF}}_{\text{rel}}$ changes of $\approx$ 100~meV can be observed since the `simultaneity' of the alternating sample scans was not given due to the big temporal WF gradient. This stabilizes at a pressure of around $10^{-3}$~mbar. What follows is a steady decrease of the WF of both samples, but with a stable WF difference of $\approx$ 20~meV. The scans were stopped overnight ($\approx$ 12 h). Continuation of scans at the end pressure of $\approx{}10^{-5}$ mbar showed stable WF conditions at a WF difference of $\approx$ 10 meV. From these investigations we deduce: In going to UHV conditions inside $a$SPECT one has to assume an additional uncertainty of 10 meV in the WF differences of the AP and DV electrodes measured under ambient conditions.

\subsubsection{Influence of the temperature differences on WF}
The work function exhibits a small dependence on the temperature \cite{Rah2015,Hoelzl1979}. We have measured temperatures at the electrode system in several places close to our electrodes during the off-line beam time in 2012. Temperatures varied between 80 K and 130 K, \textit{i.e.} by $\Delta{}T$ = 50 K. Using the formalism from \cite{Rah2015} based on first principles we can deduce a maximum work function difference of $\Delta{}\overline{\text{WF}}_T$ = 10 meV between the DV and AP electrode. This is consistent with an older phenomenological method \cite{Hoelzl1979} and constitutes an additional uncertainty to the measurement accuracy.

\section{$\beta$-decay electrons within the proton region}
\label{sec:ebg}

The background contribution within the proton region and its possible dependence on the retardation voltage $U_{\text{AP}}$ must be known precisely in order to extract the integral proton spectrum from the measured count rates at the respective voltage settings. At $U_{\text{AP}} = 780$ V, none of the decay protons can pass the analyzing plane which gives us a direct measure of this background provided the background shows no $U_{\text{AP}}$ dependence. In this case, it simply enters as a count rate offset in the integral proton spectrum, which can be considered as a free fit parameter ($c_{\text{bg}}$) in the fit function of the $\chi^2$ minimization and which only slightly affects (correlation) the extracted value of $a$. The small peak visible in the 780 V spectrum of Fig.~\ref{fig:detspec} is caused by ionized rest gas. Its $U_{\text{AP}}$ dependence was extracted from a measurement procedure discussed in detail in section~\ref{sec:bg}.

With $\leq 0.5$ cps, this background constitutes only a small fraction of the total background ($\approx 6$ cps) originated from the low-energetic $\beta-$decay electrons within the proton region (cf. Fig.~\ref{fig:pulseheight}). In order to investigate whether the contamination level of $\beta-$decay electrons varies for the different voltage settings, the $U_{\text{AP}}$ dependence was measured between ADC channel 150 and 170 of the pulse height spectrum (cf. Fig.~\ref{fig:detspec}). The selected integration window was placed sufficiently above the upper integration limit (ADC channel 120) of the proton region to ensure that pile-up events with their possible $U_{\text{AP}}$ dependency are negligibly small (section~\ref{sec:uld}).

\begin{figure}
\includegraphics[width=\linewidth]{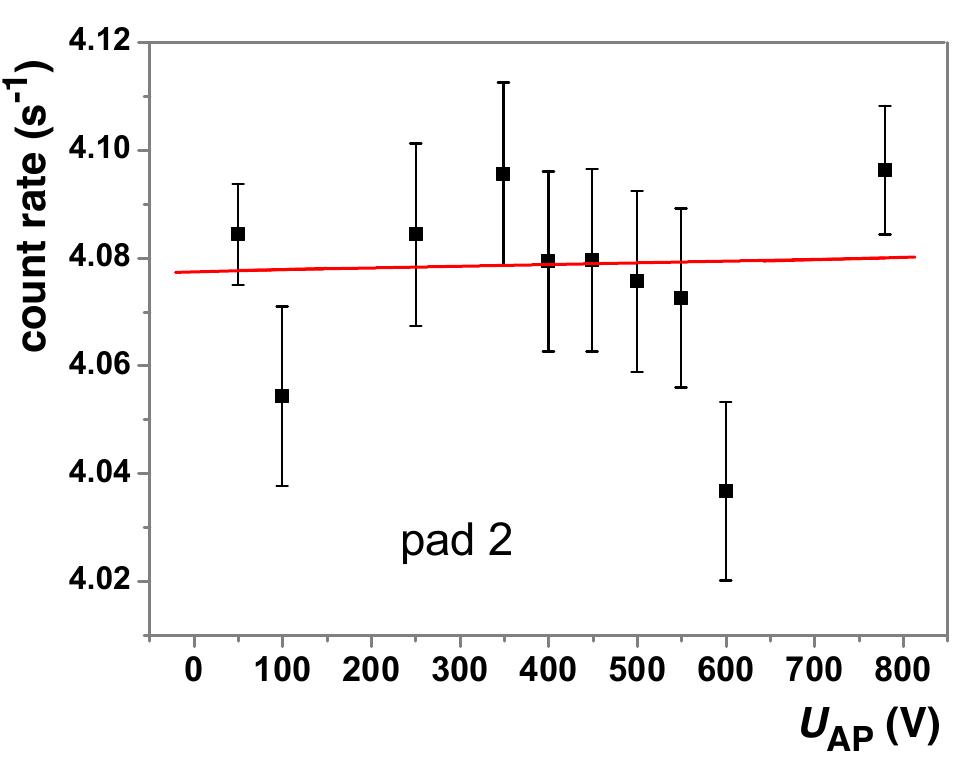}
\caption{$U_{\text{AP}}$ dependence of the integral count rate (weighted average over all configuration runs) of low-energetic $\beta-$decay electrons in the background window between ADC channel 150 and 170. The straight line fit to the data (red curve) gives $m_\beta = (0.32 \pm 1.81) \times 10^{-5}$ cps/V for the slope and $b = (4.078 \pm 0.008)$ cps for the intercept. Both numbers are used to extract a possible voltage dependence of the $\beta-$electron rate in the proton region (cf. Eq.~(\ref{eq:betarate})). \label{fig:electronbackg}}
\end{figure}

Figure~\ref{fig:electronbackg} shows the averaged integral count rates (pad 2) from all configuration runs within the chosen background window as a function of the applied retardation voltage. From a straight line fit to the data, the slope ($m_\beta$) indicates a possible $U_{\text{AP}}$ dependence of the $\beta-$decay electron background giving $m_\beta = (0.32 \pm 1.81) \times 10^{-5}$ cps/V $< 1.81 \times 10^{-5}$ cps/V. The obtained upper limit can be compared to the measured retardation voltage-dependent background in the proton region stemming from the ionized rest gas peak (cf. Fig.~\ref{fig:backgroundretvolt} a), which caused a systematic change in the $a$ value of $-1.4$ \% (cf. Table~\ref{tab:relchanges}). For our estimation given here, it is sufficient to approximate its functional dependence by a straight line with $m_{\text{bg}} \approx 3 \times 10^{-4}$ cps/V. From this, one can deduce that the $\beta-$decay electrons in the proton region may affect $a$ by the negligibly small value
\begin{equation}
\label{eq:betarate}
\left| \frac{\Delta a}{a} \right| < \frac{m_\beta}{m_{\text{bg}}} \cdot \frac{6 \text{ cps}}{4 \text{ cps}} \cdot \left| \left(- 1.4 \; \% \right) \right| \approx 0.1 \; \% \; .
\end{equation}
A possible voltage dependence of the entire $\beta-$electron events above ADC channel 150 was also investigated. Again, no significant shift in the $a$ value could be deduced.

For negatively-charged particles, $U_{\text{AP}}$ is in fact an acceleration voltage. The phase space of the $\beta-$decay electrons in the flux tube ranging from the DV region to the detector is not affected by the depth of the potential well after the electrons have passed through. Above the vertical height $z \approx 1.8$ m of the $a$SPECT magnet (cf. Fig.~\ref{fig:aspectb}), only the magnetic mirror effect and the electrostatic potential of the upper E$\times$B drift and detector electrode are effective, which have an influence on the phase space acceptance. However, these experimental field settings were not changed during the entire beam time in 2013. Therefore, the number of $\beta-$electrons reaching the detector as well as their energy distribution and angle of impingement remain unaffected, in particular no dependence on the applied retardation voltage $U_{\text{AP}}$ is to be expected.

\section{Proton recoil spectrum}
\label{sec:prrec}

In the following we document the proton recoil spectrum used in the fit function, in which we largely refer to the paper of F. Gl\"uck et al. \cite{glu1995}. Recoil-order effects and radiative corrections are neglected as the purpose of this is only to estimate a systematic correction in our fit to config~2b due to a small unwanted beam polarization.  We mark this with an asterisk (*) in the respective expressions.  The differential proton recoil spectrum $W^*(T, a, c)$ in case of a finite neutron polarization, $P$, is given by
\begin{equation}
\label{eq:recoil}
W^*(T, a, c) = \omega_{\text{p}}^*(T, a) + P\cdot{}\omega_{\text{ps}}^*(T, a)\cdot{}\cos{}\vartheta
\end{equation}
where $\vartheta$ is the angle between neutron spin and proton momentum and $T$ is the kinetic energy of the proton. $c$ is the abbreviation for the expression $P\cdot{}\cos{}\vartheta$. The respective spin-dependent and spin-independent terms $\omega_{\text{ps}}^*(T, a)$ and $\omega_{\text{p}}^*(T, a)$ can be expressed as
\begin{equation}
\label{eq:b2}
\omega_{\text{ps}}^*(T, a) = +\frac{1}{8}\cdot{}\left(A+B\right)\cdot{}\left(F_{\text{max}}(T) - F_{\text{min}}(T)\right)
\end{equation}
and
\begin{equation}
\label{eq:b3}
\omega_{\text{p}}^*(T, a) = w_{\text{max}}(T, a) - w_{\text{min}}(T, a)
\end{equation}
with $\left(A + B\right) = -4\cdot{}\lambda/\left(1+3\lambda^2\right)$ or $\left(A + B\right) = \sqrt{1+2\cdot{}a-3\cdot{}a^2}$ using Eq.~(\ref{eq:lambda}).

\begin{figure}
\includegraphics[width=\linewidth]{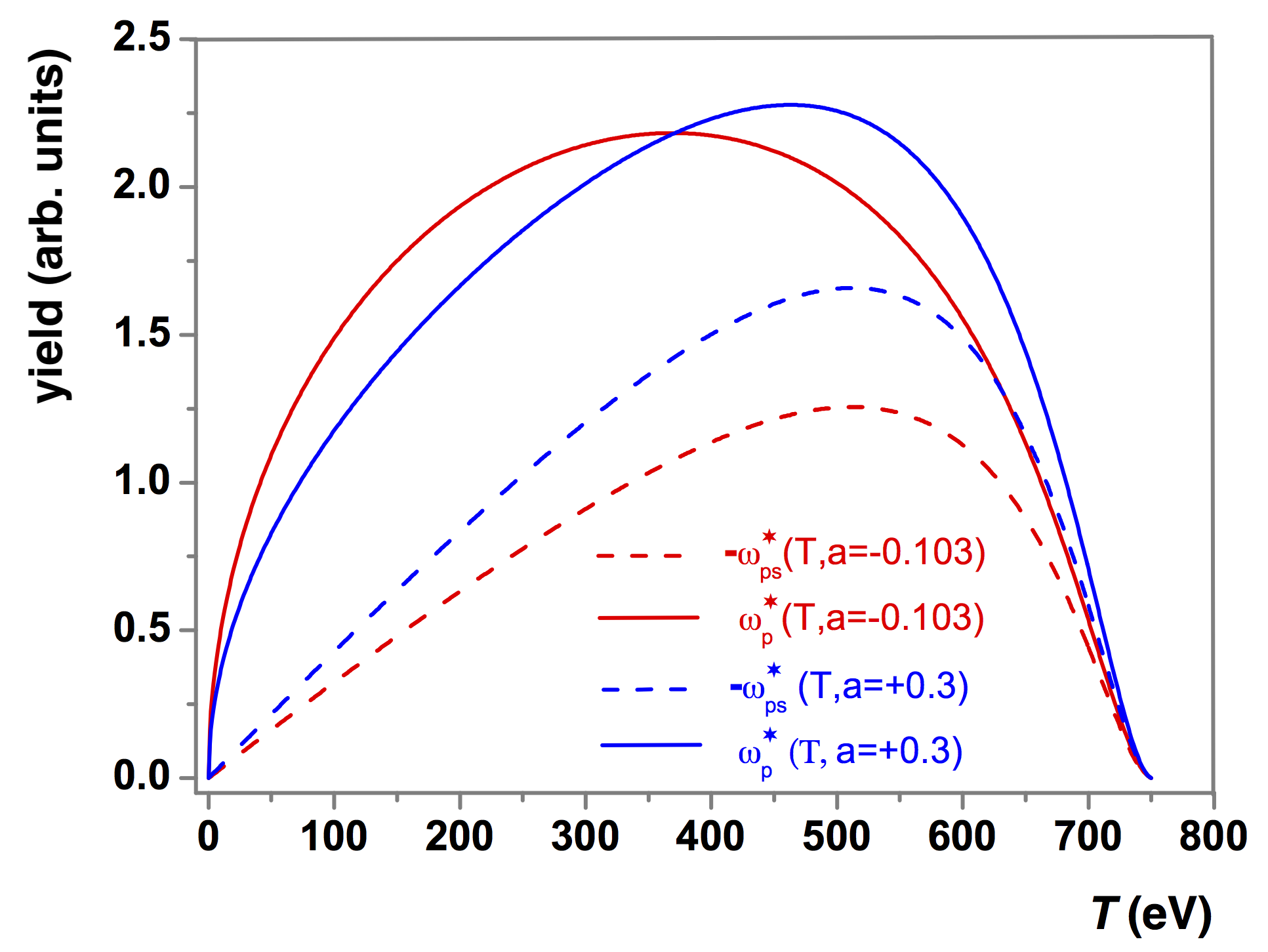}
\caption{Spin-independent $\omega_{\text{p}}^*(T, a)$ and the spin-dependent component $\omega_{\text{ps}}^*(T, a)$ of the differential proton recoil spectrum $W^*(T, a, c)$ from Eq.~(\ref{eq:recoil}) for $a_{\text{ref}} = -0.103$ and for an extreme value of $a = + 0.3$. Note that $\omega_{\text{ps}}(T, a)$ is negative, but for the yield we have $W^*(T, a, c) > 0$ for all $\cos{}\vartheta$ values since $|\omega_{\text{ps}}^*(T, a)| < |\omega_{\text{p}}^*(T, a)|$. The yield is given in (a.u.). To get prefactors for the absolute numbers, see \cite{glu1995}. \label{fig:wps}}
\end{figure}

By defining the following constants:
\begin{eqnarray}
\label{eq:b4}
\Delta &=& m_n - m_p = 1.293318\times{}10^6 \; \text{eV}, \nonumber \\
m_e &=& 0.5109989\times{}10^6 \; \text{eV}, \nonumber \\
m_n &=& 939.5654\times{}10^6 \; \text{eV}, \nonumber \\
T_m &=& \left(\Delta^2 - m_e^2\right)/\left(2\cdot{}m_n\right),
\end{eqnarray}
and further the $T$-dependent terms:
\begin{footnotesize}
\begin{eqnarray}
\label{eq:b5}
p&=& \sqrt{2\cdot{}(m_n-\Delta)\cdot{}T + T^2} \; \text{(proton momentum)}, \nonumber \\
E_{\text{min}}(T) &=& \frac{1}{2}\cdot{}\left(\Delta - T - p(T) + \frac{m_e^2}{\Delta - T - p(T)}\right), \nonumber \\
E_{\text{max}}(T) &=& \frac{1}{2}\cdot{}\left(\Delta - T + p(T) + \frac{m_e^2}{\Delta - T + p(T)}\right), \nonumber \\
x_{\text{min}}(T) &=& 2\cdot{}E_{\text{min}}(T) - \Delta , \nonumber \\
x_{\text{max}}(T) &=& 2\cdot{}E_{\text{max}}(T) - \Delta ,
\end{eqnarray}
\end{footnotesize}
we obtain for
\begin{eqnarray}
\label{eq:b6}
F_{\text{max(min)}} &=& \frac{\Delta{}\cdot{}x^3_{\text{max(min)}}}{3\cdot{}p} \nonumber \\
&-& \frac{m_e^2\cdot{}x^2_{\text{max(min)}}}{2\cdot{}p} \nonumber \\
&-& \Delta{}\cdot{}p\cdot{}x_{\text{max(min)}} ,
\end{eqnarray}
and
\begin{eqnarray}
\label{eq:b7}
w_{\text{max(min)}} &=& \frac{1}{2}\cdot{}\left(1+a\right)\cdot{}E^2_{\text{max(min)}} \nonumber \\
&\cdot&\left(\Delta - \frac{2}{3}\cdot{}E_{\text{max(min)}} \right) \nonumber \\
&+& a\cdot{}m_n\cdot{}E_{\text{max(min)}}\cdot{}\left(T-T_m\right) \; .
\end{eqnarray}

Figure~\ref{fig:wps} shows the differential spectra $\omega_{\text{p}}^*(T, a)$ and $\omega_{\text{ps}}^*(T, a)$ for $a = a_{\text{ref}} = - 0.103$ and $a = + 0.3$. As a test of these equations, we computed with Eqs. (\ref{eq:recoil}~-~\ref{eq:b7}) the integrated proton asymmetry $\alpha_p$ defined by Eq. (4.27) in Ref. \cite{glu1995}, using the $\lambda = -1.26$ value. We got $\alpha_p = 0.2402$; to compare with the $\alpha_p = 0.2404$ value in Ref. \cite{glu1995}, the small difference is due to the slightly different approximations of the two calculations.

\section{Likelihood profiling versus marginalization using Markov chain Monte Carlos}
\label{sec:markovmc}

We extract our result, the value and error of the $\beta-\bar{\nu}_e$ correlation coefficient $a$ by means of a global fit with $m = 68$ fit parameters. This vast number of fit parameters is necessary to precisely characterize our setup and different settings for the data runs and to get a handle on all systematic effects and their correlations concerning the determination of $a$. Beside many measured data sets representing integral proton energy spectra under different conditions, also Monte Carlo results are included in the global data set in order to determine parameters, \textit{e.g.}, below-threshold losses of the proton detector (cf. section~\ref{sec:lld}), which are otherwise not accessible. Because of this quite heterogeneous data set and the large number of fit parameters, two questions arise: How can we make sure that the overall likelihood is maximal for our final value of $a$ and is there a way to calculate the probability density function (PDF) of $a$ which results from our data set without any further assumption. Because of the statistical error of our data, the measured ones as well as the Monte Carlo calculations, are in good approximation normally distributed, $\chi^2$-fitting is the preferable method to obtain maximum likelihood, because the logarithm of the likelihood function $\mathcal{L}(\theta)$ is related to the $\chi^2(\theta)$ function:
\begin{equation}
\label{eq:normaldist}
\ln\left(\mathcal{L}(\theta)\right) = -\frac{1}{2}\chi^2(\theta), \; \; \text{if} \; \mathcal{L}(\theta) \; \text{normally distributed.}
\end{equation}
Once we had set up our $\chi^2$-fitter within Mathematica, we profit from the fact that Mathematica has already implemented sophisticated methods to find a global minimum, \textit{e.g.}, differential evolution \cite{storn1997} or stimulated annealing (\cite{press1992} and paragraph 11.4.1 of \cite{gregory2005}). We verified that all methods reached the same $\chi^2_{\text{min}}=\chi^2(\hat{\theta})$ for the same fit parameter set $\hat{\theta}$. Now we had to face the problem that calculating the partial derivatives $\partial{}\chi^2(\theta)/\partial{}\theta_i$ is not feasible for numerical reasons, but those are needed for the standard method to calculate the Fisher information matrix $I=M_c^{-1}$, which is the inverse of the correlation matrix $M_c$ (see Eqs. (11.8) and (11.9) of \cite{gregory2005}). From the definition of $I$ in combination with Eq.~(\ref{eq:normaldist}) we get from a Taylor expansion at $\hat{\theta}$:
\begin{eqnarray}
\ln\left(\mathcal{L}(\theta)\right) & \approx & \ln\left(\mathcal{L}(\hat{\theta})\right) - \frac{1}{2}\left(\theta - \hat{\theta}\right)I{}\left(\theta - \hat{\theta}\right) \nonumber \\
& \approx & -\frac{1}{2}\chi^2\left(\hat{\theta} \right) - \frac{1}{2}\left(\theta - \hat{\theta}\right)I_{\hat{\theta}}\left(\theta - \hat{\theta}\right) \nonumber \\
& \approx & -\frac{1}{2}\chi^2(\theta)
\end{eqnarray}
As an approximation of $I_{\hat{\theta}}$, we calculate $K_{1\sigma}$, a symmetrical $(m\times{}m)$ curvature matrix which defines the $m$-dimensional parabola representing $\chi^2(\theta)$ from $\chi^2$ values in the $1\sigma$-neighborhood of $\hat{\theta}$. To determine the $m^2/2 + m/2 = 2346$ parameters of $K_{1\sigma}$, $k = 9385$ values of $\chi^2(\theta_i)$ were calculated on a $m$-dimensional grid over the fit parameter space in the vicinity of $\hat{\theta}$ and finally this data set was fitted to a $m$-dimensional parabola defined by $K_{1\sigma}$ and $\hat{\theta}$. A measure how well the likelihood function is normally distributed (nd) is given by $\sigma^2_{\text{nd}}$, the average quadratic deviation of $\chi^2(\theta)$ from the parabolic shape
\begin{footnotesize}
\begin{equation}
\sigma^2_{\text{nd}}=\frac{1}{k}\sum^k_{i = 1}\left(\chi^2(\theta_i) - \chi^2(\hat{\theta}) - \left(\theta_i - \hat{\theta}\right)K_{1\sigma}\left(\theta_i - \hat{\theta}\right)\right)^2
\end{equation}
\end{footnotesize}

\begin{figure}
\includegraphics[width=\linewidth]{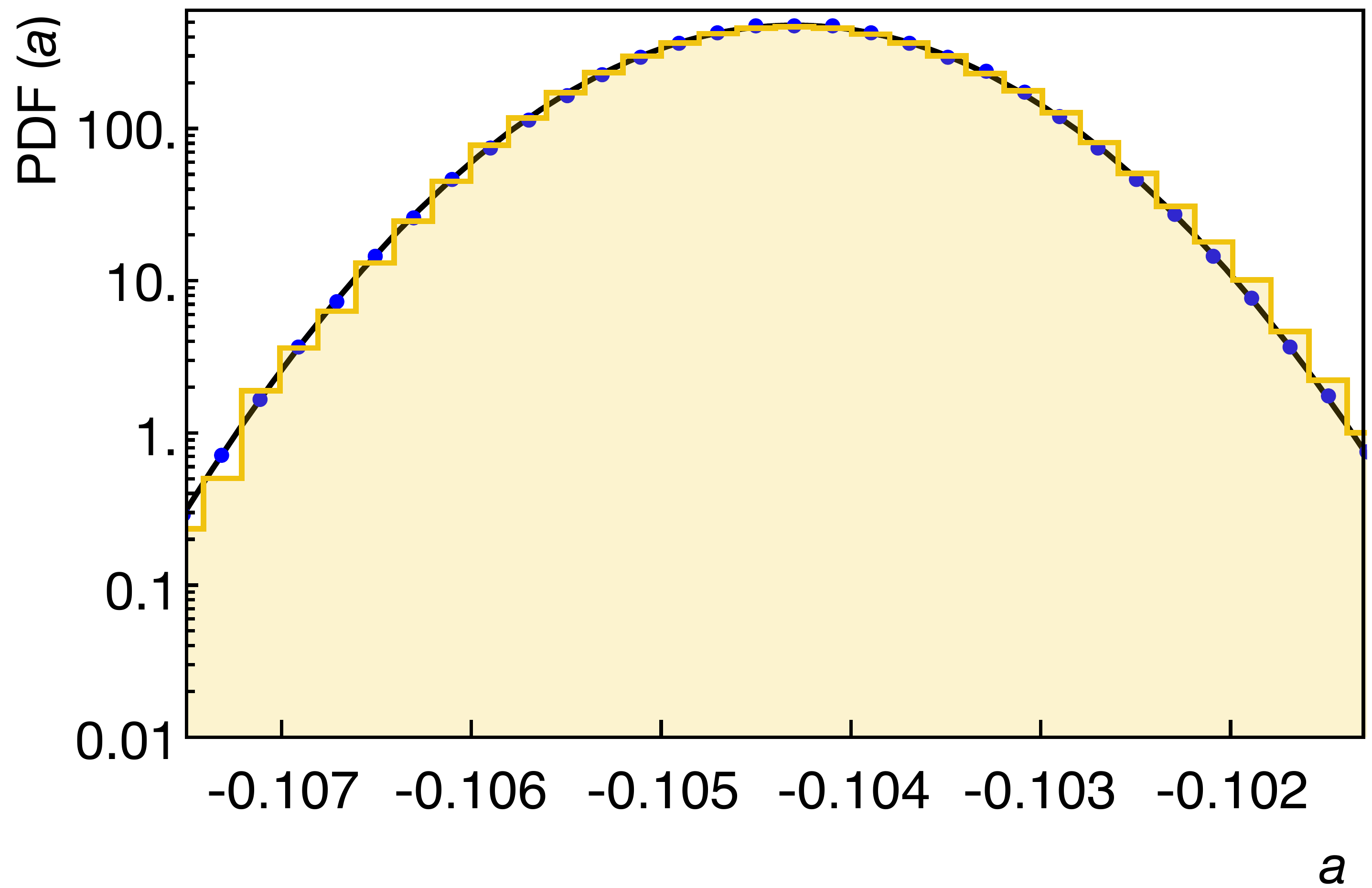}
\caption{PDF of $a$ determined by maximum likelihood profile (blue points), also called projection method, and by a histogram (yellow) of a MCMC data set which is also called marginalization method.
	%Solid line: Normal distribution using our final fit result for $a$.
	Normal distribution using our final fit result for $a$ (based on the scaled error, see Fig.~\ref{fig:ideogramcorr}).  \label{fig:pdfdist}}
\end{figure}

A normally distributed likelihood function results in $\sigma^2_{\text{nd}} = 0$, while $\sigma^2_{\text{nd}} < 10^{-4}$ indicates that on average, the relative deviation is less than 1~\% from the normal distribution, which we regard as a good approximation. We obtained $\sigma^2_{\text{nd}} = 6.8\times{}10^{-6}$, a result which justifies in retrospect our assumption regarding the normal distribution. The calculation of the PDF of $a$ shown as blue points in Fig.~\ref{fig:pdfdist} is the result of maximum likelihood profiling sometimes called projection method (paragraph 11.3.2 of \cite{gregory2005}). For further investigation of higher-order correlations between parameters, profiling can be used in principle but is out of reach with respect to computing time. Therefore, as an alternative to classical statistical approaches, we followed the Bayesian approach and performed Markov chain Monte Carlo (MCMC) calculations (chapter 12 of \cite{gregory2005}) by implementing the Metropolis-Hastings algorithm \cite{hastings1970, metropolis1953} in combination with an exponential sampler of the likelihood function. After the Marcov chain values have entered a high probability region (burn in phase), the phase space density of the Marcov chain values in our fit parameter space is proportional to the PDF of the likelihood function. Therefore, a multidimensional histogram of a MCMC data set shows directly the full picture of correlations without any assumptions regarding the PDFs of the parameters involved and/or the order of their correlations. This reasoning also holds for one dimension as shown in Fig. \ref{fig:pdfdist}, in which the PDF of $a$ is derived from a histogram of a MCMC data set in comparison to the maximum likelihood profile. Further we derived the correlation matrix of the likelihood function at maximum from the correlation analysis of a MCMC data set. Within the statistical error of 2~\%, the coefficients are identical to the ones of the classical approach, which also proves that our likelihood function is normally distributed in all its parameters and therefore the classical and Bayesian approach lead to the same result. Furthermore, we used the MCMC method in combination with heating, \textit{i.e.,} incorporating a temperature parameter (temperature factor of 1000, see paragraph 12.4 of \cite{gregory2005}) followed by a slow annealing process. The functional dependence of the sampler from the temperature steps is similar in case of annealing in the MCMC context as in the case of classical $\chi^2$ minimization with the stimulated annealing method. Nevertheless, how the sampler result is used is quite different and therefore these methods are unequal and the fact that we got the same global maximum and the same variance of the likelihood function is an independent verification.

\end{appendix}

\bibliography{ms.bib}

\end{document}